\documentclass[a4paper,11pt]{article}
\pdfoutput=1 
\usepackage{jcappub} 

\usepackage{color}
\usepackage{amsmath}
\usepackage{amsfonts}
\usepackage{graphicx}
\usepackage{amssymb}
\usepackage{float}
\usepackage{comment}
\usepackage[dvipsnames]{xcolor}
\definecolor{refs}{RGB}{245,156,74}

\newcommand{\dd}{{\rm d}}

\newcommand{\Lag}{\mathcal{L}}
\newcommand{\ham}{\mathcal{H}}
\newcommand{\mK}{\mathcal{K}}
\newcommand{\mY}{\mathcal{Y}}
\newcommand{\mS}{\mathcal{S}}
\newcommand{\mQ}{\mathcal{Q}}
\newcommand{\mI}{\mathcal{I}}

\newcommand{\ssl}{\mathfrak{sl}}

\newcommand{\Yb}{\bar{Y}}
\newcommand{\cP}{c_{\rm P}}
\newcommand{\cA}{c_{\text{A}}}
\newcommand{\mP}{m_{\rm P}}
\newcommand{\mA}{m_{\text{A}}}

\newcommand{\Ap}{A^+}
\newcommand{\Am}{A^-}
\newcommand{\ap}{a^+}
\newcommand{\am}{a^-}

\newcommand{\Bp}{B^+}
\newcommand{\Bm}{B^-}
\newcommand{\bp}{b^+}
\newcommand{\bm}{b^-}

\newcommand{\tA}{C}
\newcommand{\tB}{D}

\newcommand{\rs}{r_{\rm s}}

\newcommand{\be}{\begin{equation}}
\newcommand{\ee}{\end{equation}}
\newcommand{\bea}{\begin{eqnarray}}
\newcommand{\eea}{\end{eqnarray}}

\renewcommand{\bf}[1]{{\textbf{#1}}}

\newcommand{\Fd}{\tilde{F}}

\newcommand{\LL}{\ell(\ell+1)}

\begin{document}

\rightline{CERN-TH-2022-204}

\title{Polarisability and magnetisation of electrically $K$-mouflaged objects: \\
the  Born-Infeld ModMax case study}

\author[a]{Jose Beltr\'an Jim\'enez,}
\author[a]{Dario Bettoni,}
\author[b,c]{Philippe Brax.}

\affiliation[a]{Departamento de F\'isica Fundamental and IUFFyM, Universidad de Salamanca, E-37008 Salamanca, Spain}
\affiliation[b]{Institut de Physique Th\'eorique, Universit\'e Paris-Saclay, CEA, CNRS, F-91191 Gif-sur-Yvette Cedex, France.}
\affiliation[c]{CERN, Theoretical Physics Department, Geneva, Switzerland.}

\emailAdd{jose.beltran@usal.es}
\emailAdd{bettoni@usal.es}
\emailAdd{philippe.brax@ipht.fr}

\abstract{We consider a family of non-linear theories of electromagnetism that interpolate between Born-Infeld  at small distances and the recently introduced ModMax at large distances. These models are duality invariant and feature a $K-$mouflage screening in the Born-Infeld regime. We focus on computing the static perturbations around a point-like screened charge in terms of two decoupled scalar potentials describing the polar and the axial sectors respectively. Duality invariance imposes that the propagation speed of the odd perturbations goes to zero as fast as the effective screened charge of the object, potentially leading to strong coupling and an obstruction to the viability of the EFT below the screened radius. We then consider the linear response to external fields and  compute the electric polarisability and the magnetic susceptibility. Imposing regularity of the perturbations at the position of the particle, we  find that the polarisability for the odd multipoles vanishes whilst for the magnetisation  Born-Infeld emerges as  the only theory with vanishing susceptibility for even multipoles. The perturbation equations factorise in terms of ladder operators connecting  different multipoles. There are  two such ladder structures for the even sector: one that acts as an automorphism between the first four multipoles and another one that connects multipoles separated by four units. When requiring a similar ladder structure for the odd sector, Born-Infeld  arises again as the unique theory. We use this ladder structure to relate the vanishing of the polarisability and the susceptibility to the values of  conserved charges. Finally  the perturbation equations correspond to a supersymmetric quantum mechanical system such that the polar sector can be described in terms of Schr\"odinger's equations with four generalised hyperbolic P\"osch-Teller potentials whose eigenfunctions are in correspondence with the multipoles.}

\date{\today}

\date{\today}
\maketitle
\newpage
\section{Introduction}

Non-linear extensions of Maxwell's electromagnetism have a long history and they have been extensively studied (see e.g. \cite{Boillat:1970gw,Plebanski:1970zz,Sorokin:2021tge}). Paradigmatic examples of non-linear electrodynamics are provided by the old idea of Born and Infeld \cite{Born:1933pep,Born:1934gh} in an attempt to regularise the self-energy of the electron\footnote{These ideas have also been considered within gravity theories as an attempt to regularise black hole or cosmological singularities (see \cite{BeltranJimenez:2017doy} for a review on these attempts.)} or the Euler-Heisenberg Lagrangian \cite{Heisenberg:1936nmg} obtained as the 1-loop effective action resulting from integrating out a massive fermion. High energy physics such as string theory could also be at the origin of non-linear electrodynamics in the low energy manifestation of open strings around D-branes \cite{Fradkin:1985qd,Gibbons:2001gy}. 
These theories have been used in many applications for cosmology, black hole physics, etc. A distinctive property of these theories is that, under general assumptions, the electric interaction between charged particles is modified at short distances so that the electric attraction or repulsion is suppressed with respect to the Maxwellian one. This is nothing but an example of a screening mechanism \`a la $K$-mouflage or kinetic screening \cite{Babichev:2009ee}. In fact, although not under this name, these screening mechanisms largely precede those for scalar fields that have been exhaustively used for cosmological applications within the context of dark energy and modified gravity theories in recent years \cite{Joyce:2014kja,Brax:2021wcv}. A common drawback of the kinetic screening for scalar fields is that screened solutions seem to come hand in hand with the superluminality of the perturbations \cite{Joyce:2014kja}. Interestingly, the analogous screening within non-linear electromagnetism exhibits the opposite behaviour, i.e., the condition to have screening is nicely compatible with the absence of superluminal propagation. It is important to clarify that, although we talk about non-linear electromagnetism, we do not necessarily refer to the usual electromagnetic interaction of the standard model, i.e., the non-linear theories that we consider could be some dark electromagnetic sector. This approach has been pursued recently where the dark matter component of the Universe is provided with a dark electromagnetic force featuring this screening \cite{BeltranJimenez:2020tsl,BeltranJimenez:2020csl}. This could have interesting cosmological applications for structure formation and could even alleviate the Hubble tension \cite{BeltranJimenez:2021imo}.\footnote{See also \cite{Kaloper:2009nc} for an earlier proposal of a net dark charge albeit not involving any screening mechanism.}

Amongst all electromagnetic theories, Maxwell's is very special, not only because it is a linear theory, but because it features two interesting symmetries, namely: conformal and duality invariance. When non-linearly deforming Maxwell's theory, these symmetries are generally broken, thus losing some interesting properties of Maxwell's theory. Although this might not be detrimental\footnote{The breaking of these symmetries might even be desirable in some cases. For instance, the conformal invariance of Maxwell electromagnetism prevents the possibility of generating primordial magnetic fields during inflation, so one would want to break such a symmetry during inflation to be able to generate primordial magnetic fields. On the other hand, duality invariance is associated to the conservation of the helicity so the breaking of duality invariance is desirable in order to generate chirality.}, it is theoretically appealing to uncover the existence of non-linear electromagnetism theories that preserve one or both of these symmetries. It has been known for a long time that there is a family of non-linear electromagnetic theories that share the property of duality invariance \cite{Gibbons:1995cv}, among which we can find the Born-Infeld one. In a recent work, it has also been shown that there is one non-linear electromagnetism that is able to preserve both symmetries, duality and conformal invariance, which has been dubbed ModMax electromagnetism \cite{Bandos:2020jsw}. Its properties have been studied and some extensions have also been worked out \cite{Bandos:2020hgy,2021PhLB..82236633K,Babaei-Aghbolagh:2022uij,Lechner:2022qhb}.

In this work we will focus on a family of theories based on these two remarkable theories: the Born-Infeldised ModMax models \cite{Bandos:2021rqy}. The Born-Infeld and ModMax theories are limits of these models. They always preserve duality invariance and restore conformal invariance in the limiting case where the ModMax theories are selected. 
We will discuss the presence or absence of screening mechanisms within this family of theories as well as its efficiency, having in mind possible applications for cosmology and astrophysics. Once the existence of such screened solutions is assessed, we will compute the quadratic action governing the dynamics of the perturbations around such background solutions to study their behaviour. In particular, we obtain the propagation speeds and masses that are relevant for potential problems of superluminalities, strong coupling or the formation bound states. Furthermore, we will compute the linear response of screened objects to external stimuli. One interesting property that we uncover is that the polarisability or susceptibility can vanish depending on the underlying model of non-linear electromagnetism. For instance for the Born-Infeldised ModMax models, which interpolate between the ModMax theories and Born-Infeld, we find that the polarisability in the polar (even) sector vanishes for odd multipoles. This extends to the axial case in the Born-Infeld case only. We notice that this vanishing is related to the appearance of  conserved charges  in  the perturbation equations although the relationship does not seem to be one to one. On the other hand, the vanishing of the polarisability or the susceptibility is related to the existence of ladder operators which allow to construct the modes of the perturbation equations from the monopole and dipole solutions. In particular, we find that the states constructed by the raising operators are regular and satisfy the boundary conditions for the electric or magnetic fields. In the cases where the polarisabilities or the susceptibilities do not vanish, the tower of states cannot be constructed with appropriate regularity properties.  

This paper is arranged as follows. In a first section \ref{sec:K} we introduce $K$-mouflage in non-linear electromagnetism. Then we focus on the perturbations around an electric spherical solution in section  \ref{sec:pert}. This allows us to calculate the linear response to an external field at infinity in section \ref{sec:LinearResponse}. We notice that the polarisability and the susceptibility can vanish for the Born-Infeldised theories for odd multipoles and relate the existence of a ladder of conserved charges ultimately related to an underlying ladder structure in the perturbation equations and the existence of conserved charges, see section \ref{sec:ladder}. We then discuss our results and conclude in section \ref{sec:conc}. Two technical appendices \ref{app:polardeco} and \ref{app:xlad}, where the second order action is fully derived and where the ladder structure is considered from an alternative point of view complement our discussion. 

\vspace{0.2cm}
\bf{Conventions:} The field strength of the gauge field is $F_{\mu\nu}=\partial_\mu A_\nu-\partial_\nu A_\mu$. The dual is defined as $\Fd^{\mu\nu}=\frac12 \epsilon^{\mu\nu\alpha\beta} F_{\mu\nu}$. The electric and magnetic components are $E_i=F_{0i}$ and $B_i=\Fd_{0i}$. We will work with mostly plus signature for the metric.\\

\newpage
\section{Vector $K$-mouflage}
\label{sec:K}
We will start by reviewing some generalities about non-linear electromagnetism and the related  screening mechanisms. We will then turn to the particular non-linear electromagnetic theories that will be the focus of this work.

\subsection{Generalities}

Let us consider a theory for an Abelian gauge spin-1 field $A_\mu$ described by the Lagrangian
\be
\Lag=\mK(Y,Z)\,,
\ee
with $Y=-\frac14 F_{\mu\nu} F^{\mu\nu}$ and $Z=-\frac14 F_{\mu\nu} \Fd^{\mu\nu}$ the two independent Lorentz invariants which can be expressed in terms of the electric and magnetic components as $Y=\frac12(\vec{E}^2-\vec{B}^2)$ and $Z=\vec{E}\cdot \vec{B}$ \footnote{In this work we will make a pure classical analysis. In this respect, the non-linear dependence on $Y$ and $Z$ will come in with some scale $\Lambda$ that controls the classical non-linearities. Since quantum corrections are expected to enter with derivatives of the field strength $\partial^\ell F^n$ \cite{Brax:2016jjt} , there is a regime where classical non-linearities can be relevant within the regime of validity of the EFT. In this regime we can have $F_{\mu\nu}\sim\Lambda$ as long as $\partial_\mu\ll\Lambda$ \cite{BeltranJimenez:2020tsl}.}. 
The field equations for the gauge field are
\be
\nabla_\nu\Big(\mK_Y F^{\mu\nu}+\mK_Z \Fd^{\mu\nu}\Big)=J^\mu\,,
\ee
where we have added the current $J^\mu$ as a source term. In addition to the dynamical equations, we have the corresponding Bianchi identities (satisfied off-shell) derived from gauge invariance $\nabla_\mu \tilde{F}^{\mu\nu}=0$. 
Let us consider now a static $J^\mu$ with compact support, i.e., $J^\mu=(\rho,\vec{0})$, so that the equations reduce to 
\bea
\nabla\cdot\Big(\mK_Y\vec{E}+\mK_Z\vec{B}\Big)=\rho\,,\\
\nabla\times\Big(\mK_Y\vec{B}-\mK_Z\vec{E}\Big)=0\,.
\label{eq:eomEB}
\eea
As  well-known, these equations can be written as Maxwell's equations inside a medium with an electric displacement $\vec{D}$ and a magnetic intensity $\vec{H}$ given by
\bea
\vec{D}=\frac{\partial \mK}{\partial\vec{E}}=\mK_Y\vec{E}+\mK_Z\vec{B}\,,\\
\vec{H}=-\frac{\partial \mK}{\partial\vec{B}}=\mK_Y\vec{B}-\mK_Z\vec{E}\,,
\eea
so the equations \eqref{eq:eomEB} can be written as:
\bea
\nabla\cdot\vec{D}=\rho\,,\\
\nabla\times\vec{H}=0\,.
\label{eq:eomDH}
\eea
From these equations, we can see that a static source could, in principle, also generate a magnetic field\footnote{At a more speculative level, one could also generate electric fields from a purely magnetic monopole without being a dyon.}. It is not difficult to see, however, that assuming a vanishing magnetic field is consistent if we require parity invariance since this imposes a $\mathbb{Z}_2$ symmetry with respect to $Z$, i.e., $\mK$ can only depend on $Z^2$. In that case $\mK_Z=2Z\partial \mK/\partial Z^2$ that vanishes identically for $\vec{B}=0$. For this purely electric configuration the equations simplify to
\be
\nabla\cdot\Big(\mK_Y\vec{E}\Big)=\rho\,.
\label{eq:Erho}
\ee
In a spherically symmetric situation $\rho=\rho(r)$, we can use Gauss' theorem to integrate this equation as
\be
\mK_Y\vec{E}=\frac{q}{4\pi r^2}\hat{r}\,,
\ee
with $q=\int\rho \dd^3x=4\pi\int \rho(r) r^2\dd r$ the total charge. Now the screening is easy to understand. If $\mK$ is an analytic function of $Y$ such that $\mK(Y)\sim Y$ at large distances $r\to\infty$ ($Y\rightarrow 0$) we have that
\be
E\simeq\frac{q}{4\pi r^2}\,\quad\quad r\rightarrow \infty\,,
\ee
i.e., the Maxwellian result. As we approach the object, the non-linear terms become more relevant and deviations with respect to the $1/r^2$ behaviour are expected. If, for the sake of simplicity, we assume that $\mK=Y(1+Y^n/\Lambda^{4n})$ deep inside the non-linear region, we can introduce the screening radius defined as
\be
\rs^4=(n+1)^{1/n}\frac{q^2}{32\pi^2\Lambda^4}
\ee
and the electric field in this region acquires the following profile
\be
E\simeq\left(\frac{r}{\rs}\right)^{\frac{4n}{2n+1}}\frac{q}{4\pi r^2}\,,
\ee
where we clearly see the suppression factor for $r\ll \rs$, provided $n>1/2$. The consistency of the EFT relies on the fact that $\rs\gg\Lambda^{-1}$ since the quantum corrections are expected to become important at the scale $r_{\rm QC}\sim\Lambda^{-1}$.
The paradigmatic behaviour of Born-Infeld electromagnetism (see Sec. \ref{Sec:MMBItheory}) where the electric field becomes constant below $\rs$ is recovered in the limit $n\rightarrow\infty$. In general, there can be several branches of solutions and it is important to guarantee that the asymptotically Coulombian branch at $r\rightarrow \infty$ can be continuously connected with the screened branch below the screening radius. 

It is worth noticing that it is possible to solve the problem for an arbitrary configuration of static charges in which case the source is given by
\be
\rho=\sum_aq_a\delta(\vec{r}-\vec{r}_a)\,.
\ee
For this configuration we can integrate \eqref{eq:Erho} to obtain
\be
\mK_Y(\vec{E}^2)\vec{E}=\frac{1}{4\pi}\sum_a\frac{q_a}{\vert\vec{r}-\vec{r}_a\vert^3} \left(\vec{r}-\vec{r}_a\right)\,,
\label{eq:exactsolnonlinear}
\ee
that can be used to obtain the electric field, provided this equation can be inverted. This expression can be used to study the effect of nearby charges on the screening of a given object, although we will not delve into this interesting subject here. Instead, we will now proceed to introduce the family of non-linear electromagnetisms that we will study.

\subsection{The Born-Infeldised ModMax theory}
\label{Sec:MMBItheory}

In this work we will consider the class of non-linear electromagnetic theories described by the following Lagrangian:
\be\label{eq:BIMOMA_Lag}
\mK_{\rm MMBI} = 
 \Lambda^4\left[1 - \sqrt{
   1 - \frac{2}{\Lambda^4}  \left(\cosh\gamma\; Y + \sinh\gamma\; \sqrt{Y^2 + Z^2}\right) - \frac{1}{\Lambda^8}
    Z^2}\;\right],
\ee
where $\gamma$ is a dimensionless parameter and $\Lambda$ the scale of non-linearities. This theory is a generalisation of Born-Infeld theory that was obtained in \cite{Bandos:2020jsw}, although only in its Hamiltonian form. The Lagrangian was explicitly given in \cite{Bandos:2020hgy}. The above Lagrangian has two relevant regimes, namely:
\begin{itemize}
    \item $\gamma\ll1$. In this regime, the Lagrangian reduces to the usual Born-Infeld theory \cite{Born:1933pep,Born:1934gh}:
\be
\mK_{\rm BI}=\Lambda^4\left(1-\sqrt{1-\frac{2Y}{\Lambda^4}-\frac{Z^2}{\Lambda^8}}\right)\,.
\label{eq:BIlagrangian}
\ee
   This theory is duality invariant and has a number of interesting physical properties (absence of birefringence, no shock waves, etc.) that make it the only exceptional non-linear electromagnetism.

    \item $\Lambda\rightarrow\infty$. In this regime, the Lagrangian reads
    \be\label{eq:modmax_lag}
\mK_{\rm MM}=\cosh\gamma\, Y+\sinh\gamma\,\sqrt{Y^2+Z^2}\,,
\ee
    that has been shown to arise as the only non-linear electromagnetism sharing the same symmetries as Maxwell's theory, namely: conformal and duality invariance  \cite{Bandos:2020jsw}. This Lagrangian can be obtained as a $T\bar{T}$ deformation of the Maxwell Lagrangian \cite{Babaei-Aghbolagh:2022uij}.
\end{itemize}
Although the conformal invariance in \eqref{eq:BIMOMA_Lag} only arises as an approximate symmetry in the ModMax regime with $\gamma\ll1$, the duality invariance remains an exact symmetry and we will see that this has important consequences for the perturbations. The breaking of conformal invariance is important to obtain solutions with screening where the screening radius is related to $\Lambda^{-1}$ as we will see in section \ref{sec:conformal}. The solution around a spherically symmetric object of charge $Q$ is 
\be
E=\frac{e^{-\gamma/2}\Lambda^2}{\sqrt{1+x^4}}
\ee
where we have introduced the radial variable $x\equiv r/\rs$ with the following definition of the screening radius:
\be\label{eq:rlambdaBIMOMA}
\rs\equiv\frac{e^{-\gamma/4}\sqrt{Q}}{\Lambda}.
\ee
At small distances we have
\be
E(x\ll1)\simeq e^{-2\gamma}\Lambda^2
\ee
which is the typical behaviour in Born-Infeld theory where the electric field saturates to a constant value. The ModMax correction appears as the $\gamma-$re-dressing of the saturated electric field. On the other hand, the asymptotic behaviour is
\be
E(x\gg1)\simeq \frac{e^{-\gamma/2}\Lambda^2}{x^2}
\ee
that coincides with the ModMax solution.
The solution then interpolates between the ModMax behaviour at large distances and a Born-Infeld solution at small distances, in both cases with a $\gamma-$redressing. Notice that the ModMax regime coincides with the usual Maxwellian behaviour (up to the $\gamma-$redressing), which is a consequence of the conformal invariance. Thus, in the pure ModMax theory, there is a {\it global} screening determined by $\gamma$. As mentioned above, the breaking of conformal invariance introduced by the Born-Infeldisation of the ModMax Lagrangian  allows one to screen the small distances as compared to the large distances.

\begin{figure*}[ht]
\centering
\includegraphics[width=0.33\linewidth]{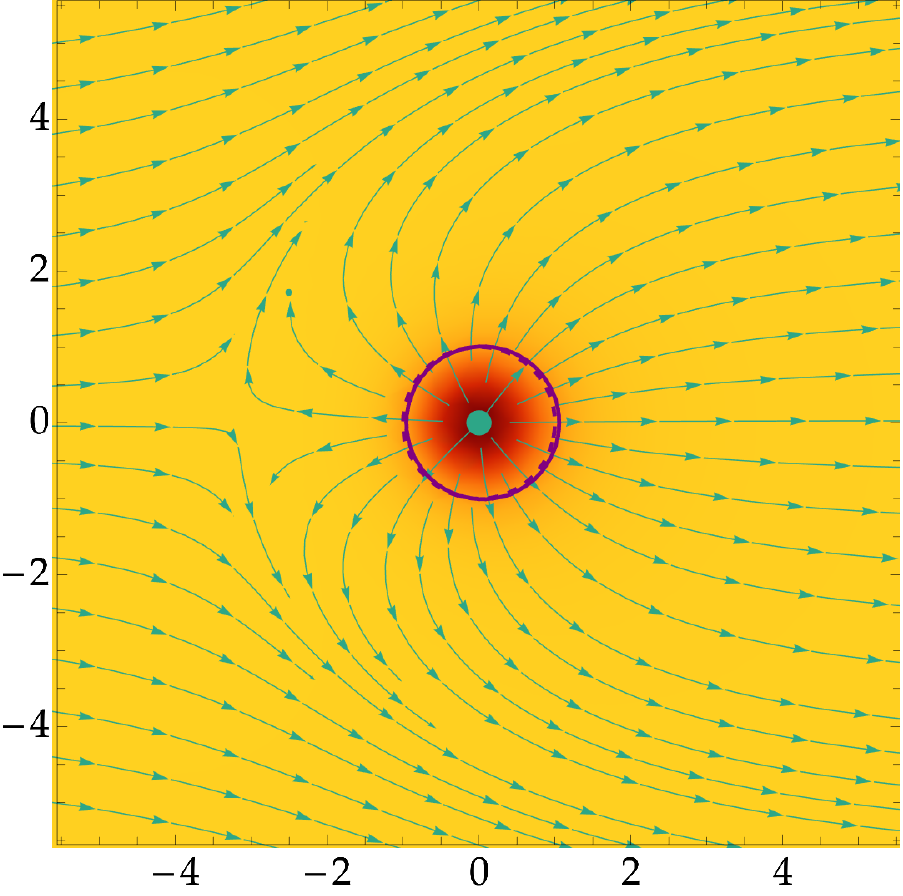}\includegraphics[width=0.33\linewidth]{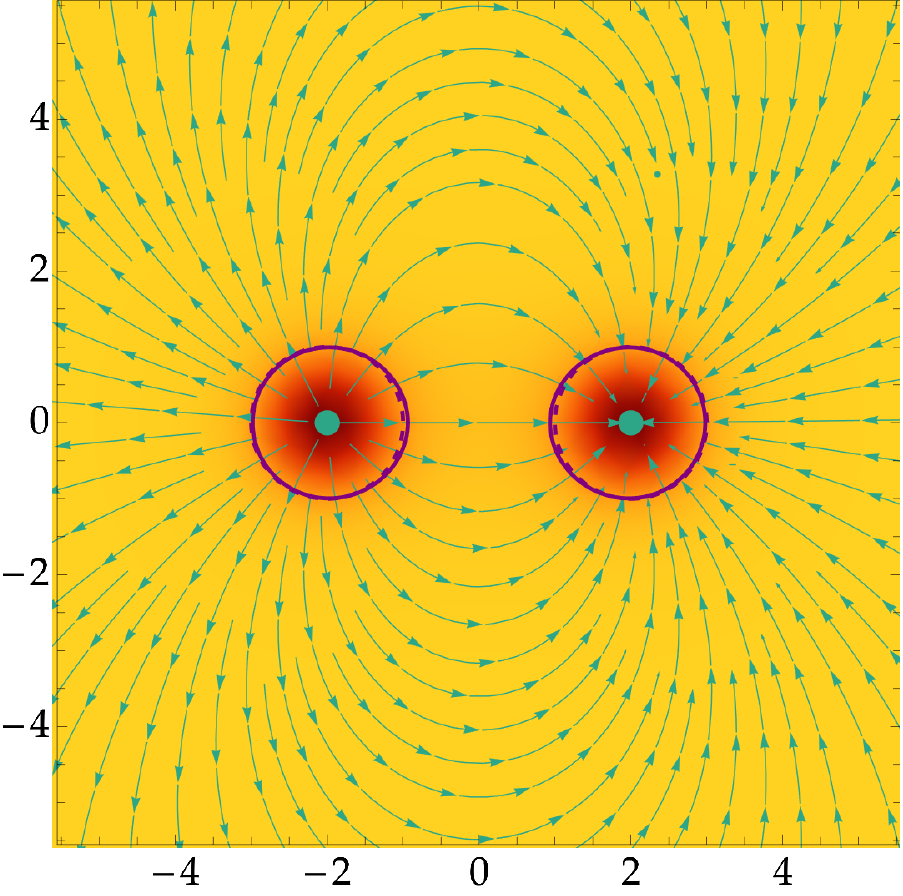}\includegraphics[width=0.33\linewidth]{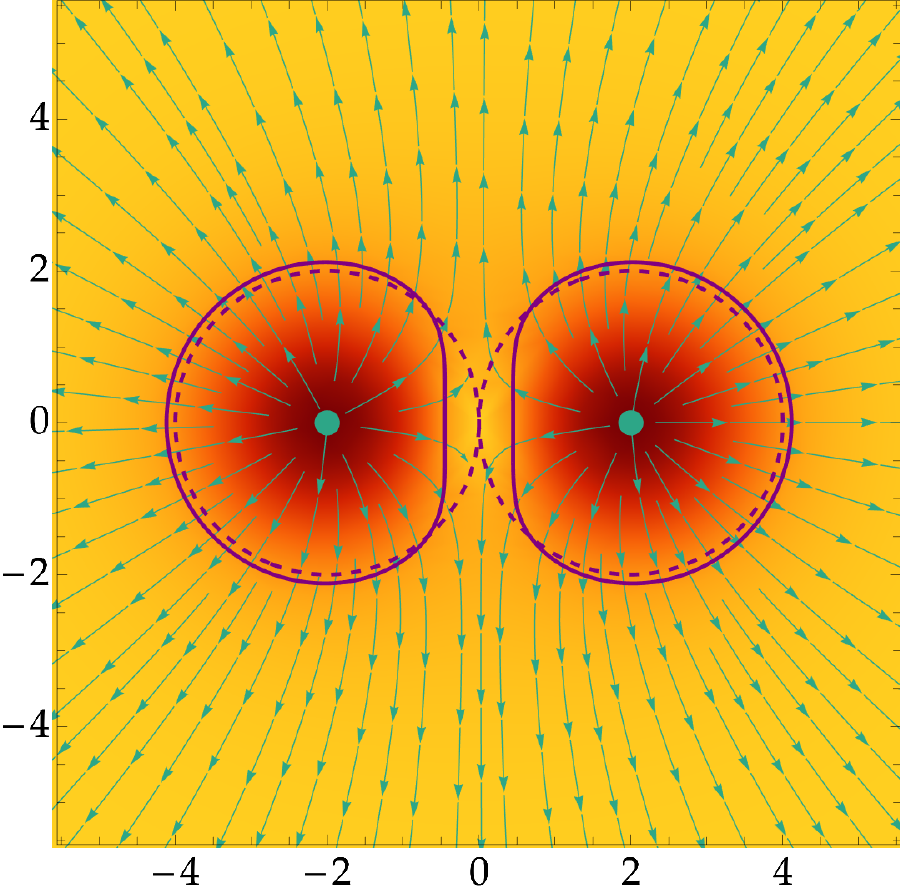}
\includegraphics[width=0.33\linewidth]{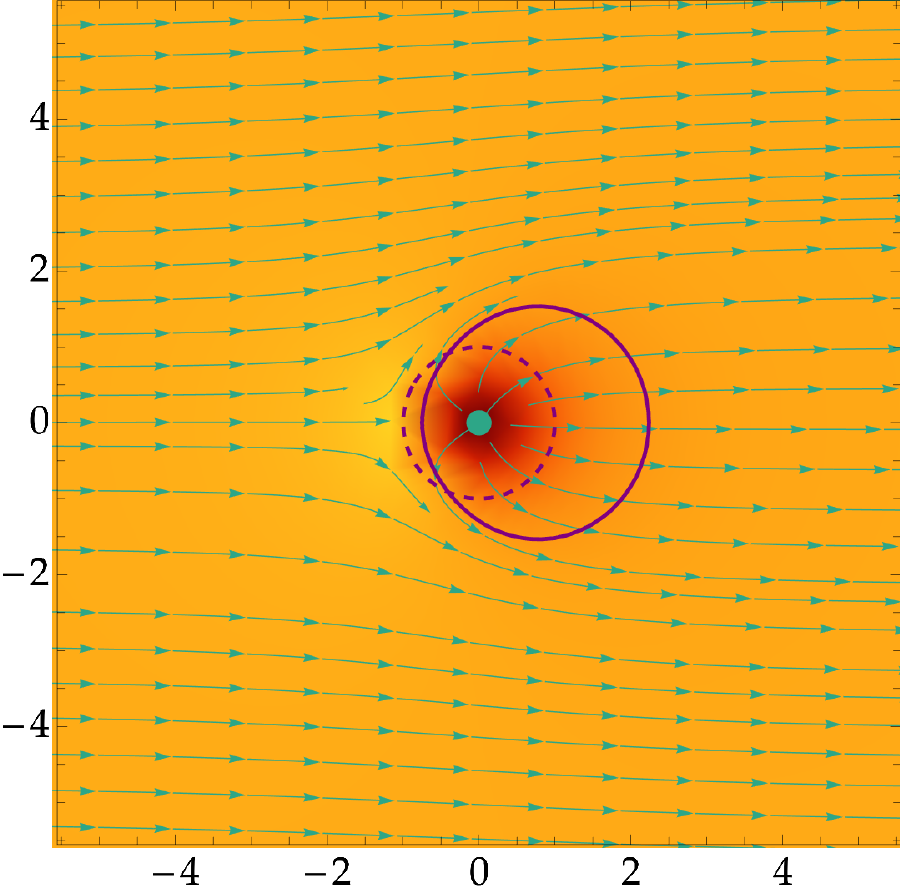}\includegraphics[width=0.33\linewidth]{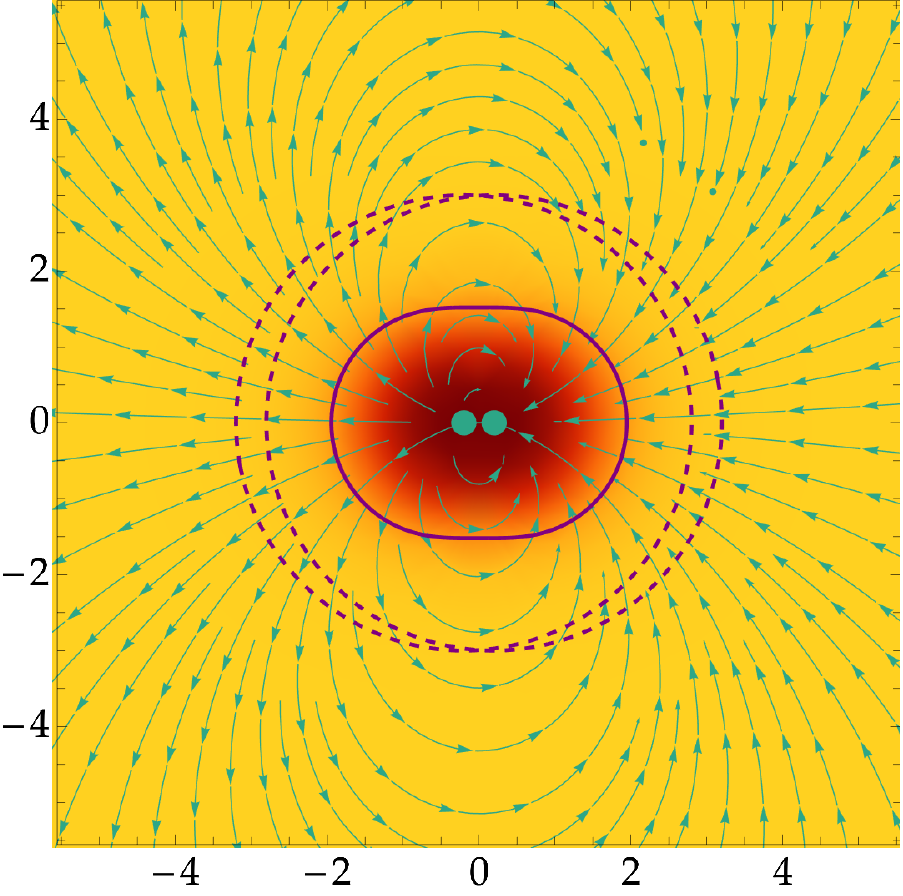}\includegraphics[width=0.33\linewidth]{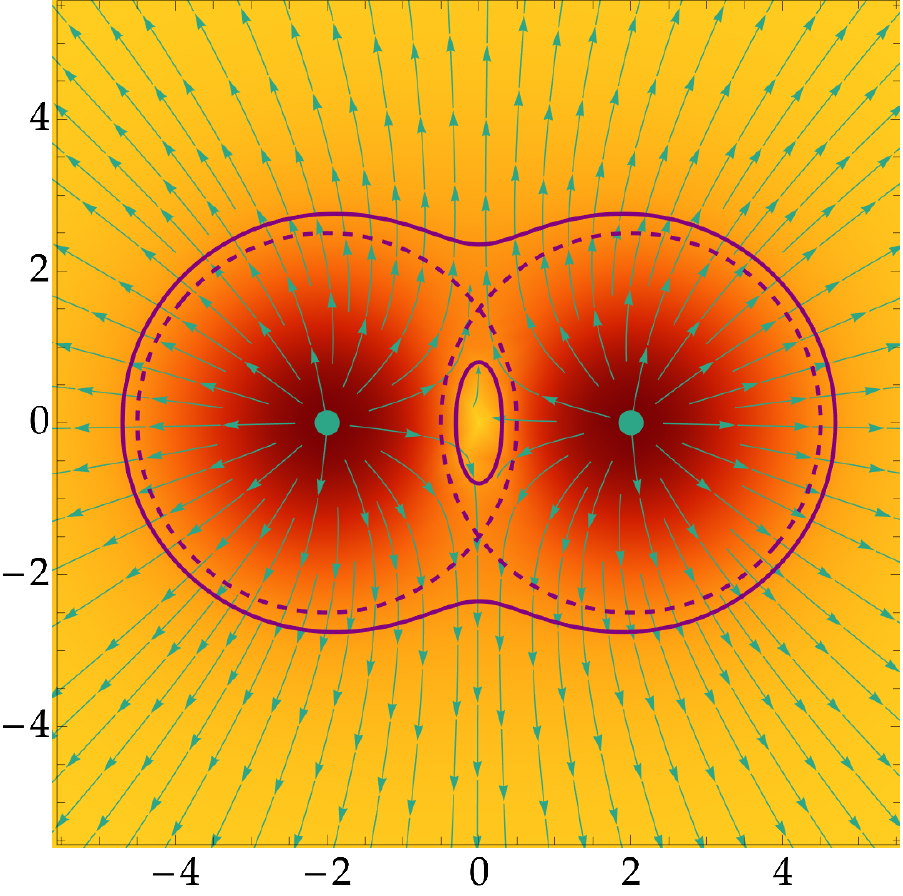}
\hspace{1.5cm}\includegraphics[width=0.33\linewidth]{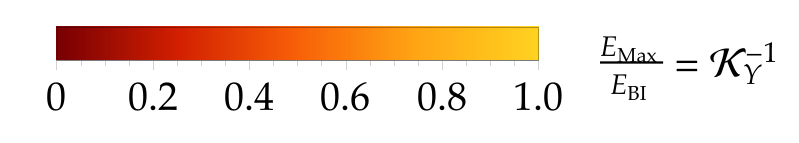}
\caption{Electric field created by several charge configurations for Born-Infeld electromagnetism. The colour represents the screening factor with the colour-code shown in the legend. The dashed-purple contours denote the particle screening radius and the solid purple contours delimitate the screening region inside of which the intensity of the Maxwellian electric field is larger than that of the Born-Infeld theory. {\bf{Left panels}}: Electric field created by a point particle in a small (upper) and large (lower) external constant field.  {\bf{Middle panels}}: a dipole with the charges outside (upper) and inside (lower) their respective screening radii. {\bf{Right panels}}: Two alike charges outside (upper) and inside (lower) their respective screening radii. We see how the repulsion between the charges deforms the screening region and even its topology changes when their screening radii overlap. This would open a possibility to probe this type of interaction in very specific positions between alike charges. For instance, in the cosmological scenario explored in \cite{BeltranJimenez:2020csl,BeltranJimenez:2020tsl,BeltranJimenez:2021imo}, this effect could be tested between galaxies or galaxy clusters dominated by charged dark matter.}
\label{Fig:electricfields}
\end{figure*}

An interesting property of the Born-Infeldised ModMax theory is that we can straightforwardly invert
\eqref{eq:exactsolnonlinear} and obtain the solution for a distribution of charges as
\be
\vec{E}=\frac{e^{-\gamma}\vec{D}}{\sqrt{1+e^{-\gamma}\vec{D}^2/\Lambda^4}}
\ee
with
\be
\vec{D}=\frac{1}{4\pi}\sum_a\frac{q_a}{\vert\vec{r}-\vec{r}_a\vert^3} \left(\vec{r}-\vec{r}_a\right).
\ee
Furthermore, for this theory there is only one branch (unlike generic non-linear electromagnetisms where several branches can exist). Again, we see that in regions where $\vec{D}^2\ll \Lambda^4$ we recover the ModMax regime, while the regions where $\vec{D}$ is not small (e.g. near any of the charges), the electric field saturates. In Fig. \ref{Fig:electricfields} we have plotted the electric field for several configurations of charges and for the pure Born-Infeld theory ($\gamma=0$). This exact solution makes an ideal starting point to test several phenomenological and observational consequences of these non-linear electromagnetism. We leave this route for future work and  will now proceed to the main goal of this paper.

\section{Perturbations}
\label{sec:pert}

A natural and pertinent question to ask is how perturbations behave around the screened solutions. This will allow us to analyse their reliability and physical relevance. We will see that, unlike the usual $K$-mouflage models for scalar fields where the screened branch leads to superluminal propagation, for the spin-1 fields the screen branch precisely guarantees subluminal propagation so that it avoids the usual obstructions for a standard Wilsonian local and Lorentz invariant UV completion. This is not very surprising since we know that Born-Infeld electrodynamics admits a UV completion in string theory. In this respect, screening based on spin-1 fields seems to exhibit a better theoretical behaviour than their scalar counterpart. In this section we will derive the equations for the perturbations for an arbitrary non-linear electromagnetic theory so our results will be completely general. We will only commit to the family of Born-Infeldised ModMax theories in the subsequent sections. This will permit us to  signal clearly the distinctive properties of this family of theories among all the non-linear electromagnetism theories. 

We can decompose the field strength in terms of the electric and the magnetic components with respect to a comoving observer $u^\mu$ as follows
\be
F_{\mu\nu}=2E_{[\mu} u_{\nu]}+\epsilon_{\mu\nu\alpha\beta} u^\alpha B^\beta.
\ee
If we now consider perturbations around a spherically symmetric and static electric background, the quadratic action can be written as\footnote{We recall that we are imposing parity invariance so $\mK_Z=0$ for the electric background configuration. This is the case for the theories considered in this work. Removing this requirement would result in additional terms in the quadratic action like, e.g., the quasi-topological term  $\Lag^{(2)}\supset\mK_Z\delta\vec{E}\cdot\delta\vec{B}$.}
\be
\mS^{(2)}=\frac12\int\dd^3x\dd t\sqrt{-g}\left[\Big( \mK_Y+2Y \mK_{YY} \Big)\delta E_r^2+\mK_Y \delta E_{\Omega}^2-\Big( \mK_Y-2Y \mK_{ZZ} \Big)\delta B_r^2-\mK_Y \delta B_{\Omega}^2\right].
\ee
From this expression we can immediately read off the condition for the absence of ghosts as $\mK_Y>0$ to avoid angular ghosts and $\mK_Y+2Y \mK_{YY}>0$ to avoid radial ghosts. If we further require the absence of Laplacian instabilities, we obtain the additional requirement $\mK_Y>2Y\mK_{ZZ}$.

We can also obtain the propagation speeds (of high frequency modes) from the above quadratic action, which will depend on the direction of propagation as well as on the polarisation of the wave. For a wave travelling in the radial direction, so the electric and magnetic fields oscillate in the transverse angular directions, the propagation speed is given by the ratio of the coefficients of the angular magnetic $\delta B_{\Omega}^2$ and electric $\delta E_{\Omega}^2$ components respectively, i.e., 
\be
c_r^2=1
\ee
so radial modes propagate at the speed of light. This is expected because those waves are oblivious to the radial profile of the background configuration. On the other hand, for waves propagating along the angular directions we have two different speeds depending on whether the electric or the magnetic field oscillates along the radial direction. These speeds are respectively given by
\bea\label{eq:speeds_general}
c_{P}^2&=&\frac{\mK_Y}{\mK_Y+2Y\mK_{YY}},\nonumber\\
c_{A}^2&=&1-\frac{2Y\mK_{ZZ}}{\mK_Y},
\eea
where $P$ and $A$ refer to the polar and axial nature of these waves. Although this derivation of the propagation speeds might appear somewhat hand wavy, in Appendix \ref{app:polardeco} we perform the full derivation of the quadratic action for the two physical degrees of freedom where we obtain the same propagation speeds and the axial and polar nature of the modes is also more apparent. Notice that we have not committed to any theory so far so the above expressions are general. We can now see how the screening is compatible with sub-luminal propagation. In general, we expect the screening factor $1/\mK_Y$ to be a monotonically growing function of $r$ so it is reasonable to impose 
\be
\partial_r\mK_Y^{-1}=-\frac{\mK_{YY}}{\mK_Y^2}\partial_rY>0.
\ee
For the electric background we consider we have $Y=\frac12 E^2$ which is a monotonically decreasing function so $\partial_rY<0$. We thus conclude that we must have $\mK_{YY}>0$ which then implies that $c_P^2<1$. Incidentally, having $\mK_{YY}>0$ together with $\mK_{Y}>0$ guarantees the absence of angular ghost as well. For the axial sector we cannot conclude anything from this analysis (for instance $\mK_{ZZ}$ can be either positive or negative in a generic theory without affecting the background), but we will show below (see Eq. \eqref{eq:dualityconstraint}) that duality invariance imposes $c_A^2=\mK_Y^{-2}$ so it is given by the screening factor. Since this factor interpolates between $0$ and $1$, we find that, for duality invariant theories, $c_A^2<1$. This shows the nice compatibility between screening and subluminal propagation, in contrast to the scalar $K$-mouflage where the screening in turn leads to superluminalities. We should bear in mind that we have given a general argument to motivate how screening is compatible with sub-luminalities, but this is not a proof that {\it any} non-linear electromagnetism featuring screening will avoid sub-luminal propagation.

If the theory admits an asymptotic region $r\gg \rs$ where $\mK\sim Y$, both angular speeds become
\bea
c_{P}^2&\simeq&1-\frac{2Y\mK_{YY}}{\mK_Y},\\
c_{A}^2&\simeq&1-\frac{2Y\mK_{ZZ}}{\mK_Y},
\eea
i.e., they approach the speed of light as it corresponds to the Maxwell theory. In our Born-Infeldised ModMax theory, there is no Maxwell regime and this results in a $\gamma-$suppression of the asymptotic propagation speed for the axial modes. Since in this work we are interested in static perturbations, we will now proceed to the derivation of the relevant equations for the perturbations by neglecting the time-dependence from the onset.

\subsection{Perturbations around static screened objects}

If we consider a static and spherically symmetric configuration, the first order perturbation equations together with the Bianchi identities reduce to\footnote{Notice that there is no source term in these equations since it has been taken into account in the background. At the level of the action, as we show in appendix \ref{app:polardeco}, the background equations being satisfied, the perturbed action is of second order implying that the perturbed equations of motion are linear with no source terms.  This is consistent with the idea of studying how the electromagnetic fields respond to external perturbations which will be taken as the values of the perturbations at infinity.}
\bea
\nabla\cdot\delta\vec{D}=0\,,\quad\nabla\times\delta\vec{H}=0\,\label{eq:deltaD}\,,\
\nabla\cdot \delta\vec{B}=0\,,\quad\nabla\times\delta\vec{E}=0,\label{eq:deltaB}
\eea
where
\bea
\delta\vec{D}=\mK_Y\delta\vec{E}+\mK_{YY}\big(\vec{E}\cdot\delta\vec{E}\big)\vec{E}\,,\\
\delta\vec{H}=\mK_Y\delta\vec{B}-\mK_{ZZ}\big(\vec{E}\cdot\delta\vec{B}\big)\vec{E}\,.
\eea
These equations imply that we can introduce two scalar potentials $\phi$ and $\psi$ as
\be
\delta\vec{E}=-\nabla \phi\,,\quad\delta\vec{H}=\nabla\psi\,.
\ee
Let us notice that the transformation properties of $\vec{E}$ and $\vec{B}$ translate into $\phi$ and $\psi$ actually being a scalar and a pseudo-scalar respectively. Due to these different transformation properties and the fact that parity is not broken, they will decouple at linear order so we can treat them separately.

\subsubsection{Polar sector}

The first equation in \eqref{eq:deltaD} can be expressed as:
\be
\partial_r\left(\frac{r^2\mK_Y}{c_P^{2}}\phi'\right)+\mK_Y\nabla_\Omega^2\phi=0\,. 
\label{eq:phi}
\ee
In view of this equation, it is convenient to introduce the following master variable 
\be
\Phi\equiv \frac{r^2\mK_Y}{c_P^{2}}\phi'\,.
\label{eq:DefPhi}
\ee
that satisfies
\be
\partial_r\Phi+\mK_Y\nabla_\Omega^2\phi=0\,.
\label{eq:relPhiphi}
\ee
The variable $\Phi$ is very simply related to the gauge-invariant perturbation of the electric field as follows: 
\be
\delta E_r=-\phi'=-\frac{c_P^{2}}{r^2\mK_Y}\Phi\,,
\ee
so it represents a useful and more physical quantity than $\phi$. If we take the partial derivative w.r.t. to $r$ and use \eqref{eq:DefPhi} we obtain
\be
\Phi''-\partial_r\ln\mK_Y\Phi'+\frac{c_P^2}{r^2} \nabla_\Omega^2\Phi=0\,.
\ee
We can alternatively express this equation by re-scaling the field $\Phi\to\sqrt{\mK_Y}\Phi$ to get rid of the first derivative term. Thus, if we decompose into spherical harmonics $\Phi=\sum_{\ell,m}\sqrt{\mK_Y}\Phi_\ell(r) Y_{\ell,m}$ we finally obtain
\be\label{eq:PhiPolar}
\Phi_\ell''-m_P^2\Phi_\ell=0\,,
\ee
with
\be
m_P^2=\frac{c_P^2\ell(\ell+1)}{r^2}+\frac14(\partial_r\ln\mK_Y)^2-\frac12\partial^2_r\ln\mK_Y\,.
\ee
In this derivation we have exploited the spherical symmetry of the background to get rid of the dependence on $m$ of the multipole components $\Phi_\ell$ so we can easily perform the sums over $m$ or use this symmetry to set $m=0$. The relation to the multipoles of the electric field perturbation is then
\be
\delta E_{r,\ell}=-\frac{c_P^{2}}{r^2\sqrt{\mK_Y}}\Phi_\ell\,,
\ee
a relation that we will exploit later when imposing boundary conditions. Now, let us turn to the axial sector.

\subsubsection{Axial sector}

The equation for the axial scalar potential $\psi$ can be obtained from the definition of $\delta\vec{H}$ and the Bianchi identity $\nabla\cdot\delta\vec{B}=0$. We first express $\delta\vec{B}$ in terms of $\nabla\psi$ as
\be
\delta\vec{B}=\frac{1}{\mK_Y}\left(\nabla\psi+\frac{\mK_{ZZ}}{\mK_Y-2Y\mK_{ZZ}}\vec{E}\cdot\nabla\psi\vec{E}\right)\,.
\label{eq:Btopsi}
\ee
Then, the Bianchi identity leads to
\be
\partial_r\left(\frac{r^2}{\mK_Yc_A^2}\psi'\right)+\frac{1}{\mK_Y}\nabla_\Omega^2\psi=0\,. 
\ee
This equation is the same as \eqref{eq:phi} with the replacements $\mK_Y\to1/ \mK_Y$ and $c_P^2\to c_A^2$. Thus, we can follow the same procedure to obtain the equation 
\be
\Psi''+\partial_r\ln\mK_Y\Psi'+\frac{c_A^2}{r^2} \nabla_\Omega^2\Psi=0\,,
\ee
with
\be
\Psi\equiv \frac{r^2}{\mK_Yc_A^{2}}\psi'\,.
\label{eq:DefPsi}
\ee
We can now decompose into spherical harmonics as $\Psi=\sum_{\ell,m}\frac{\Psi_\ell(r)}{\sqrt{\mK_Y}} Y_{\ell,m}$ to  obtain finally
\be\label{eq:PsiPolar}
\Psi_\ell''-m_A^2\Psi_\ell=0
\ee
with
\be
m_A^2=\frac{c_A^2\ell(\ell+1)}{r^2}+\frac14(\partial_r\ln\mK_Y)^2+\frac12\partial^2_r\ln\mK_Y\,.
\ee
Notice that the squared masses $m^2_{P,A}$ of the two types of perturbations are simply obtained by flipping the sign of the $\partial_r^2 \log{\cal K}_Y$ term. This will be significant when studying the symmetries associated to these equations in the Born-Infeldised ModMax case.

\subsection{Electromagnetic duality}
\label{sec:duality}

A remarkable property of Maxwell's electromagnetism is its duality invariance that is a symmetry under SO(2) rotations whose associated conserved charge gives the conservation of helicity for photons\footnote{This conservation can be broken,  however, at the quantum level via an anomaly \cite{Agullo:2016lkj,Agullo:2018nfv}.}. It is well-known that electromagnetic duality is a property shared by a certain family of non-linear theories of electromagnetism among which we can find the Born-Infeld theory and the ModMax theories \cite{Gibbons:1995cv,Gaillard:1997rt,Hatsuda:1999ys}. This symmetry can be understood as an invariance under the U(1) transformation
\bea
\vec{D}+i\vec{B}\to e^{i\vartheta}(\vec{D}+i\vec{B}),\\
\vec{E}+i\vec{H}\to e^{i\vartheta}(\vec{E}+i\vec{H})\,,
\eea
where $\vartheta$ is the transformation parameter. This invariance implies that duality invariant theories must fulfil the following constraint
\be
\vec{D}\cdot\vec{H}=\vec{E}\cdot\vec{B}\,.
\ee
This constraint gives rise to the following condition on the the Lagrangian:
\be
\Big(\mK_{Y}^2-\mK_{Z}^2\Big) Z-2\mK_Y \mK_Z Y=Z\,.
\label{eq:Dualityfull}
\ee
This condition is trivially satisfied for Maxwell's theory with $\mK_Z=0$ and $\mK_Y=1$ and it also holds for the Born-Infeld Lagrangian given in \eqref{eq:BIlagrangian}. Our goal here is to obtain the constraints that duality invariance impose on the quadratic action of the perturbations around the screened solution. To that end, we can expand $Y=\Yb+\delta Y$ and $Z=\delta Z$ in the duality constraint \eqref{eq:Dualityfull}. At zeroth order we find
\be
-2\Yb \mK_Y\mK_Z=0\,.
\label{eq:zerothCond}
\ee
This is trivially satisfied for all parity-preserving theories around the electric background so it does not give any new constraint. At first order we obtain the condition:
\be
-2\Yb \mK_Y\mK_{YZ}\delta Y+\mK_Y\Big(\mK_Y-2\Yb \mK_{ZZ}\Big)\delta Z=\delta Z\,,
\ee
where we have used the zeroth order condition \eqref{eq:zerothCond}. If we impose that this is satisfied off-shell, we obtain the two conditions:
\be
\mK_{YZ}=0\quad \text{and}\quad 2\Yb \mK_{ZZ}=\frac{\mK_Y^2-1}{\mK_Y}\,.
\label{constDuality}
\ee
Again, the first one is trivially satisfied for parity-preserving theories, while the second one gives a non-trivial relation between the derivatives of the Lagrangian evaluated on the background. Higher orders in the perturbations give a hierarchy of relations between higher order derivatives of $\mK$. The only place where $\mK_{ZZ}$ appears is for the propagation speed of the transverse modes. When using the duality constraint, we obtain\footnote{This constraint can also be obtained form the condition $\vec{D}\cdot\vec{H}=\vec{E}\cdot\vec{B}$ to satisfy duality invariance. At first order around our background configuration, this relation reduces to $\vec{D}\cdot\delta\vec{H}=\mK_Y(\mK_Y-2Y\mK_{ZZ})\vec{E}\cdot\delta\vec{B}$ that gives \eqref{eq:dualityconstraint}.}
\be
c_A^2=1-\frac{2Y\mK_{ZZ}}{\mK_Y}=\frac{1}{\mK_Y^2},
\label{eq:dualityconstraint}
\ee
in agreement with the result for Born-Infeld. This result shows that any duality invariant theory that reduces to Maxwell in the small field limit, has a potential strong coupling problem for screened backgrounds in the sense that the more efficient the screening is, the smaller the propagation speed is. In fact, we can write the following relation between the effective coupling to charged matter and the propagation speed
\be
c_A=\frac{q_{\rm eff}}{q}
\label{eq:cAqeff}
\ee
that explicitly shows how the screening mechanism leading to a decoupling of charges $q_{\rm eff}\ll q$ comes hand in hand with a small propagation speed of the same order and, therefore, a tighter coupling from this sector. This behaviour might hint at the usual strong/weak coupling regimes for dual theories. The situation is however improved with respect to the scalar field case where super-luminalities are unavoidable.

\subsection{Conformal invariance}
\label{sec:conformal}
In addition to the duality invariance discussed in the preceding section, Maxwell's electromagnetism features another symmetry that is rooted in the massless nature of the photon, namely: conformal invariance. Imposing duality invariance restricts the class of non-linear electrodynamics, but further requiring conformal invariance uniquely selects the ModMax theory. \footnote{A family of theories where conformal invariance is retained but not duality invariance has been explored in \cite{Denisova:2019lgr}.} The presence of conformal invariance can be unveiled in different manners. Perhaps the most direct one is from the tracelessness of the corresponding energy-momentum tensor. At the level of the Lagrangian $\mK(Y,Z)$, conformal invariance can be imposed by factorising it as $\mK(Y,Z)=Y\mathcal{F}(Y/Z)$ since $Y$ is conformally invariant and so is the ratio $Y/Z$. This factorisation implies the non-trivial constraint
\be
\mK=Y\mK_Y+Z\mK_Z
\label{eq:conformalinvariance}
\ee
that is satisfied for both Maxwell and ModMax. Since the energy-momentum tensor for a general non-linear electromagnetism is given by
\be
T_{\mu\nu}=\mK_YF_{\mu\alpha}F_\mu{}^\alpha+g_{\mu\nu}(\mK-Z\mK_Z)
\ee
we see that its trace
\be
T=4(\mK-Y\mK_Y-Z\mK_Z)
\ee
indeed vanishes for theories satisfying \eqref{eq:conformalinvariance}.
Perturbing around an electric background and imposing the condition for conformal invariance $\delta T=0$, we find the following constraint
\be 
{\cal K}_{YY}=0
\label{eq:conformalconstraint}
\ee
that must be satisfied. This constraint is trivially satisfied by Maxwell's electromagnetism, but it is also non-trivially satisfied by the ModMax theories. Since the polar sector has the propagation speed
\be
c_{P}^2=\frac{\mK_Y}{\mK_Y+2Y\mK_{YY}}
\ee
the constraint from conformal invariance implies that the polar sector always propagates at the speed of light $c_P^2=1$, while the propagation speed of the axial sector is not affected by this constraint. For our Born-Infeldised ModMax theory, this means that the polar sector will propagate at the speed of light in the asymptotic region $r\rightarrow\infty$ where it approaches the ModMax regime. 

The constraint \eqref{eq:conformalconstraint} has another consequence for the screening mechanism. If we compute the radial derivative of the screening factor $\mK_Y$ for the electric background, we obtain $\partial_r\mK_Y=\mK_{YY}\partial_rY$ that vanishes on-shell for conformally invariant theories. This means that we need to break conformal invariance to have a genuine $K$-mouflage screening where the field is suppressed below a certain radius. This is natural since the very appearance of the screening scale implies the breaking of conformal invariance. Thus, for conformally invariant theories we can have, at most, a global screening like for the pure ModMax theory.

\section{Static linear response}
\label{sec:LinearResponse}
Equipped with the equations for the perturbations in both the even and odd sectors \eqref{eq:PhiPolar} and \eqref{eq:PsiPolar}, we will proceed to computing the static linear response for the theories of interest in this work. The electric polarisability of the pure Born-Infeld theory has been obtained in \cite{Chruscinski:1997pe,Chruscinski:1997rw}. Here we expand that result to the general Born-Infeldised ModMax theory (although for the polar sector there is no difference) and by also computing the magnetic susceptibility. Furthermore, we carry out a more exhaustive discussion of the physical results. Before going into the details of the specific models, we will rewrite the equations in terms of the dimensionless variable $x=r/\rs$, where $\rs$ is the screening scale as introduced in \eqref{eq:rlambdaBIMOMA}. Then, for a generic non-linear electromagnetism  we have
\begin{eqnarray}\label{eq:differential_general}
\Phi_\ell'' - m^2_\Phi \Phi_\ell &=&0\,,\\
\Psi_\ell'' - m^2_\Psi \Psi_\ell &=&0\,,
\end{eqnarray}
where now the primes refers to differentiation with respect to the $x$ variable and we have introduced the dimensionless masses
\begin{eqnarray}\label{eq:masses_general}
m_\Phi^2 & = & \frac{c_P^2\ell(\ell+1)}{x^2}+\frac14(\partial_x\ln\mK_Y)^2-\frac12\partial^2_x\ln\mK_Y\,,\\
m_\Psi^2 & = &\frac{c_A^2\ell(\ell+1)}{x^2}+\frac14(\partial_x\ln\mK_Y)^2+\frac12\partial^2_x\ln\mK_Y\,,
\end{eqnarray}
where the speeds of sound are defined in Eq.~\eqref{eq:speeds_general}. We should already note that these equations resemble a couple of Schr\"odinger equations where the masses play the role of the corresponding potentials. Furthermore, the terms that depend on the non-linearities (i.e., those determined by $\mK$) precisely generate two potentials that form a super-symmetric quantum mechanical system (see e.g. \cite{Cooper:1994eh}) where $\partial_x\ln\mK_Y$ plays the role of a superpotential. We will come back to this resemblence in Sec.~\ref{sec:superLadder} and \ref{sec:PoshlTeller}. We also explore it in Appendix \ref{app:xlad}.

\subsection{The pure ModMax theory}
\label{subsec:ModMax}

Despite the singular character of the ModMax theory described by \eqref{eq:modmax_lag}, the constraint due to duality given in \eqref{constDuality} still holds. The propagation speeds have a constant profile and are simply given by
\be\label{eq:ModMaxSpeeds}
\cP^2=1,\quad\quad\cA^2=e^{-2\gamma}\,.
\ee
Since this theory produces a constant redressing of the electric charge, we see that \eqref{eq:dualityconstraint} still holds. Furthermore, the axial mode is always luminal, while the polar modes are subluminal for $\gamma>0$ and superluminal for $\gamma<0$ in agreement with the result found in \cite{Bandos:2020jsw}. The solutions for the perturbations around the spherically symmetric electric background are easy to obtain. For the polar modes they are exactly the same as in Maxwell's theory, while the axial modes are corrected by the $\gamma-$redressing.
\bea\label{eq:ModMax_solutions}
\Phi_\ell&=&A_\ell r^\ell+B_\ell r^{-(\ell+1)}\\
\Psi_\ell&=&\tA_{\ell} r^{n_+}+\tB_{\ell} r^{n_-}
\eea
with
\be
n_{\pm}=\frac12\left(1\pm\sqrt{1+4e^{-2\gamma}\ell(\ell+1)}\right).
\ee
Thus, the polar sector exhibits the usual growing $r^\ell$ and decaying $r^{-(\ell+1)}$ solutions of Maxwell's electromagnetism, whereas the axial sector also exhibits a growing and a decaying modes but corrected by the $\gamma-$re-dressing. As a matter of fact, the multipoles with $\ell\lesssim e^{\gamma}$ have
\be
n_{+}\simeq\left[1+e^{-2\gamma}\ell(\ell+1)\right],\quad 
n_{-}\simeq -e^{-2\gamma}\ell(\ell+1)\ll1
\ee
so the two solutions reduce to a linearly growing mode and a nearly constant mode, both of which become independent of $\ell$. This theory does not have a screening scale, since all scales are screened with the global $\gamma-$redressing. In particular, the electric field does not get regularised at the position of the particle so the situation is qualitatively similar to usual Maxwell's theory. Since we are interested in studying the effects coming from the non-linearites, we will not consider this case here and we will proceed to its Born-Infeldised version directly, where the background field is regular at the position of the particle so we can impose appropriate boundary conditions there that will in turn affect how the system responds to external fields.

\subsection{ModMax Born-Infeldised}
\label{subsec:ModMaxBI}

\begin{figure*}
\includegraphics[width=0.49\linewidth]{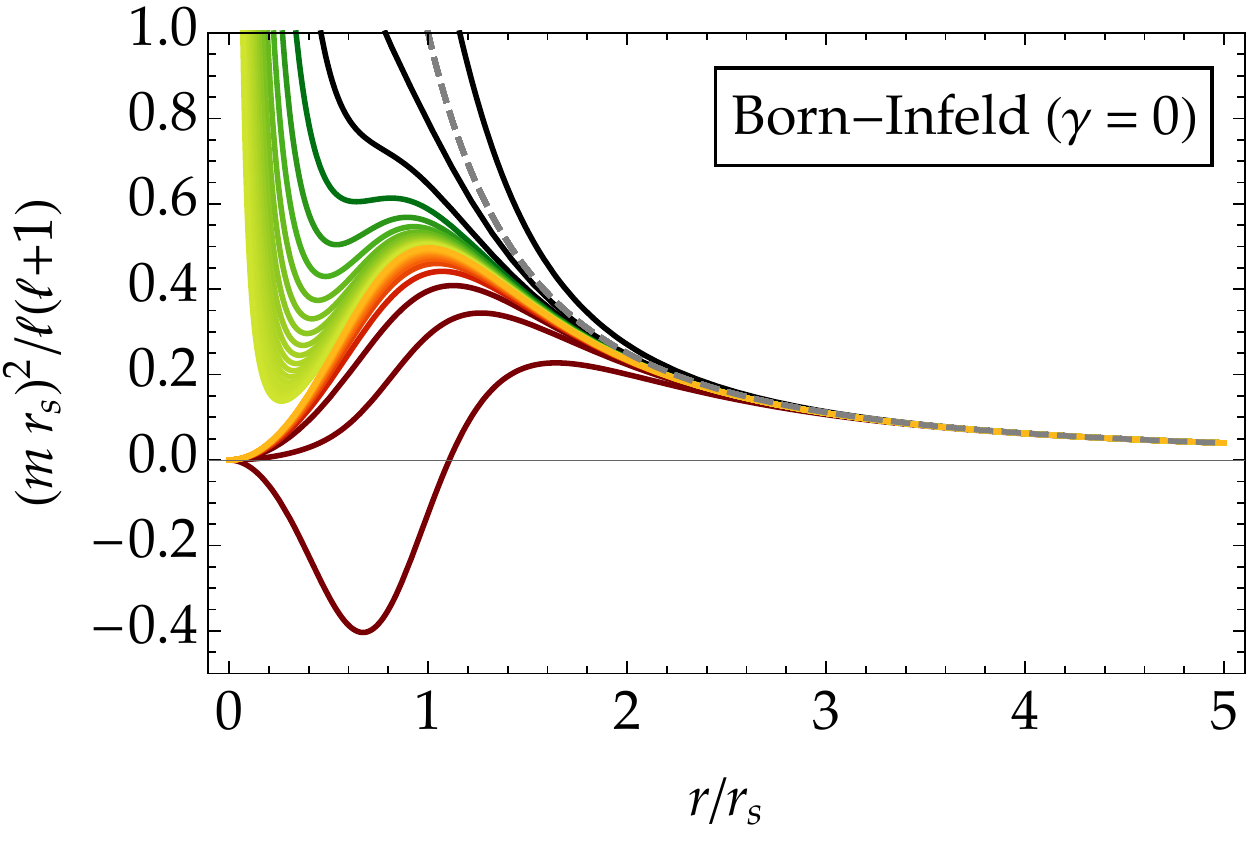}
\includegraphics[width=0.49\linewidth]{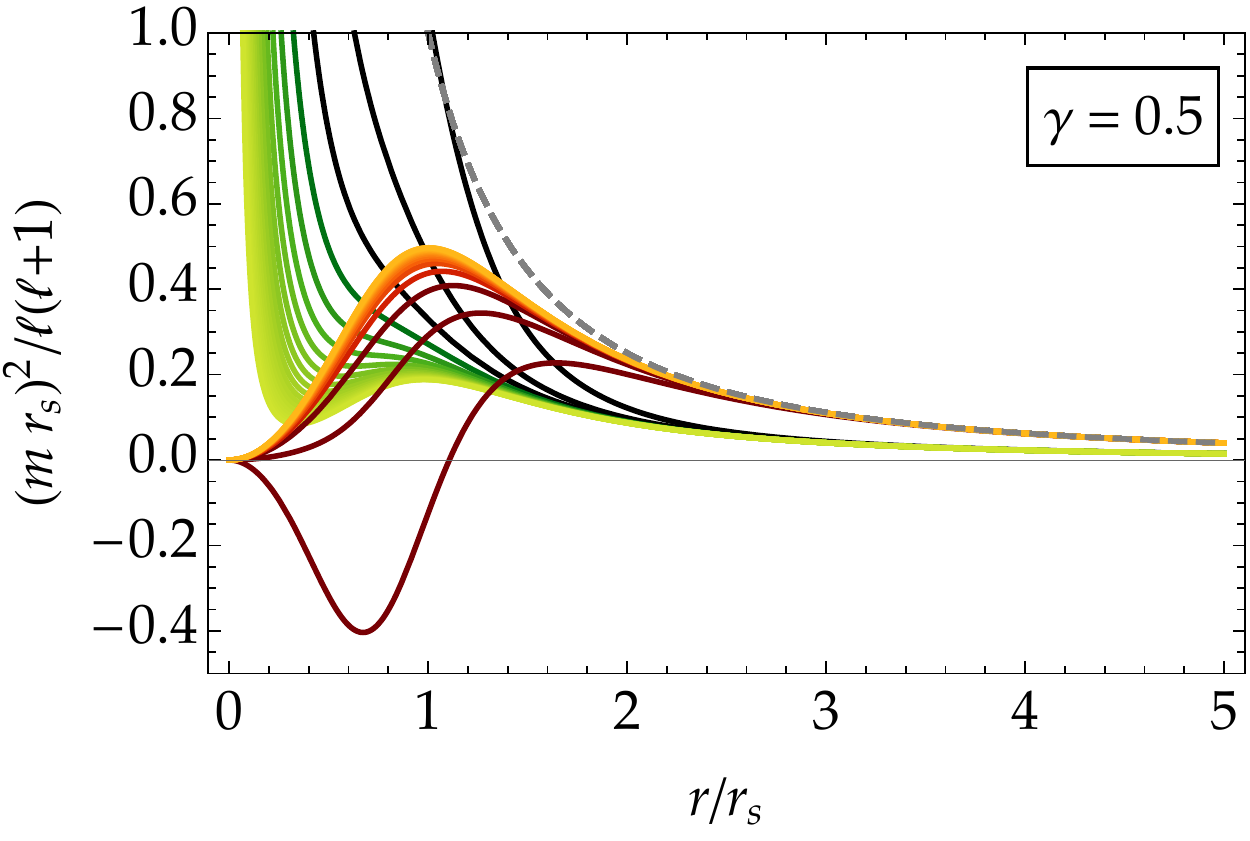}
\caption{In this plot we show the effective masses for both the axial (yellow-green) and polar (orange-red) sectors for Born-Infeld (left) and Born-infeldised ModMax (right) models. For both sectors we plot the masses for multipoles  from $\ell=1$ (darker) to $\ell=20$ (lighter). We observe how the polar sector does not depend on $\gamma$ and that the masses go to zero as $r\to 0$ and reproduce the Maxwell behaviour (dashed gray) in the asymptotic region. For the axial sector, the masses diverge at the origin and only recover the Maxwellian asymptotic behaviour in the pure Born-Infeld case. Furthermore, as we increase $\gamma$ the masses are suppressed with respect to Born-Infeld.}
\label{fig:mases}
\end{figure*}

In this case the Lagrangian is given by \eqref{eq:BIMOMA_Lag} and the propagation speeds read
\be\label{eq:BIModMaxSpeeds}
\cP^2=e^{2\gamma}\cA^2=\frac{x^4}{1+x^4}\,
\ee
while the masses are given by
\bea
\mP^2&=&\frac{1}{\rs^2}\frac{x^2}{1+x^4}\left[\ell(\ell+1) -\frac{5}{1+x^4}\right],\\
\mA^2&=&\frac{1}{\rs^2}\frac{x^2}{1+x^4}\left[e^{-2\gamma}\ell(\ell+1) +\frac{2+5x^4}{x^4(1+x^4)}\right]\,
\eea
and are shown in Fig. \ref{fig:mases}.
At large distances $x\gg1$, we obtain the ModMax regime with
\be
\cP^2=e^{2\gamma}\cA^2\simeq1,\quad\mP^2\simeq \frac{\ell(\ell+1)}{r^2},\quad
\mA^2\simeq\frac{e^{-2\gamma}\ell(\ell+1)}{r^2}\,,
\ee
while at short distances however we find the typical BI behaviour
\be
\cP^2=\cA^2\simeq \left(\frac{r}{\rs}\right)^4,\quad\mP^2\simeq\frac{1}{\rs^2}\Big[\ell(\ell+1)-5\Big]x^2,\quad\mA^2\simeq\frac{2}{r^2}\,.
\ee
In the following we will show that the perturbation equations for both sectors can be put in the form of hypergeometric equations so we can solve them analytically.

\subsubsection{Polar (Even) sector}
\label{subsubsec:polar sector}

Let us start with the simpler polar sector. In order to analyse the solutions, it is convenient to work with the rescaled field
\be
\tilde \Phi_\ell \equiv (1+x^4)^{1/4} \Phi_\ell
\ee
 and the radial variable $z=-x^4$. In terms of these quantities the equation can be recast in the form
\be\label{eq:polarhyperg}
z(1-z)\tilde\Phi_\ell''+\frac{3-z}{4}\tilde\Phi_\ell'+\frac{\ell(\ell+1)-2}{16}\tilde\Phi_\ell=0
\ee
that is nothing but the hypergeometric equation
\be
z(1-z)u''(z)+\Big[c-(a+b+1)z\Big]u'(z)-abu(z)=0
\ee
with parameters $a=-(\ell+2)/4$, $b=(\ell-1)/4$ and $c=3/4$. Since two independent solutions\footnote{Since we have $c=3/4$ for our hypergeometric equation, these two solutions are independent for all the multipoles. This contrasts with, e.g., the perturbations of a Schwarzschild black hole where the parameter $c$ depends on the multipole $\ell$ and some degenerate cases appear.} are given by the hypergeometric function $\;_2F_1(a,b,c;z)$ and $z^{1-c}\;_2F_1(1+a-c,1+b-c,2-c;z)$, transforming back to our original master variable $\Phi_\ell$, we have the general solution:
\be
\Phi_\ell=\frac{1}{(1+x^4)^{1/4}}\left[A_{\ell}  \;_2F_1\left(-\frac{\ell+2}{4},\frac{\ell-1}{4},\frac{3}{4}, -x^4\right)+B_{\ell}\;x  \;_2F_1\left(-\frac{\ell+1}{4},\frac{\ell}{4},\frac{5}{4},-x^4\right)\right].
\ee
At small scales, the solution is
\be
\Phi_\ell\simeq A_{\ell}\left[1+\frac{1}{12}\Big(\ell(\ell+1)-5\Big)x^4\right]+B_{\ell}\left[x+\frac{1}{20}\Big(\ell(\ell+1)-5\Big)x^5\right].
\ee
We will impose a boundary condition so that the electric field remains regular at the origin\footnote{This boundary condition makes sense for Born-Infeld-like theories where the electric field remains finite at the position of the particle. Since the class of theories we are considering reduce to Born-Infeld near the origin, this is an appropriate boundary condition. For other non-linear electromagnetisms where the electric field still diverges at the origin, even if there is an efficient $K$-mouflage, the boundary condition should be re-considered.}. This boundary condition also guarantees that the perturbation theory does not break down since the background field saturates to a finite value as we have seen above. The perturbed electric field is computed as
\be
\delta\vec{E}=-\nabla\phi=-\partial_r\phi\;\hat{r}-\frac{1}{r}\nabla_\Omega\phi=-\sum_{\ell,m}\left[\phi'_{\ell}\;\hat{r}+\frac{\phi_{\ell}}{r}\nabla_\Omega Y_{\ell,m}\right].
\ee
We need to relate $\phi_\ell$ with our master variable $\Phi_\ell$. The derivative of $\phi_\ell$ is obtained directly from the definition of $\Phi$ (taking into account the re-scaling of its multipolar expansion) so 
\be
\phi'_\ell=\frac{c_P^2}{r^2\sqrt{\mK_Y}}\Phi_\ell\simeq A_\ell x^3+B_\ell x^4\,,
\ee
where we have used that $c_P^2\simeq x^4$ and $\mK_Y\simeq 1/x^2$ for $x\ll1$. In order to compute the angular component of the electric field, we will use \eqref{eq:relPhiphi} to write 
\be
\partial_r\Psi=-\mK_Y\nabla^2_\Omega\phi\Rightarrow \phi_\ell=\frac{\partial_r\Big(\sqrt{\mK_Y}\Phi_\ell\Big)}{\mK_Y\ell(\ell+1)}\simeq\frac{-A_\ell+\frac{1}{5}\big(\ell(\ell+1)-5\big)x^5}{\ell(\ell+1)}.
\ee
The perturbed electric field near the origin is then given by
\be
\delta\vec{E}\simeq-\sum_{\ell,m}\left[\Big(A_\ell+B_\ell x\Big)x^2\;\hat{r}+\frac{1}{\ell(\ell+1)}\left(-\frac{A_\ell}{x}+\frac{1}{5}\big(\ell(\ell+1)-5\big)x^4 \right)\nabla_\Omega Y_{\ell,m}\right]\,.
\ee
so requiring regularity of the electric field $\delta \vec{E}$ at the origin imposes $A_{\ell}=0$. Notice that the singular term comes from the angular component of the electric field. Since the monopole does not have any angular component, we need to treat it separately (also apparent from the fact that the resulting expression diverges for $\ell=0$). The monopole represents a re-scaling of the background charge and, as such, the associated electric field is expected to remain finite. To see it explicitly, we can notice that the monopolar contribution to the perturbation of the electric field is simply
\be
\delta \vec{E}_{\ell=0}=-\frac{c_P^2}{r^2\sqrt{\mK_Y}}\Phi_0\hat{r}\,.
\ee
Near the origin, we have
\be
\delta \vec{E}_{\ell=0}(x\to0)\simeq-x^3\Big(A_0+B_0x\Big)\hat{r}\,,
\ee
while at infinity
\be
\delta \vec{E}_{\ell=0}(x\to\infty)\simeq\frac{-A_0x+B_0}{x^2}\hat{r}\,.
\ee
Both modes $A_0$ and $B_0$ remain finite at the origin, as expected, but the mode $A_0$ grows with respect to the background electric field at infinity. The mode $B_0$ however has the same $1/r^2$ tail as the background configuration (the usual Maxwelian behaviour) so it is this mode the one that corrects the background charge. We therefore disregard $A_0$ for the monopole. Thus, the solutions with the appropriate boundary conditions at the origin for all the multipoles, including the monopole, read
\be
\Phi^{\text{reg}}_\ell=B_{\ell}\;\frac{x}{(1+x^4)^{1/4}}  \;_2F_1\left(-\frac{\ell+1}{4},\frac{\ell}{4},\frac{5}{4},-x^4\right)\,.
\label{eq:regPolar}
\ee
The constants $B_\ell$ are fixed by the amplitude of the external perturbation that is not relevant for our computation of the polarisability so we do not need to specify them. For the solutions with these boundary conditions, the asymptotic behaviour at infinity is
\be
\Phi^{\text{reg}}_\ell\simeq B_{\ell}\left[\frac{\Gamma\left(\frac{5}{4}\right)\Gamma\left(-\frac{2\ell+1}{4}\right)}{\Gamma\left(-\frac{\ell+1}{4}\right)\Gamma\left(\frac{5-\ell}{4}\right)}x^{-\ell}+\frac{\Gamma\left(\frac{5}{4}\right)\Gamma\left(\frac{2\ell+1}{4}\right)}{\Gamma\left(\frac{\ell}{4}\right)\Gamma\left(\frac{\ell+6}{4}\right)}x^{\ell+1}\right]
\ee
so we can read off the polarisability as the ratio of the coefficients of the decaying and growing modes \footnote{The definition of the polarisability can depend on the quantity employed to define it. We define the polarisability from the asymptotic behaviour of the variable $\Phi$ because it is related to the radial electric field as $\Phi\propto r^2 \delta E_r$ and so it corresponds to the definition in terms of the asymptotic behaviour of a physical quantity. Had we used the potential $\phi$ instead, we would have obtained a factor $-\frac{\ell}{\ell+1}$ of difference because of the relation $\delta E_r=-\phi'(r)$ that brings down a factor $-\ell$ for the decaying solution and a factor $\ell+1$ from the growing mode. This explains the difference with respect to the result found in \cite{Chruscinski:1997rw}.}
\be
\alpha_\ell=\frac{\Gamma\left(-\frac{2\ell+1}{4}\right)\Gamma\left(\frac{\ell}{4}\right)\Gamma\left(\frac{\ell+6}{4}\right)}{\Gamma\left(-\frac{\ell+1}{4}\right)\Gamma\left(\frac{5-\ell}{4}\right)\Gamma\left(\frac{2\ell+1}{4}\right)}\rs^{2\ell+1}.
\label{eq:alphal1}
\ee
The polarisability parameterically grows as the volume of the $(2\ell+1)$-dimensional screened sphere, ${\mathcal V}_{{\rm s},2\ell+1}=\frac{\pi^{\ell+1/2}}{\Gamma(\ell+\frac32)}\rs^{2\ell+1}$, and it vanishes in the Maxwellian limit for which $\rs=0$, i.e., the screened sphere shrinks to zero.\footnote{This also applies for the pure ModMax theory where $\rs$ is also zero despite the global screening due to $\gamma$.} The polarisability can be alternatively written as
\be
\alpha_\ell=\frac{2^{-(2\ell+1/2)}(\ell+1)(\ell+2)}{\pi^{\ell+1}}\frac{\Gamma\left(-\frac{2\ell+1}{4}\right)\Gamma\left(\ell-1\right)\Gamma\left(\ell+\frac32\right)}{\Gamma\left(\frac{2\ell+1}{4}\right)}{\mathcal V}_{{\rm s},2\ell+1}\cos\left(\frac{\ell\pi}{2}\right).
\label{eq:alphal2}
\ee
This expression shows that the polarisability vanishes for odd multipoles above the dipole (see also Fig. \ref{Fig:PolandMag}). The vanishing of the polarisability for the $\ell=3$ and $\ell=5$ multipoles was already noticed in \cite{Chruscinski:1997rw}, although it was incorrectly stated that $\alpha_\ell\neq0$ for the remaining multipoles. The reason is that those only correspond to the first poles in the $\Gamma$ functions in the denominator of \eqref{eq:alphal1}, but there are additional poles corresponding to non-positive integers of their arguments, which give all odd multipoles above the dipole, as it is apparent from \eqref{eq:alphal2}. For the dipole, the divergent  factor $\Gamma\left(\ell-1\right)$ prevents the vanishing of $\alpha_1$ and we obtain $\alpha_1=\sqrt{\frac{\pi}{2}}\Gamma(\frac{1}{4})/\Gamma(\frac{3}{4})\rs^3\simeq3.71\rs^3$, that recovers the result found in \cite{Chruscinski:1997pe}, barring the different definition of the polarisability. This result shows how the dipolar polarisability parameterically grows as the 3-dimensional volume of the screened region, which is in analogy to the dipolar polarisability of a conducting sphere in an external homogeneous electric field.

We can understand the vanishing of the the polarisability for these modes from the properties of the regular solution $\Phi^{\text{reg}}$. We know that the hypergeometric function reduces to a finite polynomial for non-positive integer $a$ or $b$. If this is the case, the regular solution does not have an asymptotically decaying mode and, thus, the polarisability vanishes. The hypergeometric function in our regular solution \eqref{eq:regPolar} has parameters $a=-(\ell+1)/4$ and $b=\ell/4$. Since $b$ cannot be negative, the hypergeometric function will become polynomial whenever $a$ is a non-positive integer, i.e., for $\ell=4k-1$ for $k=1,2,3...$.\footnote{We exclude the value $k=0$ because that would lead to $\ell=-1$ that is not physical.} In this case, we can use the expansion
\be
 \;_2F_1\left(-k,\frac{\ell}{4},\frac{5}{4},-x^4\right)=\sum_{n=0}^k(-1)^n\binom{k}{n}\frac{(\ell/4)_n}{(5/4)_n}(-x)^{4n}\,,
\ee
with $(\cdot)_n$ the Pochhammer symbol. Since the pre-factor in \eqref{eq:regPolar} is a purely growing function, we see that this regular solution is also purely growing and this explains why we obtain vanishing polarisability for those modes, since the coefficient of the would-be decaying mode is zero. This accounts for the vanishing of $\alpha_\ell$ with $\ell=3,7,11,\dots$. Alternatively, we can use the identity $\;_2F_1(a,b,c;z)=(1-z)^{c-a-b}\;_2F_1(c-a,c-b,c;z)$ to express the regular solution in the following equivalent form:
\be
\Phi^{\text{reg}}_\ell=B_{\ell}\;x\left(1+x^4\right)^{5/4}  \;_2F_1\left(\frac{\ell+6}{4},\frac{5-\ell}{4},\frac{5}{4},-x^4\right)\,.
\label{eq:regPolar2}
\ee
The prefactor is again a growing function so the decaying mode must come from the hypergeometric function. Thus, if the hypergeometric function is polynomial, there will not be any decaying mode and, thus, the polarisability will vanish. This will happen if $b=\frac{5-\ell}{4}=-(k'-1)$ for a strictly positive integer $k'$, i.e., for $\ell=4k'+1$ so we have $\alpha_\ell=0$ for $\ell=5,9,13,\dots$. This series completes the previous one to comprise all odd multipoles with $\ell>1$ for which the polarisability vanishes. Later we will relate this vanishing of the polarisability for the odd modes with a hidden ladder structure and the nature of conserved charges.

Let us finally give the polarisability for large angular momentum $\ell\gg1$. If all the $\Gamma$ functions remain regular, i.e., avoiding the even multipoles above the dipole as discussed above, we can express the polarisability in the remarkably simple form
\be
\left(\frac{\alpha_\ell}{\rs^{2\ell+1}}\right)_{\ell\gg1}\simeq 2^{-\ell}
\label{eq:asymptalpha}
\ee
that shows how the polarisability for high multipoles is exponentially suppressed, i.e., they exhibit a strong resistance to being polarised. We will see later that the same asymptotic form is obtained for the magnetic susceptibility in Born-Infeld.

Since the polar sector is not sensitive to the value of $gamma$, all the Born-Infeldised ModMax theories share the same behaviour as the pure Born-Infeld theories. The differences will appear in the axial sector as we show next.

\subsubsection{Axial (Odd) sector}
\label{subsubsec:axial}

For the axial sector it is convenient to perform the field redefinition
\be
\tilde \Psi_\ell\equiv \frac{x}{(1+x^4)^{1/4}} \Psi_\ell
\ee
so, in terms of the variable $z\equiv-x^4$, the equation for the multipoles takes the form of the following hypergeometric equation:
\begin{equation}
    z(1-z) \tilde \Psi_\ell''+\frac{1-3z}{4}\tilde \Psi_\ell'+\frac{e^{-2\gamma} \ell(\ell+1)}{16}\tilde \Psi_\ell=0\,.
    \label{eq:PolarHyperequ}
\end{equation}
Transforming back to the original variables the solution for the axial modes is thus given by
\be
\Psi_\ell=\left(1+x^4\right)^{1/4}\left[\frac{A_{\ell}}{x}  \;_2F_1\left(a_{-},a_{+},\frac{1}{4}, -x^4\right)+B_{\ell}\;x^2  \;_2F_1\left(b_-,b_+,\frac{7}{4},-x^4\right)\right]\,,
\label{eq:solPsigen}
\ee
with
\be
a_\pm=\frac{-1\pm\sqrt{1+4\ell_{\rm eff}}}{8},\quad b_\pm=\frac{5\pm\sqrt{1+4\ell_{\rm eff}}}{8}\,,
\label{eq:apmbpm}
\ee
and we have introduced $\ell_{\rm eff}=e^{-2\gamma} \ell(\ell+1)$. In the limit of large $\gamma$ (let us recall that we are assuming positive $\gamma$ for causality reasons) where $\ell_{\rm eff}\ll 1$ the dependence on the angular momentum $\ell$ is very mild. In this regime, it is straightforward to see from the equation \eqref{eq:PolarHyperequ} that there is an approximately {\it conserved} charge given by
\be
\mathcal{Q}_\ell=(-z)^{1/4}\sqrt{1-z}\frac{\dd\tilde{\Psi}_\ell}{\dd z}\,.
\ee
These quantities are approximately conserved in the sense that $\frac{\dd\mathcal{Q}_\ell}{\dd z}=\mathcal{O}(\ell_{\rm eff})$. Thus, the solution can be written as 
\be
\tilde\Psi_\ell(\gamma\gg1)\simeq C_\ell+\mathcal{Q}_\ell\int_z^0\frac{\dd
z}{(-z)^{1/4}\sqrt{1-z}},
\ee
where $C_\ell$ and $\mathcal{Q}_\ell$ are determined by the boundary conditions and they contain all the dependence on $\ell$. This solution of course reproduces \eqref{eq:solPsigen} with $a_-=-1/4$, $a_+=0$, $b_-=1/2$, $b_+=3/4$. The constants $C_\ell$ correspond to the trivial charge solution $\mathcal{Q}_\ell=0$ and describes the contribution from the origin, i.e., the particle. In Sec. \ref{sec:charges} we will discuss in more detail the existence of conserved charges for the perturbations and will show that, in fact, there are hierarchies of exactly conserved charges for all the multipoles. For now, let us notice that the introduced charge is exactly conserved for the monopole.

Let us go back to the general case and discuss the boundary conditions for our solutions. As for the axial perturbations, we will require regularity at the origin as one of our boundary conditions, this time for the magnetic field perturbation. At short distances $x\ll1$ we have
\be
\Psi_\ell\simeq \frac{A_{\ell}}{x}+B_{\ell} x^2\,,
\ee
so we see that, again, the $A_\ell$ modes seem more prone to a singular behaviour at the origin and, thus, it should be set to zero. Indeed, this is the case for the magnetic field. To see it more explicitly, let us first notice that the magnetic field perturbation is expressed in terms of the axial scalar potential as given in \eqref{eq:Btopsi}, that can be more conveniently written as
\be
\delta \vec{B}=\frac{1}{c_A^2\mK_Y}\partial_r\psi\hat{r}+\frac{1}{r\mK_Y}\nabla_\Omega\psi\,.
\label{eq:deltaB2}
\ee
The radial component can be easily expressed in terms of our axial master variable from its definition
\be
\partial_r\psi=\frac{r^2}{\mK_Y c_A^2}\Psi\Rightarrow \partial_r\psi_\ell=\frac{r^2}{\mK_Y^{3/2} c_A^2}\Psi_\ell\,.
\ee
The angular component in \eqref{eq:deltaB2} can be computed from the Bianchi identity directly as
\be
\partial_r\Psi=-\frac{1}{\mK_Y}\nabla^2_\Omega\psi\Rightarrow \psi_\ell=\frac{\mK_Y}{\ell(\ell+1)}\partial_r\left(\frac{\Psi_\ell}{\sqrt{\mK_Y}}\right).
\ee
With these relations, we can express the magnetic field in terms of our master variable as:
\be
\delta \vec{B}=\sum_{\ell,m}\left[\frac{\Psi_\ell}{r^2\sqrt{\mK_Y}}Y_{\ell,m}\hat{r}+\frac{1}{r\ell(\ell+1)}\partial_r\left(\frac{\Psi_\ell}{\sqrt{\mK_Y}}\right)\nabla_\Omega Y_{\ell,m}\right].
\label{eq:deltaBtoPsi}
\ee
Using now the behaviour of $\Psi_\ell$ near the origin, we find
\be
\delta \vec{B}\simeq\sum_{\ell,m}\left[\left(\frac{A_\ell}{r^2}+B_\ell r\right)Y_{\ell,m}\hat{r}+\frac{1}{r\ell(\ell+1)}\partial_r\left(A_\ell+B_\ell r^3\right)\nabla_\Omega Y_{\ell,m}\right].
\label{eq:deltaBsmallx}
\ee
In this case, already the radial component presents a divergent behaviour at the origin for the modes $A_\ell$, while $B_\ell$ are regular. The angular component however remains regular at the origin for both modes. For the monopole\footnote{This monopolar contribution for the axial sector would be associated to the response to magnetic monopoles and we include it for completeness.} contribution a similar argument shows that regularity at the origin requires $A_{\ell=0}=0$. Thus, the regular solution for all modes is given by
\be
\Psi^{\rm{reg}}_\ell=B_{\ell}\;x^2\left(1+x^4\right)^{1/4}  \;_2F_1\left(b_+,b_-,\frac{7}{4},-x^4\right)\,,
\label{eq:solPsigen1}
\ee
The asymptotic behaviour is found to be
\be
\Psi_\ell\simeq B_{\ell}\left[\frac{\Gamma\left(\frac{7}{4}\right)\Gamma\left(b_--b_+\right)}{\Gamma\left(b_-\right)\Gamma\left(\frac{7-4b_+}{4}\right)}x^{3-4b_+}+\frac{\Gamma\left(\frac{7}{4}\right)\Gamma\left(b_+-b_-\right)}{\Gamma\left(b_+\right)\Gamma\left(\frac{7-4b_-}{4}\right)}x^{3-4b_-}\right].
\label{eq:chilg}
\ee
Again, we can compute the magnetic susceptibility as the ratio of the coefficients of the decaying and growing modes, so it is given by
\be
\chi_\ell=\frac{\Gamma\left(b_+\right)\Gamma\left(b_--b_+\right)\Gamma\left(\frac{7-4b_-}{4}\right)}{\Gamma\left(b_-\right)\Gamma\left(b_+-b_-\right)\Gamma\left(\frac{7-4b_+}{4}\right)}r_{\rm s}^{\Delta b}\,,
\label{eq:chigengl}
\ee
with $\Delta b =\sqrt{1+4\ell_{\rm eff}}/4$.  In the limit of large $\gamma$ (i.e., $\ell_{\rm eff}\ll1$), the above expression reduces to
\be
\left(\frac{\chi_\ell}{\rs^{\Delta b}}\right)_{\gamma\gg1}= -\frac{\Gamma^2\left(\frac{3}{4}\right)}{\sqrt{\pi}}+\mathcal{O}(\ell_{\rm eff}),
\ee
that is independent of $\ell$. This stems from the $\ell$-independence of the equation for the perturbations in this regime as discussed above.

\begin{figure}[ht]
\includegraphics[width=0.49\linewidth]{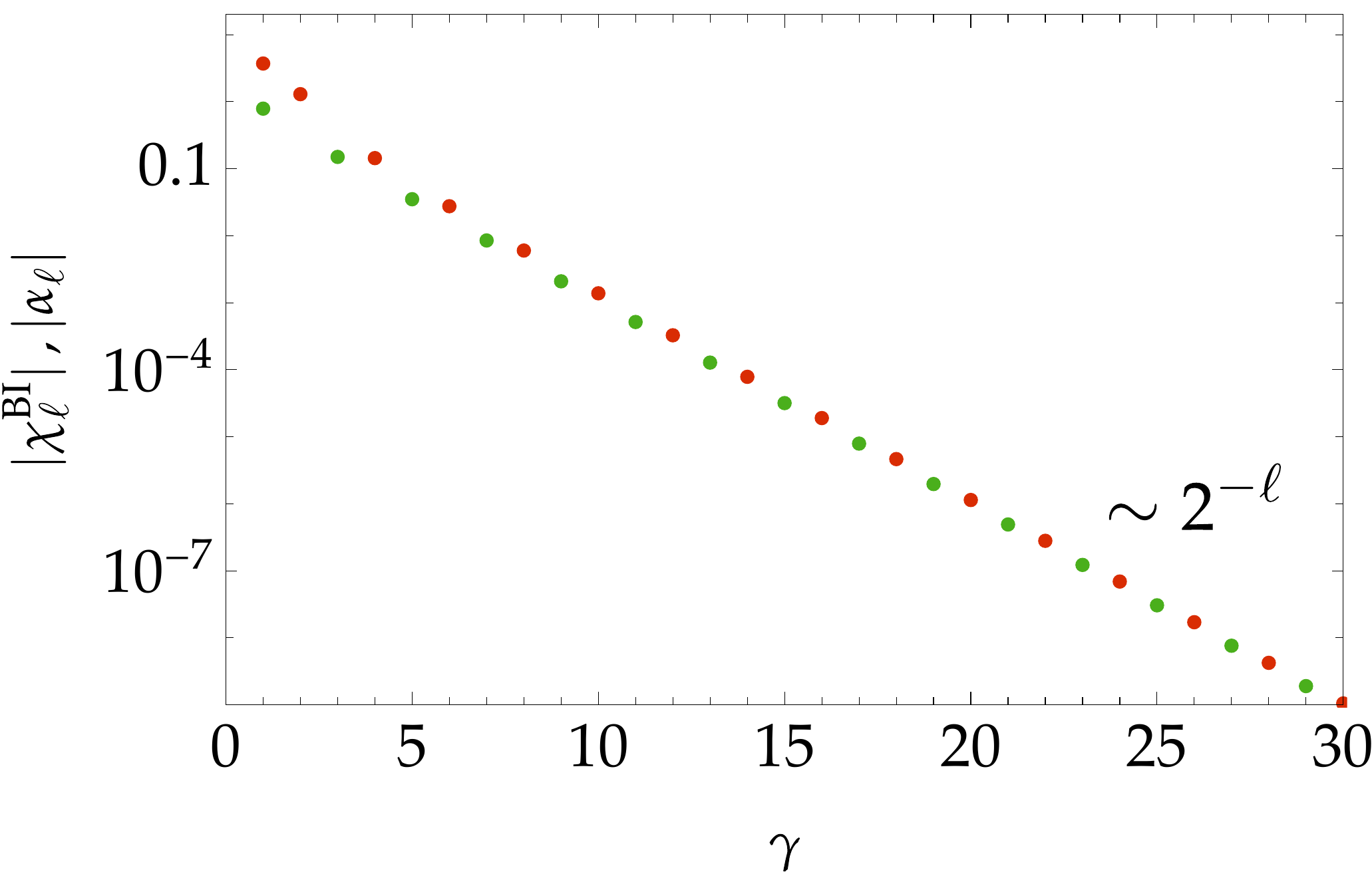}
\includegraphics[width=0.49\linewidth]{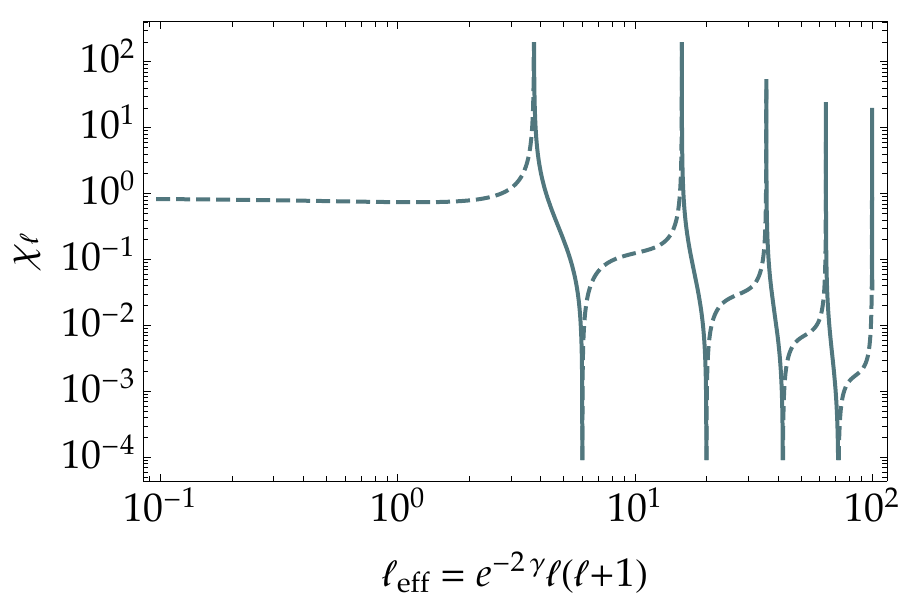}
\includegraphics[width=0.49\linewidth]{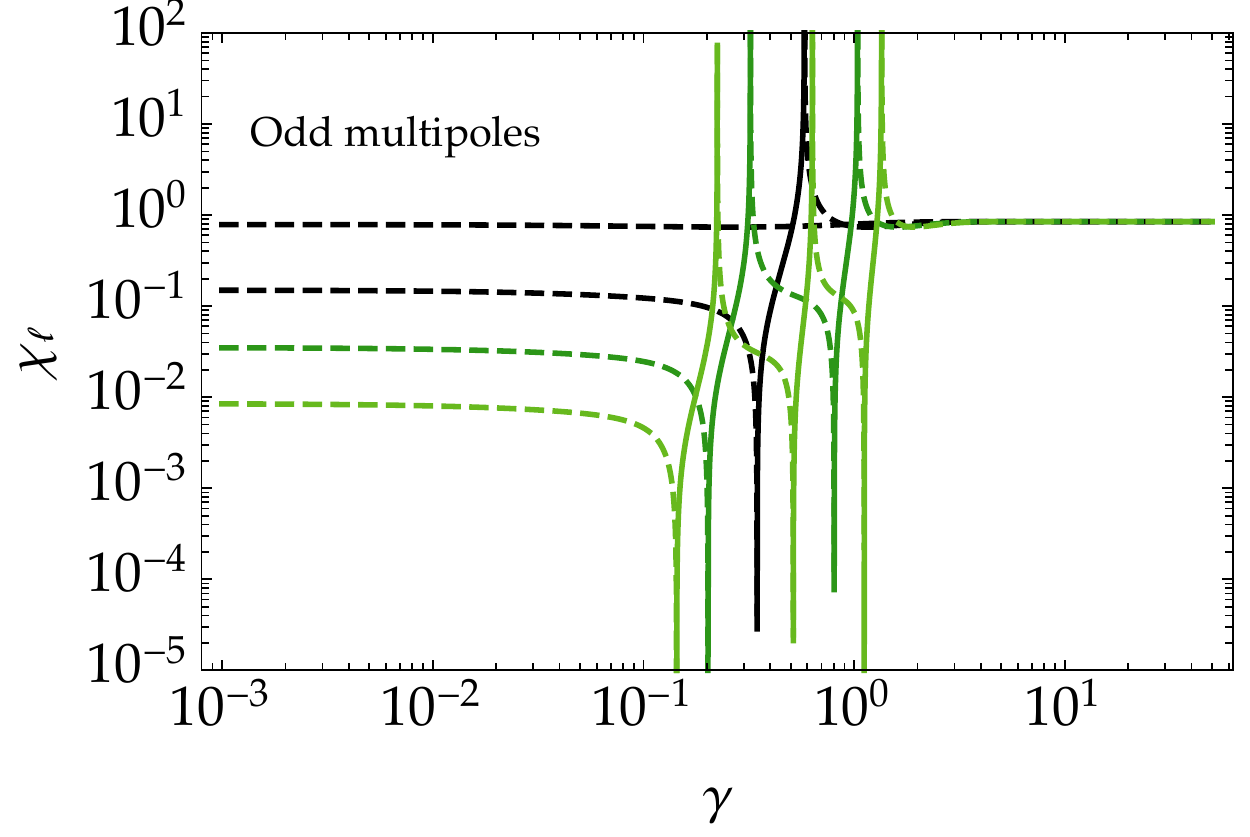}
\includegraphics[width=0.49\linewidth]{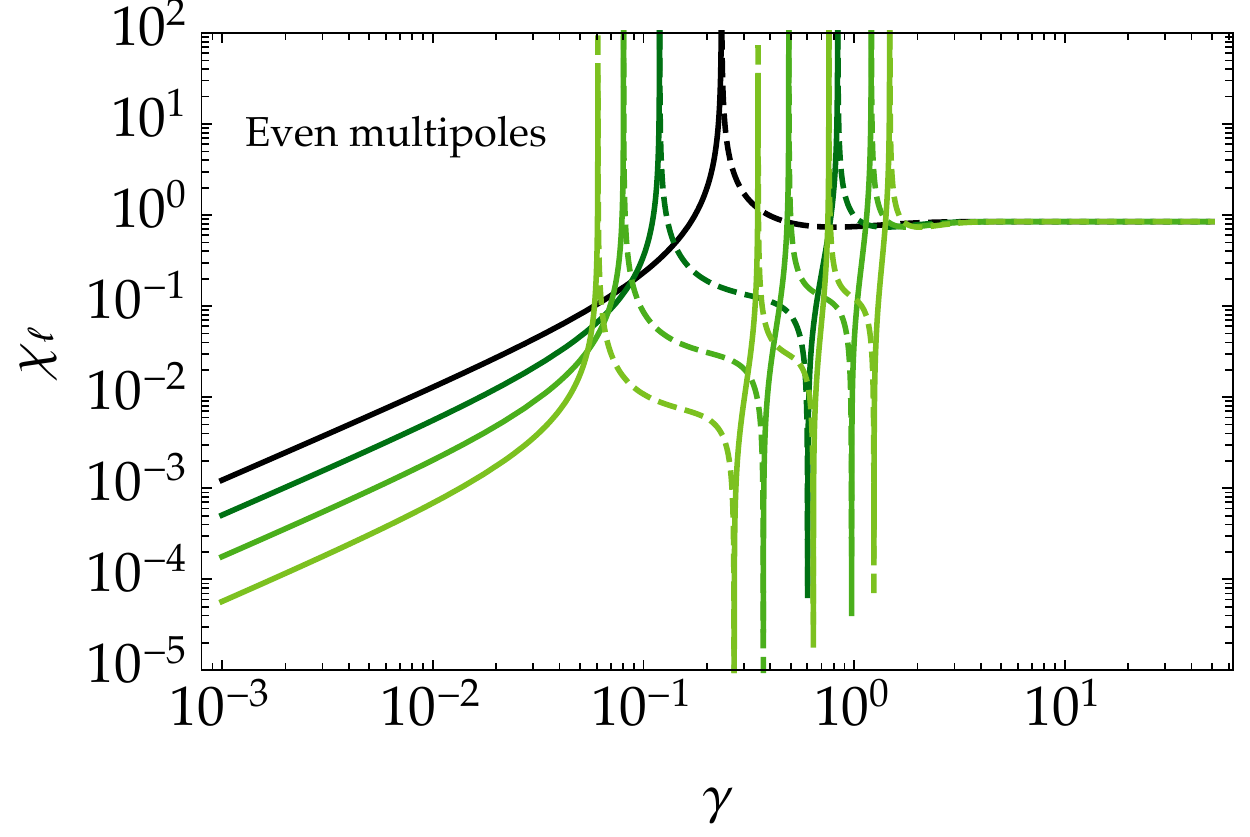}
\caption{%
\textbf{Upper left panel}: Electric polarisability (red) and magnetic susceptibility (green) as a function of $\ell$ for the pure Born-Infeld theory (although $\alpha_\ell$ is the same for all the Born-Infeldises ModMax theories). We can see the asymptotic behaviour $2^{-\ell}$. \textbf{Upper right panel}: magnetic susceptibility as a function of the continuous variable $\ell_{\rm eff}$ for Born-Infeldised ModMax. Solid and dashed denote positive and negative values respectively. In this plot we can see the $\ell_{\rm eff}-$independence for small $\ell_{\rm eff}$ as well as the spikes corresponding to those theories exhibiting vanishing and diverging susceptibilities as discussed in the main text. \textbf{Bottom}: Dependence of the magnetic susceptibility on $\gamma$ for odd (left) and even (right) multipoles. Increasing values of $\ell$ go from bottom to upper. We corroborate how all the multipoles approach the asymptotic value $\chi_\ell\simeq -\Gamma^2(\frac34)/\sqrt{\pi}$ that is independent of both $\gamma$ and the multipole. On the other hand, for small values of $\gamma$ we observe that the even multipoles go to zero as $\chi_\ell\propto \ell \gamma$, while the odd multipoles go to a constant value, as it corresponds for Born-Infeld.
}
\label{Fig:PolandMag}
\end{figure}

As for the polarisability, the magnetisation will present zeros for parameter values corresponding to poles of the $\Gamma-$functions appearing in the denominator of \eqref{eq:chigengl}. Since we have an additional parameter $\gamma$ that can take continuous values, it is guaranteed that $\chi_\ell$ will have zeros. Furthermore, for the same reason, there will be values for which $\chi_\ell$ diverges corresponding to the poles of the numerator in \eqref{eq:chigengl}. This behaviour can be seen in Fig. \ref{Fig:PolandMag}. and it can also be understood analytically. Since $b_+-b_-=\frac14\sqrt{1+4\ell_{\rm eff}}$ is positive, the parameters for which the susceptibility vanishes can be easily computed as
\bea
b_-=n&\Rightarrow&\ell_{\rm eff}=2 \left(3 - 10 n + 8 n^2\right)\equiv f(n)\,\\
\frac{7-4b_+}{4}=m&\Rightarrow&\ell_{\rm eff}=  4 \left(5 - 9 m + 4 m^2\right)\equiv g(m),
\eea
with $n,m$ non-positive integers. Since both polynomials $f$ and $g$ are positive for non-positive values of their arguments, we can always find a value of $\gamma$ for which the susceptibility vanishes for a given value of $\ell$. As a matter of fact, there is an infinite family of values of $\gamma$ for which a given multipole vanishes. This family is found from
\be
e^{-2\gamma}=\frac{f(n)}{\ell(\ell+1)},\quad e^{-2\gamma}=\frac{g(m)}{\ell(\ell+1)}.
\label{eq:zeoreschilg}
\ee
An interesting feature of the existence of these two series is that actually there is a family of multipoles with vanishing magnetic susceptibility. Let us assume that we have a multipole $\ell_*$ for which $\chi_{\ell_*}$ vanishes corresponding to a certain value of $n=n_*$ in \eqref{eq:zeoreschilg}. This will give a rational value for the corresponding value of $e^{-2\gamma_*}$. Then, we can find other multipoles $\ell_*'$ with vanishing susceptibility corresponding to some values $n'_*$ and/or $m'_*$ provided one of the two following conditions holds:
\be
e^{-2\gamma_*}=\frac{f(n'_*)}{\ell'_*(\ell'_*+1)}\,,\quad e^{-2\gamma_*}=\frac{g(m'_*)}{\ell'_*(\ell'_*+1)}\,.
\ee
for some non-positive integers $n'_*$ and $m'_*$. Since $e^{-2\gamma_*}$ is a rational number and so are the right hand sides of the above equations, solutions may exist although the number of solutions is not determined. In fact, the new multipole with vanishing magnetisation must satisfy either
\be
\ell_*'=\frac12\left[-1+\sqrt{1+4\frac{f(n_*)}{f(n'_*)}\ell_*(\ell_*+1)}\right]
\label{eq:solellstarprimef}
\ee
or
\be
\ell_*'=\frac12\left[-1+\sqrt{1+4\frac{f(n_*)}{g(m'_*)}\ell_*(\ell_*+1)}\right]
\label{eq:solellstarprimeg}
\ee
so we need the quantities inside the square roots be a perfect square. In general, this only allows for some solutions. Let us illustrate it with an example. Let us impose to have vanishing magnetisation for the dipole $\chi_1=0$ and choose the value of $\gamma$ so this happens for $n=0$ and we have $e^{-2\gamma_*}=3$. Then, it is straightforward to check that the same value of $e^{-2\gamma_*}$ is obtained for the multipoles $\ell=76$ and $\ell=285$ corresponding to $m=-32$ and $n=-123$ respectively. 

A singular case occurs when imposing vanishing magnetisation for the quadrupole and for $n=0$. For these values, we obtain $e^{-2\gamma}=1$, i.e., the pure Born-Infeld theory. We further obtain that \eqref{eq:solellstarprimef} and \eqref{eq:solellstarprimeg} reduce to
\be
\ell_*'=2(1-2n'_*)\quad{\text{and}}\quad
\ell_*'=4(1-m'_*),
\label{eq:vanishingchiBImn}
\ee
so we have two infinite families of multipoles with vanishing magnetic susceptibility. In fact, these two families together comprise all even multipoles. Thus, the Born-Infeld theory stands out as the most resilient against external odd perturbations since all even multipoles but the monopole exhibit perfect rigidity (at linear order). 

The vanishing of the magnetisation for Born-Infeld can be directly seen by setting $\gamma=0$ in \eqref{eq:chigengl} that dramatically simplifies to
\be
\chi^{\text{BI}}_\ell=
\frac{\Gamma\left(-\frac{2\ell+1}{4}\right)\Gamma\left(\frac{\ell+3}{4}\right)\Gamma\left(\frac{\ell+5}{4}\right)}
{\Gamma\left(-\frac{\ell-2}{4}\right)\Gamma\left(-\frac{\ell-4}{4}\right)\Gamma\left(\frac{2\ell+1}{4}\right)}
\rs^{(2\ell+1)/4}\,.
\label{eq:chiBIgeneral1}
\ee
From this expression we can see how the magnetic susceptibility vanishes for all even multipoles in Born-Infeld in a more direct manner. Similarly to the electric polarisability, the zeros of the above expression coincide with the poles of the $\Gamma-$functions in the denominator that occur when their arguments are some negative integers. This occurs when either $\ell-2=4k$ or $\ell-4=4k$ for $k=1,2,3,\dots$, that are the two series obtained in \eqref{eq:vanishingchiBImn} and which, together, scan all even modes. A perhaps more transparent form of $\chi^{\text{BI}}_\ell$ is the following
\be
\chi^{\text{BI}}_\ell=\frac{2^{-(4\ell+1)/4}}{\ell\sqrt{\pi}}
\frac{\Gamma\left(-\frac{2\ell+1}{4}\right)\Gamma\left(\ell+2\right)}
{\Gamma\left(\frac{2\ell+1}{4}\right)}
\sin\left(\frac{\ell\pi}{2}\right)\rs^{(2\ell+1)/4}\,,
\label{eq:chiBIgeneral2}
\ee
that makes more apparent the vanishing of the magnetisation for even multipoles above the monopole. As for the polarisability, the vanishing of the magnetic susceptibility can be understood from the modes that turn the hypergeometrical functions in the regular solutions into polynomials. Since the analysis fully parallels the one performed for the polar sector, we will not repeat it here. Let us however mention that the vanishing of $\chi_\ell^{\rm BI}$ for even multipoles relates to the existence of a ladder structure and the nature of conserved charges in the subsequent Sections.

We can also take the limit of small $\gamma$ and large (even) angular momentum in \eqref{eq:chilg} to obtain
\be
\left(\frac{\chi_\ell}{\rs^{\Delta b}}\right)\simeq\pi 2^{-(\ell+1)}\ell\gamma
\ee
which shows how the vanishing of the magnetization for the even multipoles only occurs in the Born-Infeld theory, while in the general Born-Infeldised ModMax, the magnetisation acquires a correction due to $\gamma$. Although this result is only valid for small $\gamma$, it shows once again the remarkable properties of Born-Infeld theory among generic non-linear electromagnetisms.

Finally we can take the limit of large angular momentum $\ell\gg1$ limit for Born-Infeld to obtain the simple law
\be
\left(\frac{\chi^{\rm BI}_\ell}{\rs^{(2\ell+1)/4}}\right)_{\ell\gg1}\simeq 2^{-\ell}
\ee
valid for the odd modes. This expression for the asymptotic magnetisation coincides with the one obtained for the polarisability in \eqref{eq:asymptalpha}, although for the even modes in that case. This shows that the polarisability and the magnetisation in Born-Infeld follow the same asymptotic law for large $\ell$ and for alternating multipoles where the corresponding quantity does not vanish (see Fig. \ref{Fig:PolandMag}). This simple relation between the electric polarisability and the magnetic susceptibility can also be deduced from the following remarkable relation that holds for all multipoles:
\be
\frac{\chi^{\text{BI}}_\ell}{\alpha_\ell}=\frac{\ell-1}{\ell+2}\tan\left(\frac{\ell\pi}{2}\right)\,.
\ee
This relation makes apparent the alternating of the multipoles with vanishing polarisability and susceptibility as it corresponds to the zeros and singular points of the tangent function.

Let us notice that the numerator is never singular because $(2\ell+1)/4$ is never an integer so the Born-Infeld theory does not possess multipoles that are infinitely deformed by external perturbations. This is however not the case for the general expression \eqref{eq:chigengl} with an arbitrary $\gamma$. Since both $b_+$ and $(7-4b_-)/4$ are positive, the only possible diverging factor comes in \eqref{eq:chigengl} from $\Gamma(b_--b_+)$ whose poles are
\be
b_--b_+=-n\Rightarrow \ell_{\rm eff}=\frac14\left(16n^2-1\right)
\ee
with $n$ a non-positive integer. Thus, we can fix $\gamma$ to have a divergent magnetisation for a given multipole $\ell$ as
\be
e^{-2\gamma}=\frac{(4n+1)(4n-1)}{4\ell(\ell+1)}.
\ee
The diverging character of the magnetisation can be associated to an unbounded linear response to the external field and, thus, these particular theories are expected to be prone to instabilities. From the perspective of the solutions, this divergent response is due to the absence of decaying modes, which is in line to the presence of instabilities. However, before definitely concluding the unstable character of these theories, a more careful analysis should be performed.

The relations between the polarisability and the magnetisation for Born-Infeld may be traced back to the duality invariance. Of course, this is not all the story because the generic Born-Infeldised ModMax theories are duality invariant but do not exhibit the same properties. Instead, it seems to be the coincidence of the propagation speeds in both sectors (that is related to the absence of birefringence in these theories) that leads to the  remarkable properties of Born-Infeld. However, duality invariance may be enough to explain why the vanishing of the polarisability and the magnetisation occurs for odd and even modes in the polar and the axial sectors respectively.

So far, the vanishing of the electric polarisability and the magnetic susceptibility for certain multipoles have been obtained by direct computation of the solutions with the appropriate boundary conditions. We have shown how the vanishing of the static linear responses for certain multiples can be traced back to the corresponding hypergeometric functions reducing to polynomials. In the remaining of this work we will delve deeper into the special properties of the hypergeometric functions that conform the space of solutions for the perturbations and unveil novel manners to understand the vanishing of the polarisability and magnetisation from a more physical point of view. In particular, we will construct ladder operators connecting different multipoles and we will use them to generate a hierarchy of conserved charges.

\section{Ladder structure}
\label{sec:ladder}

In this section we aim at finding ladder operators for the space of solutions of the perturbation equations that will connect different $\ell-$modes. The ladder structure will serve to obtain a hierarchy of symmetries and their corresponding conserved charges from the obvious conserved quantities that are obtained for the monopole and the dipole for the axial and polar sectors respectively. The factorization method that we will employ for the identification of the ladder operators resemble the exhaustive classification carried out in the seminal work by Infeld and Hull in 1951 \cite{Infeld:1951mw}, although, as we will see, we will need to introduce some tweaks since our system of equations present some remarkable peculiarities. In our treatment, we have also taken inspiration from similar studies recently carried out within the context of de Sitter and black hole physics \cite{Lagogiannis:2011st,Compton:2020cjx,Hui:2021vcv}. However, unlike those studies, we will show the existence of two ladder structures that, in turn, do not connect adjacent $\ell-$modes. Rather, there is a wide ladder that connects modes with $\ell$ and $\ell+4$ and a narrow ladder that establishes an automorphism (or a sort of duality) for the first four modes. 

\subsection{Polar ladder}
\label{subsec:polarladder}

In order to find the ladder operators we will start from the hypergeometric form of the equations ~\eqref{eq:polarhyperg}  that we reproduce here:\footnote{In the remaining of the paper we will drop the tilde for the master variables to simplify the notation.}
\be
z(1-z)\Phi_\ell''+\frac{3-z}{4}\Phi_\ell'+\frac{\ell(\ell+1)-2}{16}\Phi_\ell=0\,.
\ee
The first step will be to introduce the following family of Hamiltonians
\be
H_{\ell}\equiv -z(1-z)\left[z(1-z)\partial^2_{z}+\frac{3-z}{4}\partial_z+\frac{\ell(\ell+1)-2}{16}\right]
\ee
whose kernels coincide with the space of solutions of our equation.\footnote{Notice that, since $z$ is non-positive, the factor $1-z$ does not introduce any new singular point. The singular point $z=0$ is the original one that we use to impose the appropriate boundary conditions.} The goal is then to find a set of operators $\Am_\ell$ and $\Ap_\ell$ that factorize these Hamiltonians as
\bea
\Am_\ell \Ap_\ell&=&H_\ell+\varepsilon_{1\ell},\nonumber\\
\Ap_\ell \Am_{\ell}&=&H_{\ell+n}+\varepsilon_{2\ell}.
\label{Eq:FactorizationH}
\eea
with $n$ some integer number and $\varepsilon_{i,\ell}$ some (in principle different) scalars, i.e., they do not contain differential operators nor do they depend on the variable $z$. In order to find such operators, we will make the following Ansatz:
\bea
\Am_\ell&\equiv& z(z-1)\partial_z+W_{1,\ell}(z),\\
\Ap_\ell&\equiv&-z(z-1)\partial_z+W_{2,\ell}(z),
\eea
with $W_{i\ell}(z)$ two functions to be determined from the factorization in \eqref{Eq:FactorizationH}. By imposing such a factorisation we then find
\bea
\Am_\ell \Ap_\ell&=&H_\ell+\frac14 z(z-1) \Big(1 - 7 z - 4 W_{1,\ell} + 4 W_{2,\ell}\Big)\partial_z\nonumber\\
&&+\left[
W_{1,\ell}W_{2,\ell} + \frac{1}{16} z(z-1)  \Big(2 - \ell(\ell+1) + 16 W'_{2,\ell}\Big)
\right] ,\\
\Ap_\ell \Am_{\ell}&=&H_{\ell+n}+
\frac14 z(z-1) \Big(1 - 7 z - 4 W_{1,\ell} + 4 W_{2,\ell}\Big)\partial_z\nonumber\\
&&+\left[
W_{1,\ell}W_{2,\ell} + \frac{1}{16} z(z-1)  \Big((2+\ell+n)(1-\ell-n) - 16 W'_{1,\ell}\Big)
\right].
\label{Eq:FactorizationH2}
\eea
that should allow to identify $\varepsilon_{i,\ell}$. Since they cannot contain differential operators, a first condition is obtained by requiring the vanishing of the coefficients of $\partial_z$ in the above expressions, that turn out to be the same. Thus, we must have
\be
W_{2,\ell}=\frac{7 z-1}{4} +  W_{1,\ell}.
\label{eq:relW1W2}
\ee
On the other hand, since $\varepsilon_{i,\ell}$ cannot depend on $z$, they must differ at most by a constant that we will denote $\varepsilon_{1,\ell}-\varepsilon_{2,\ell}\equiv C_\ell$. Thus, by subtracting the last terms in $\eqref{Eq:FactorizationH2}$ and using \eqref{eq:relW1W2} we find that the following condition must hold
\be
z(z-1)  \Big(28 + n + 2 \ell n + n^2 + 32 W'_\ell\Big)=C_\ell\,.
\ee
The solution of this differential equation is given by
\be
W_\ell(z)=-\frac{1}{32} \left(28 + n + 2 \ell n + n^2\right) z + C_\ell {\text{arctanh}}(1 - 2 z) + c_\ell\,,
\ee
with $c_\ell$ another integration constant. Regularity of the function $W_\ell$ at the origin requires $C_\ell=0$. Thus, we finally obtain the factorization:
\bea
\Am_\ell \Ap_\ell=H_\ell+\varepsilon_\ell \,,\quad\quad
\Ap_\ell \Am_{\ell}=H_{\ell+n}+\varepsilon_\ell\,,
\label{Eq:FactorizationH3}
\eea
with
\bea
\varepsilon_\ell&=&\frac{(n+2\ell-3) (n+2\ell+5) (n^2-16 )}{1024}z^2\nonumber\\
&&+\frac{  8 \ell^2 + 2 \ell (4 + n (5 - 8 c_\ell)) + n (1 + n) (5 - 8 c_\ell)-100}{128} z   
     - \frac{c_\ell}{4} + c_\ell^2\,.
\eea
Since this quantity cannot depend on $z$, additional (non-trivial) conditions are obtained from cancelling the coefficients of $z^2$ and $z$. Notice that the regularity condition imposing $C_\ell=0$ is also necessary to be able to cancel the $z-$dependence. The vanishing of the coefficient of $z^2$ leads to the three solutions $n=4$, $n=3-2\ell$ or $n=-5-2\ell$. The latter is not physical because our ladder structure requires $n>0$ and the latter solution would connect $\ell \to -5-\ell$, which is unphysical. Thus, we have the first two possibilities. On the other hand, the vanishing of the coefficient linear in $z$ finally determines $c_\ell$ so the factorization is completed. We have thus obtained two possible factorisations that we analyse in more detail in the following.

\subsubsection{Big polar ladder}
We will first study the solution with $n=4$ that gives a ladder whose steps connect $\ell$ and $\ell+4$. This choice leads to the ladder operators
\bea
\Am_\ell&\equiv& z(z-1)\partial_z-\frac{\ell+6}{4}\left(z-\frac{\ell}{2\ell+5}\right),\\
\Ap_\ell&\equiv&-z(z-1)\partial_z-\frac{\ell-1}{4}\left(z-\frac{\ell+5}{2\ell+5}\right).
\eea
that satisfy the relations
\bea
\Am_\ell \Ap_\ell=H_\ell+\varepsilon_\ell,\quad\quad
\Ap_\ell \Am_\ell=H_{\ell+4}+\varepsilon_\ell\,,
\label{eq:defApAm}
\eea
with
\be
\varepsilon_\ell=\frac{\ell(\ell+6)(\ell+5)(\ell-1)}{16(2\ell+5)^2}\,.
\ee
The obtained operators are non-local,\footnote{The non-local character we refer to here has to do with the non-polynomial $\ell$-dependence that denotes the non-local nature of the operators in the angle variables. The operators are local in the radial variable.} although in the large angular momentum limit they take the approximately local form
\bea
\Am_\ell&\simeq& z(z-1)\partial_z-\frac{\ell}{4}\left(z-\frac{1}{2}\right),\\
\Ap_\ell&\simeq&-z(z-1)\partial_z-\frac{\ell}{4}\left(z-\frac{1}{2}\right),
\eea
and they become hermitian-conjugate to each other. The non-locality of the ladder operators is not surprising and, in fact, it is a common feature. For instance, in the 3-dimensional Coulomb problem, the radial function $R(r)$ is determined by the following Hamiltonian (in appropriate units):
\be
H_{\rm C}\;R(r)\equiv-R''(r)+\left[\frac{\ell(\ell+1)}{r^2}-\frac{q}{r}\right]R(r).
\ee
This Hamiltonian admits the factorisation 
\be
H_{\rm C}=A^+_{\ell,\rm C}A^-_{\ell,\rm C}-\frac{q^2}{4\ell^2}=A^-_{\ell+1,\rm C}A^+_{\ell+1,\rm C}-\frac{q^2}{4(\ell+1)^2}
\ee
with the ladder operators
\bea
A^{\pm}_{\ell,\rm C}\equiv \mp\partial_r+\frac{\ell}{r}-\frac{q}{2\ell}\,.
\eea
The remarkable difference with our ladder is that, while in the Coulomb problem the ladder connects adjacent multipoles, our ladder climbs from $\ell$ to $\ell\pm4$.

\begin{figure}
\includegraphics[width=0.45\linewidth]{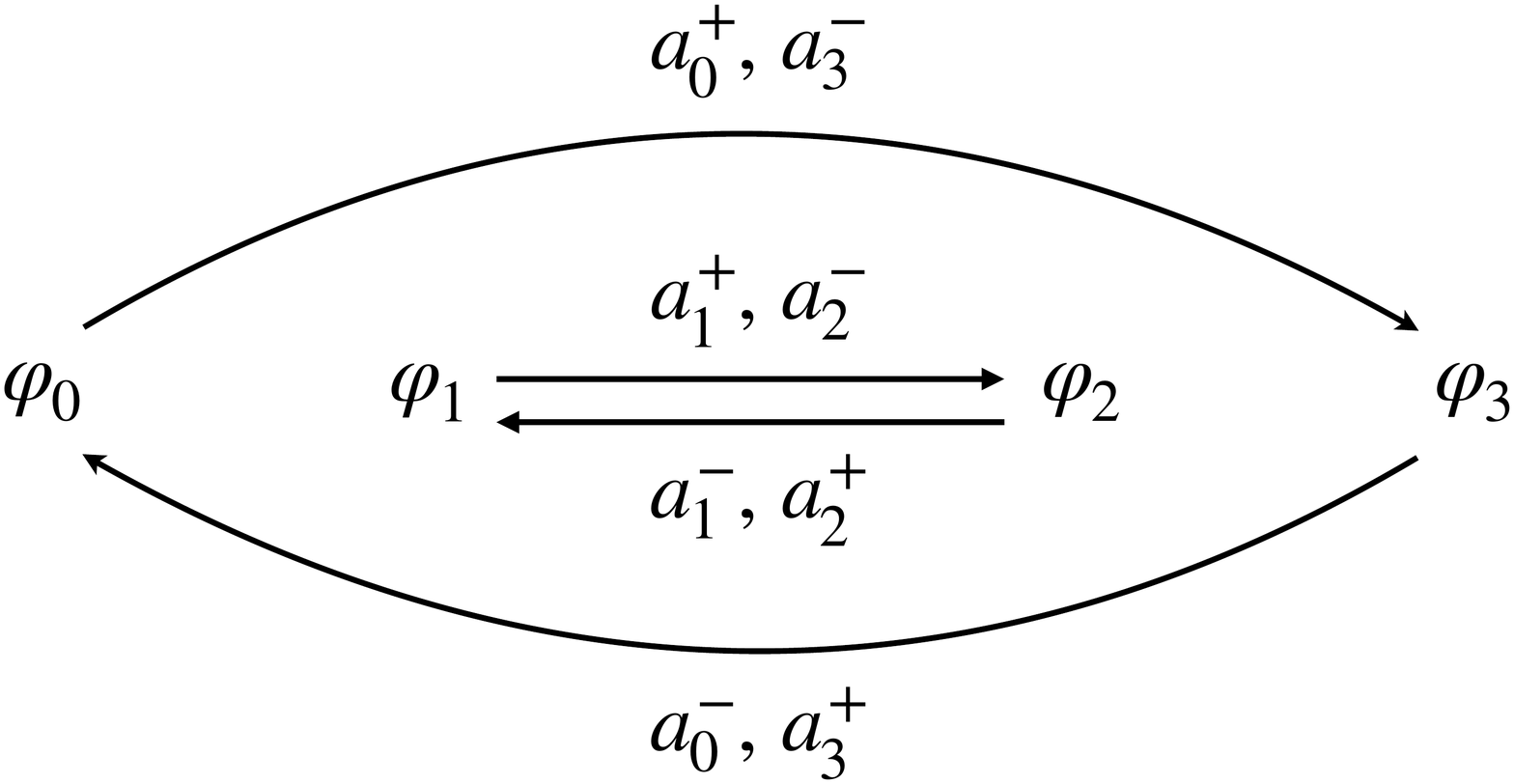}\hspace{1cm}
\includegraphics[width=0.45\linewidth]{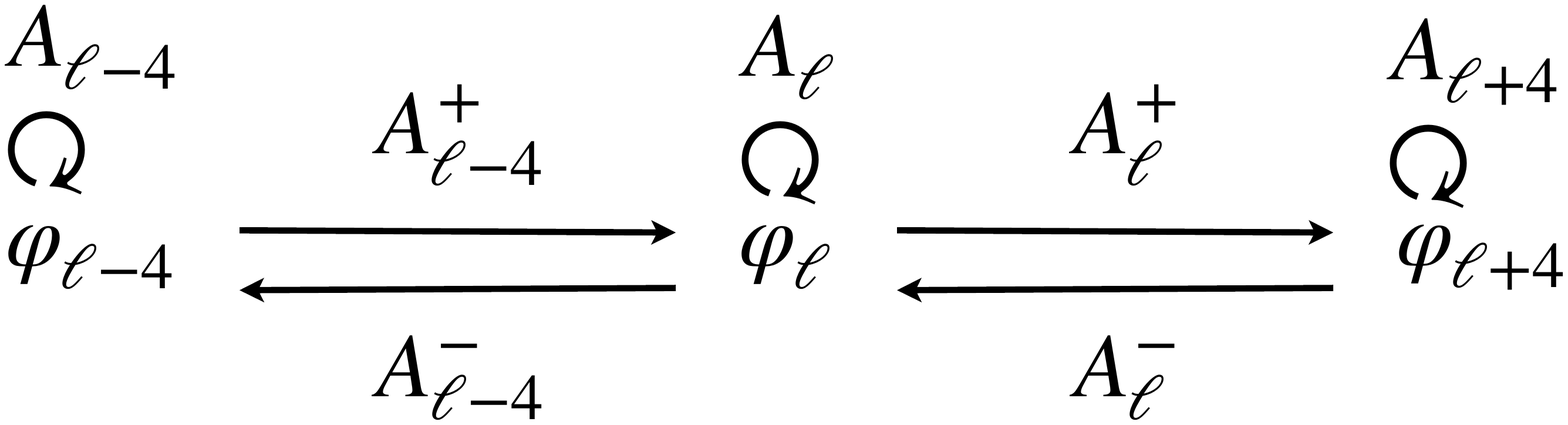}
\caption{In this Figure we show the natural action of the small (left) and big (right) ladders. The operators $a_\ell^+$ and $a_{3-\ell}^-$ act on the same spaces and their images are also in the same space and so they are redundant in the sense explained in the main text. For the big ladder, there is no analogous relation and we can define the operator $A_\ell$ to describe the difference between climbing down and then up from climbing up first and then down. Notice that, since the two ladders have different images, they commute.}
\label{Fig:LadderAlgebra}
\end{figure}

An interesting feature of the resulting factorisation  is that $\varepsilon_\ell$ vanishes for $\ell=0$ and $\ell=1$ and this means that
\be
\ker \Am_0 \Ap_0=\ker H_0\quad \text{and}\quad \ker \Am_1 \Ap_1=\ker H_1\,,
\ee
a property that will obstruct the construction of all the higher multipoles from the first ones by using the ladder operators, as we will discuss in more detail in Sec. \ref{sec:charges}. From \eqref{eq:defApAm}, we can also obtain the following useful intertwining relations
\be
\Ap_\ell H_\ell=H_{\ell+4}\Ap_\ell\,,\quad \Am_\ell H_{\ell+4}=H_{\ell}\Am_\ell\,.
\label{eq:commutationHAAd}
\ee
Thus, we have that if $H_\ell\varphi_\ell=0$, then $\Ap_\ell\varphi_\ell$ is a solution of $H_{\ell+4}$, i.e., $\Ap_\ell$ raises $\varphi_\ell$ by four $\ell-$steps. Likewise, $\Am_\ell\varphi_{\ell+4}$ solves $H_{\ell}$ if $H_{\ell+4}\varphi_{\ell+4}=0$ so $\Am_\ell$ lowers four $\ell-$steps. We thus have the natural actions (see Fig.\ref{Fig:LadderAlgebra})
\bea
\Ap_\ell:\quad\varphi_\ell\to\varphi_{\ell+4}\,,\\
\Am_\ell:\quad\varphi_{\ell+4}\to\varphi_{\ell}\,.
\eea
Using these natural actions together with \eqref{eq:commutationHAAd} allows to define the following diagonal operator:
\be
2A_\ell\equiv\Ap_{\ell-4}\Am_{\ell-4}-\Am_\ell\Ap_\ell
=\varepsilon_{\ell-4}-\varepsilon_\ell\,.
\ee
This operator measures the non-commutativity of the operations one-step-down $\to$ one-step-up and one-step-up $\to$ one-step-down. We can also compute the non-commutativity of {\it jumping} on a step (action of $A_\ell$) and then going one step up ($\Ap_\ell$) or down ($\Am_{\ell-4}$):
\bea
A_{\ell+4}\Ap_\ell-\Ap_\ell A_\ell&=&\frac12\Big(2\varepsilon_\ell-\varepsilon_{\ell-4}-\varepsilon_{\ell+4}\Big) \Ap_\ell\equiv\lambda_+(\ell)\Ap_\ell\,,\\
A_{\ell-4}\Am_\ell-\Am_\ell A_\ell&=&-\frac12\Big(2\varepsilon_{\ell-4}-\varepsilon_{\ell-8}-\varepsilon_{\ell}\Big) \Am_\ell\equiv\lambda_-(\ell)\Am_\ell\,.
\eea
We can verify the relation $\lambda_+(\ell)=-\lambda_-(\ell-4)$. In the limit of high angular momentum $\ell\gg1$, we find $\lambda_+\simeq\lambda_-\simeq-1/4$ so it becomes independent of $\ell$ in this regime.

The ladder structure unveiled above permits to organize the  multipoles into multiplets formed by
\be
(\vec{\Phi}_L)_i\equiv \Phi_{4L+i-1}
\ee
and we can introduce the operators
\be
\Big(\hat{\Ap}_{L}\Big)_{ij}\equiv \Ap_{4L+i-1}\delta_{ij}
\ee
that act on the multiplets as
\be
\hat{A}^+_L\vec{\Phi}_L=\vec{\Phi}_{L+1}\,,
\ee
which resembles the more traditional action of a ladder operator connecting adjacent multipoles.

\subsubsection{Small polar ladder}

The second solution for the factorization with $n=3-2\ell$ gives rise to a ladder with finer steps. This ladder connects $\ell$ with $\ell+n=3-\ell$. Since $\ell\geq0$, this ladder only reaches up to $\ell=3$ and provides an automorphism for the first four multipoles. The ladder operators in this case read
\bea
\am_\ell&\equiv& z(z-1)\partial_z+\frac{\ell-5}{4}\left(z-\frac{\ell+1}{2\ell-3}\right)\,,\\
\ap_\ell&\equiv&-z(z-1)\partial_z+\frac{\ell+2}{4}\left(z-\frac{\ell-4}{2\ell-3}\right)\,,
\eea
and the Hamiltonian factorises as
\bea
\am_\ell \ap_\ell=H_\ell+\varepsilon_\ell,\quad\quad \ap_\ell \am_\ell=H_{3-\ell}+\varepsilon_\ell\,,
\eea
with
\be
\varepsilon_\ell=\frac{(\ell-5)(\ell-4)(\ell+2)(\ell+1)}{16(2\ell-3)^2}\,.
\ee
As the big ladder, this small ladder operators are non-local, but its large $\ell$ limit gives the approximately local mutually hermitic-conjugate operators
\bea
\am_\ell\simeq z(z-1)\partial_z+\frac{\ell}{8}\Big(2z-1\Big)\,,\\
\ap_\ell\simeq-z(z-1)\partial_z+\frac{\ell}{8}\Big(2z-1\Big)\,,
\eea

For this small ladder we now have that $\varepsilon_\ell$ vanishes for $\ell=4$ and $\ell=5$, but this ladder is only defined for $\ell\leq3$ so we will not have any identification of kernels as we found for the big ladder. We can further obtain the useful relations
\be
\ap_\ell H_\ell=H_{3-\ell}\ap_\ell\,,\quad \am_\ell H_{3-\ell}=H_{\ell}\am_\ell\,,
\label{eq:commutationHaad}
\ee
that show how the small ladder connects the solutions of the first four multipoles so that $\ap_\ell\varphi_\ell$ and $\am_\ell\varphi_{3-\ell}$ solve the equations corresponding to $H_{3-\ell}$ and $H_{\ell}$ respectively. Furthermore, the fact that $\varepsilon_{3-\ell}=\varepsilon_\ell$, as can be explicitly checked, allows to obtain the following {\it commutation} relations:
\be
\ap_{3-\ell}\am_{3-\ell}-\am_\ell\ap_\ell=\varepsilon_{3-\ell}-\varepsilon_\ell=0\,.
\ee
In this case we obtain that these operators realise an Abelian algebra.
This small ladder satisfies the following additional relation 
\be
\ap_{3-\ell} \ap_\ell=-\am_{\ell} \ap_\ell
\ee
that has no analogue in the big ladder. We then have that, for physical solutions $\varphi_\ell$
\be
\ap_{3-\ell} \ap_\ell\varphi_\ell=-\am_{\ell} \ap_\ell\varphi_\ell=-\varepsilon_\ell\varphi_\ell\,.
\ee
This means that we only need to use one set of operators, either the plus or the minus, to move within the first four steps, while the wider ladder constructed with $\Am_\ell$ and $\Ap_\ell$ allows us to move to higher $\ell$'s. In combination, they allow, in principle, to reach any level starting from the first two levels. However, as commented above and will be shown below, the non-trivial kernels of the big ladder operators for the first multipoles represents an obstruction for this construction.

Finally, the small ladder acts as a permutation of the multiplet components of $\vec{\Phi}_0$ by means of the following anti-diagonal operator:
\be
\hat{a}^+=
\left[\begin{array}{cccc}
0 & 0 & 0 & 1 \\
0 & 0 & 1 & 0 \\
0 & 1 & 0 & 0 \\
1 & 0 & 0 & 0
\end{array}\right]
\ee
that encodes the action of $a^+_\ell$ on the elements of $\vec{\Phi}_0$.

\subsection{Axial ladder}

In order to obtain a ladder for the axial sector we will follow the very same steps as for the polar sector. We start from the hypergeometric form of the perturbation  equations Eq.~\eqref{eq:PolarHyperequ} that we reproduce here
\be
z(1-z)\Psi_\ell''+\frac{1-3z}{4}\Psi_\ell'+\frac{e^{-2\gamma}\ell(\ell+1)}{16}\Psi_\ell=0
\label{eq:AxialHyper}
\ee
and introduce the family of Hamiltonians
\be
H_\ell\equiv -z(1-z)\left[z(1-z)\partial^2_z+\frac{1-3z}{4}\partial_z+\frac{e^{-2\gamma}\ell(\ell+1)}{16}\right]
\ee
whose kernels span the space of solutions of the axial sector multipoles. We then look for operators $B_\ell^+$ and $B_\ell^-$ that factorize $H_\ell$ as
\bea
\Bm_\ell \Bp_\ell&=&H_\ell+\delta_{1\ell}\,,\nonumber\\
\Bp_\ell \Bm_{\ell}&=&H_{\ell+n}+\delta_{2\ell}\,.
\label{Eq:FactorizationHOdd}
\eea
for some integer $n$ and scalar functions $\delta_{1\ell}$ and $\delta_{2\ell}$. We will make an analogous Ansatz:
\bea
\Bm_\ell&\equiv& z(z-1)\partial_z+W_{1,\ell}(z)\,,\\
\Bp_\ell&\equiv&-z(z-1)\partial_z+W_{2,\ell}(z)\,,
\eea
and, after requiring the factorisation \eqref{Eq:FactorizationHOdd}, we obtain that the two functions $W_{1,\ell}$ and $W_{2,\ell}$ must be related as
\be
W_{2,\ell}=\frac{5z-3}{4} +  W_{1,\ell}\,,
\label{eq:relW1W2A}
\ee
while the regular solution for $W_\ell\equiv W_{1,\ell}$ is now
\be
W_\ell(z)=-\frac{1}{32} \left[20 + e^{-2\gamma}n\big(1 + 2 \ell  + n\big)\right] z + c_\ell\,.
\ee
With  this solution, we find that $\delta_{1,\ell}=\delta_{2,\ell}\equiv\delta_\ell$ is given by
\bea
\delta_\ell&=&
\frac{240+e^{-4\gamma}n^2(n+2\ell+1)^2-32e^{-2\gamma}\Big[n(n+1)+2\ell(n+\ell+1)\Big]}{1024}z^2\nonumber\\
&&+\frac{e^{-2\gamma}\Big[8 \ell^2 + n (1 + n) (7 - 8 c_\ell) + 2 \ell (4 + 7 n - 8 n c_\ell)\Big]-20}{128} z   
     - \frac{3c_\ell}{4} + c_\ell^2\,.
\eea
We can then choose $c_\ell$ to remove the term linear in $z$ as
\be
c_\ell=\frac{ 8 \ell^2 + 7 n (n+1) + 2 \ell (7 n+4)-20e^{2\gamma}}{8 n (1 + 2 \ell + n)}\,.
\ee
However, the above expression shows that only certain values of $\gamma$ lead to the desired factorisation since now we need to have 
\be
240+e^{-4\gamma}n^2(n+2\ell+1)^2-32e^{-2\gamma}\Big[n(n+1)+2\ell(n+\ell+1)\Big]=0
\label{eq:gammaaxial}
\ee
for some integer $n$. This is not sufficient however. We can solve the above equation for $e^{-2\gamma}$ so we obtain the two branches
\bea
e^{-2\gamma_\pm}&=&
\frac{4}{n^2(n+2\ell+1)^2}\Big[8n(n+1)+16\ell(n+\ell+1)\\
&&\pm\sqrt{64 \ell^2 (\ell+1)^2 + 
 64 \ell ( \ell+1) (2 \ell+1) n + (1 + 68 \ell (\ell+1)) n^2 + 
 2 ( 2 \ell+1) n^3 + n^4}\Big]\,.\nonumber
\eea
In order to have a ladder structure, the obtained value for $\gamma$ must be independent of $\ell$ so we must further impose
\be
\frac{\dd}{\dd\ell}e^{-2\gamma_\pm}=0\,.
\label{eq:gammapml}
\ee
Now, we could consider the case of an $\ell-$dependent step for the ladder, but we are seeking for a ladder of fixed step so we further require that $n$ does not depend on $\ell$. We thus solve \eqref{eq:gammapml} for some integer $n$. We find that the branch $e^{-2\gamma_-}$ does not have solutions while the positive branch $e^{-2\gamma_+}$ has one solution for $n=4$ that is in turn unique and gives $\gamma_+=0$. To show this, we can take the derivative w.r.t. $\ell$ of \eqref{eq:gammaaxial} keeping both $\gamma$ and $n$ independent of $\ell$ so we obtain the following equation:
\be
n=4e^{\gamma}
\ee
whose unique solution is in fact $\gamma=0$ and $n=4$. This solution indeed gives an $\ell-$independent value for $n$ so it fulfils all our requirements. Thus, we find that there is only one value of $\gamma$ that allows for a ladder structure with fixed step in the axial sector. Remarkably, the obtained value for the parameter $\gamma$ selects nothing other than the Born-Infeld theory. In other words, among all the Born-Infeldized ModMax theories, only Born-Infeld exhibits a ladder structure in both the polar and the axial sectors with the required properties. Furthermore, when we replace  $\gamma=0$ in \eqref{eq:gammaaxial}, the equation reduces to 
\be
(n^2-16)(n+2\ell-3)(n+2\ell+5)=0,
\ee
so we obtain the same two ladder structures as in the polar case connecting $\ell\to\ell+4$ and $\ell\to3-\ell$. This adds to the collection of remarkable properties of Born-Infeld theory. In the following we will analyse these two ladders obtained for Born-Infeld theory in detail. Most of the properties are shared with the polar ladders, so we will save repeating the same discussions and will simply quote the main expressions. There are, however, some interesting differences that will in turn be at the heart of the different behaviour of the polarisability and the magnetisation found in the preceding sections.

\subsubsection{Big axial ladder}
The explicit expressions for the big ladder are given by
\bea
\Bm_\ell&\equiv& z(z-1)\partial_z-\frac{\ell+5}{4}\left(z-\frac{\ell+3}{2\ell+5}\right),\\
\Bp_\ell&\equiv&-z(z-1)\partial_z-\frac{\ell}{4}\left(z-\frac{\ell+2}{2\ell+5}\right).
\eea
These ladder operators permit to write the factorisations
\bea
\Bm_\ell \Bp_\ell=H_\ell+\delta_\ell,\quad\quad \Bp_\ell \Bm_\ell=H_{\ell+4}+\delta_\ell,
\eea
with
\be
\delta_\ell=\frac{(\ell+5)(\ell+3)(\ell+2)\ell}{16(2\ell+5)^2}.
\ee
As for the polar case, we have the intertwining relations
\be
\Bp_\ell H_\ell=H_{\ell+4}\Bp_\ell\,,\quad \Bm_\ell H_{\ell+4}=H_{\ell}\Bm_\ell,
\label{eq:commutationHAAdAx}
\ee
from which it is immediate to obtain that $H_{\ell+4}(\Bp_\ell\psi_\ell)=0$ if $H_{\ell}\psi_\ell=0$ so the solutions for multipoles separated by four $\ell$ units can be connected via this big ladder. All the same properties and relations discussed for the big polar ladder apply to this ladder as well so we will not repeat it and we will proceed directly to small axial ladder.

\subsubsection{Small axial ladder}
For the small ladder we obtain
\bea
\bm_\ell&\equiv& z(z-1)\partial_z-\frac{\ell-4}{4}\left(z+\frac{\ell-2}{3-2\ell}\right),\\
\bp_\ell&\equiv&-z(z-1)\partial_z-\frac{\ell+1}{4}\left(z+\frac{\ell-1}{3-2\ell}\right),
\eea
that produce the factorisation
\bea
\bm_\ell \bp_\ell=H_\ell+\delta_\ell,\quad\quad \bp_\ell \bm_\ell=H_{3-\ell}+\delta_\ell,
\eea
with
\be
\delta_\ell=\frac{(\ell-4)(\ell-2)(\ell-1)(\ell+1)}{16(2\ell-3)^2}.
\ee
The same multiplets structure as in the polar case can therefore be introduced for the axial sector. In this case we also have the commutation relations
\be
\bp_{3-\ell}\bp_\ell=-\bm_\ell \bp_{\ell}=0,
\ee
so that we again have a redundancy among the operators of the small ladder. 

The difference with the polar sector is that $\delta_\ell$ vanishes now only for $\ell=0$ for the big ladder and for $\ell=1$ and $\ell=2$ for the small ladder. This means that the small ladder will have non-trivial kernels in the space of solutions and it will prevent the construction of all the higher multipole solutions starting from the lowest ones. This is distinctive of the axial sector, since the polar sector presents trivial kernels only in the big ladder, while the small ladder permits to move between the first four multipoles without obstructions. We will show these obstructions more explicitly in the next section.

Before concluding this section, we cannot resist to observe that the ladder structures that we have unveiled in both the polar and the axial sector for the Born-Infeld theory have a striking relation with the dimension of spacetime. The big ladder connects multipoles that are separated by the number of dimensions\footnote{The fact that the big ladder connects $\ell$ and $\ell+4$ is related to the fourth power introduced in the radial coordinate redefinition $z=-x^4$.} while the small ladder establishes the same relation as the Hodge dual for differential forms in four dimensions. It will be interesting to obtain the ladders in an arbitrary dimension to see if similar coincidences occur, thus signalling an underlying connection between the size of the steps of the ladders and the dimensionality of the spacetime. We will not explore this surmise any further here and will proceed to discussing the existence of conserved charges.

\subsection{Conserved charges}
\label{sec:charges}

We are now ready to undertake the construction of conserved charges. We will commence by observing how the commutation relations \eqref{eq:commutationHAAd} may serve the purpose of generating a hierarchy of conserved charges from a known one. Let us assume that we have a charge generator $\mI_\ell$ which satisfies $[\mI_\ell,H_\ell]=0$. Then, we can construct the generator $\mI_{\ell+4}\equiv \Ap_\ell \mI_\ell \Am_\ell$ that commutes with $H_{\ell+4}$, as can be seen by a direct computation:\footnote{It may be worth noticing that our family of Hamiltonians depend explicitly on $z$ and so do the conserved charges that we will unveil. It is however possible to change coordinates to avoid this issue, although other properties are more obscure in the transformed coordinates. In principle, one should check if and how the introduced Hamiltonians generate {\it time evolution} (i.e., translation along $z$ in our case) and use the appropriate condition $[\mI,H]+\partial_z\mI=0$ for conserved charges that depend explicitly on the coordinate. All these issues will not be important for us because we will use an alternative procedure to construct the conserved charges.}
\bea
[\mI_{\ell+4},H_{\ell+4}]
&=&\Ap_\ell \mI_\ell \Am_\ell H_{\ell+4}-H_{\ell+4}\Ap_\ell \mI_\ell \Am_\ell\nonumber\\
&=&\Ap_\ell \mI_\ell H_{\ell}\Am_\ell -\Ap_\ell H_{\ell} \mI_\ell \Am_\ell
=\Ap_\ell[\mI_\ell,H_\ell]\Am_\ell=0\,.
\eea
Analogously, we can construct $\mI_{3-\ell}\equiv \ap_\ell \mI_\ell \am_\ell$ that satisfies 
\be
\Big[\mI_{3-\ell},H_{3-\ell}\Big]=\ap_\ell \Big[\mI_{\ell},H_{\ell}\Big]\am_\ell=0\,.
\ee
Thus, these two hierarchies allow to construct a hierarchy of conserved charges for all multipoles from the conserved charge of a given multipole. These hierarchies of conserved charges can indeed be obtained by noticing that the equations for the monopole of the axial sector and dipole of the polar sector can be written in the form of a conservation law since in both cases the non-derivative term of the equations vanish. In particular, this means that both will admit a constant mode solution. For these cases, it is trivial to obtain a conserved charge. 

We will exploit this fact to construct the hierarchy of conserved charges and obtain them by climbing down the ladder from an arbitrary angular multipole $\ell$ until  reaching  the lower multipole $\ell'$ for which the conserved charge exists. In practice, we will have $\ell'=0,1$ and this will provide the associate charge for the higher multipoles. For instance,
if we have a conserved charge for the monopole generated by $\mQ_0$, then we can define the conserved charge at level $\ell=4k$ with $k=1,2,3,\dots$ as  $\mQ_\ell\equiv \mQ_0\Am_1\cdots\Am_{4(k-1)}$. We can proceed analogously for a charge in the dipole and we should notice that the last ladder operator should be replaced by the corresponding small ladder operator for the multipoles that are connected to $\ell=2$ and $\ell=3$ so we can eventually reach the monopole or the dipole. Let us see how this works explicitly for each sector. Before proceeding, it is convenient to make a couple of important remarks about the ladder operators. 

Near the origin, all the ladder operators take the approximate form $\pm z\partial_z+w_\ell$ with $w_\ell$ some constant that only depends on $\ell$. On the other hand, the solutions for the multipoles have the generic expression near the origin $\sim c_1+c_2z^{1/4}$ where $c_1$ and $c_2$ correspond to the singular and the regular modes respectively. The action of the ladder operators close to the origin reduces to $(\pm z\partial_z+w_\ell)(c_1+c_2z^{1/4})\sim w_\ell c_1+(w_\ell\pm1/4)c_2z^{1/4}$, i.e., the ladder operators transform regular modes into regular modes and singular modes into singular modes. On the other hand, at large $z$, the ladder operators take the asymptotic form $\pm z^2\partial_z+v_\ell z$ with $v_\ell$ some constants, while the solutions for the multipoles in this region reduce to power laws. Thus, the action of the ladder operators in this region amounts to raising one power of $z$.

\subsubsection{Polar sector}

We will start by analysing the dipole that exhibits an obvious conserved charge and we will then proceed to the monopole where, although less evident, it is also possible to find a conserved charge.

\vspace{0.5cm}
\noindent\bf{Dipole}\\

 The equation for the polar perturbations with $\ell=1$ can be written as
\be
\frac{\dd}{\dd z}\left[\frac{z^{3/4}}{\sqrt{1-z}}\frac{\dd\Phi_{1}}{\dd z}\right]=0
\ee
so it is immediate to identify a conserved charge generated by:\footnote{This generator commutes with the $\ell=1$ level Hamiltonian in the sense that $\left[\mQ_{1},\frac{H_1}{(1-z)^3\sqrt{z}}\right]=0$. Alternatively, we can define the Hamiltonian $\tilde{H}_1\equiv\frac{z^{3/4}}{\sqrt{1-z}}H_1$ with the same kernel as $H_1$ so the equation becomes $\frac{\dd\Phi_1}{\dd\rho}=0$ with the coordinate $\rho$ defined by $\frac{\dd}{\dd \rho}=\frac{z^{3/4}}{\sqrt{1-z}}\frac{\dd}{\dd z}$. Thus, $\mQ_1$ is the generator of translations along this coordinate and the conserved charge is the corresponding momentum, i.e., this is the cyclic coordinate adapted to the symmetry.}
\be
\mQ_{1}[\Phi_1]=\frac{z^{3/4}}{\sqrt{1-z}}\Phi_{1}'(z).
\ee
Recalling that the solution of the multipole near the origin behaves as $\Phi_1\simeq c_1+c_2z^{1/4}$ with $C_1$ and $C_2$ the singular and regular solutions respectively, we can obtain the relation $Q_{1}\equiv\mQ_{1}[\Phi_1]=\frac14C_2$ so we see that the regular (physical) solution carries a non-trivial charge, while the singular mode is identically annihilated by $\mQ_1$. We can then express the solution for the dipole as\footnote{We use $\varphi_\ell$ for the space of solutions of the corresponding multipole equation, while $\Phi_\ell$ denotes the multipole variable not necessarily on-shell, so $\Phi_\ell\vert_{\rm on-shell}=\varphi_\ell$.}
\be
\varphi_1=c_1+4 Q_1 z^{1/4} \;_2F_1\left(-\frac12,\frac{1}{4},\frac{5}{4}, z\right).
\ee
As explained above, the ladder operators connect regular modes to regular modes so we can, in principle, generate all the physical solutions connected with the dipole via the ladders from the dipole with a non-trivial charge and this charge should eventually determine the conserved charges of all those multipoles. There can however be some obstructions if the kernel of some ladder operators have a component on the space of solutions, i.e., if the kernels of the ladder operators and the Hamiltonians have a non-trivial intersection. We have already discussed above that this is the case, so let us see how it affects the construction of the hierarchy of charges.

By employing the small ladder we can generate the $\ell=2$ multipole as
\be
\varphi_2=\ap_1\varphi_1
\ee
that, together with the big ladder, permits to obtain all the multipoles $\Phi_{4k-2}$ with $k=1,2,3,\dots$ from the dipole. Since neither $\ap_1$ nor $A^{\pm}_1$ have kernels on the space of solutions, this path will be safe. On the other hand, climbing up with the big ladder directly from $\varphi_1$ to generate the tower of multipoles $\Phi_{4k+1}$, does exhibit this obstruction since the constant mode $c_1$ (that is associated to a trivial charge $Q_1=0$) clearly belongs to the kernel of $\Ap_1=-z (z-1)\partial_z$ so we cannot raise it with the big ladder. Fortunately, this constant mode that forms the kernel of $\Ap_1$ is the singular mode and the regular mode corresponding to a non-trivial charge $Q_1\neq 0$ gives the physical regular solution. Incidentally, this means that the image of $\Ap_1$ is the regular mode of $\Phi_5$. Since the higher $\ell$ operators in the tower have trivial kernels in the space of solutions, we can raise the relevant physical solution to all those multipoles. We can be more explicit and construct the $\ell=5$ solution as
\be
\varphi_5=\Ap_1\varphi_1=(1-z)^{3/2}z^{1/4}Q_1,
\label{eq:varphi5regular}
\ee
where the singular mode with trivial charge has been projected out and we obtain the regular solution for $\ell=5$ as associated to a non-trivial $Q_1$. Furthermore, we see that $\Ap_1$ has generated a purely growing solution which means that its polarisability will vanish due to the absence of a decaying mode. Recalling that all raising operators are linear operators with polynomial coefficients, all the higher order multipoles connected to $\ell=1$ via the repeated application of the corresponding raising operators $\Ap_{4k+1}$ will also be purely growing functions and, hence, the polarisability of the multipoles with $\ell=4k+1$ for $k=1,2,3,\dots$  will also vanish. This shows the advertised direct relation between the vanishing of the polarisability for all these multipoles and the regularity of the $\ell=1$ solution as being ascribed to having a non-trivial charge $Q_1$. 

We can now construct the two hierarchies of conserved charges connected with the dipole. The first hierarchy corresponds to the multipoles $\ell=4k-2$ with $k=1,2,3,\dots$, i.e., $\ell=2,6,10,\dots$. The conserved charges for these multipoles are defined as
\be
\mQ_{4k-2}\Phi_{4k-2}\equiv \mQ_1\am_1\Am_2\cdots\Am_{4k-6}\Phi_{4k-2}.
\ee
It is pleasant to see how the conservation of these charges permits to connect the presence of an asymptotically decaying mode for these multipoles with the value of the corresponding charge. The decaying solution behaves asymptotically as $\Phi_\ell^{\rm dec}\sim z^{(1-\ell)/4}$. Since each ladder operator essentially acts as raising one power of $z$ in this asymptotic zone, we have that $\am_1\Am_2\cdots\Am_{4k-6}\Phi_{4k-2}\sim z^k\Phi_{4k-2}\sim z^{3/4}$, which is not annihilated by $\mQ_1=z^{3/4} (1-z)^{-1/2}\partial_z$. Thus, the solutions with a decaying tail necessarily have a non-trivial charge $Q_{4k-2}\neq0$. Since the singular mode has $Q_1=0$, we conclude that the regular solutions for this tower of multipoles have decaying modes and, hence, their polarisability does not vanish. 

The second tower of conserved charges occurs for $\ell=4k+1$, i.e., $\ell=5,9,13,\dots$. Following the same procedure, we can aim at constructing the conserved charges for these multipoles as
\be
\mQ_{4k+1}\Phi_{4k+1}\equiv \mQ_1\Am_1\Am_2\cdots\Am_{4k-3}\Phi_{4k+1}.
\ee
However, an obstruction occurs again because the kernel of the lowering operator $\Am_1$ precisely corresponds to the regular sector of $\ell=5$. This is corroborated by solving the equation $\Am_1\Phi_5=0$ whose solution is $\Phi_5\propto (1-z)^{3/2}z^{1/4}$, precisely the regular solution given in \eqref{eq:varphi5regular}. This has two important consequences. Firstly, although the regular solution for $\ell=5$ expressed in \eqref{eq:varphi5regular} is nicely achievable from the regular solution of the dipole, the fact that is annihilated by $\Am_1$ impedes climbing back down. This means that the tower of multipoles connected with $\ell=5$ via the big ladder cannot climb all the way down to the dipole, but it ends at $\ell=5$. If we start from an arbitrary multipole $\ell=4k+1$ with $k=1,2,\dots$ and climb down with $\Am_\ell$, when reaching $\ell=5$, the regular solution is projected out and we end up in the singular sector of the dipole with $Q_1=0$. Since the hierarchy of charges for these multipoles is generated precisely by translating it into the dipolar charge, we conclude that $Q_{4k+1}=0$ with $k=1,2,\dots$ for the charges defined above. Let us emphasise however that the regular solutions for this tower of multipoles are connected to the non-trivial charge $Q_1\neq0$ since they can be obtained by climbing up the ladder, although the connection is not both ways. This fact stems from having $\varepsilon_1=0$ so $\Am_1\Ap_1=H_1$.\\

\noindent\bf{Monopole}\\

It is less evident to see that the monopole also has a conserved charge. To see that, we first notice that we can introduce an appropriate integrating factor to write the $\ell=0$ equation as
\be
\frac{\dd}{\dd z}\left[\frac{z^{5/4}}{\sqrt{1-z}}\frac{\dd}{\dd z}\left(\frac{\Phi_0}{z^{1/4}}\right)\right]=0,
\ee
from where it is immediate to identify the conserved charge generated by
\be
\mQ_{0}[\Phi_0]=\frac{z^{5/4}}{\sqrt{1-z}}\frac{\dd}{\dd z}\left(\frac{\Phi_0}{z^{1/4}}\right).
\ee
From the solution of the monopole near the origin $\Phi_0\simeq c_0+c_1 z^{1/4}$ we see that now it is the physical regular solution that is annihilated by the generator, while the singular mode gives a non-vanishing charge. Thus, we can express the solution in terms of $Q_0\equiv\mQ_{0}[\Phi_0]$ as
\be
\varphi_0=c_0z^{1/4}-4 Q_0 \;_2F_1\left(-\frac12,-\frac{1}{4},\frac{3}{4}, z\right).
\ee
and regularity selects the trivial charge sector in this case. It turns out that the regular mode ($Q_0=0$) spans the kernel of $\Ap_0$ so $\Ap_0$ projects out the regular mode sector. The situation is worse than what occurred for the dipole, because here we do encounter an obstruction to generate the physical solutions for the multipoles $\Phi_{4k}$ from the monopole via the big ladder. Since the ladder operators connect regular modes with regular modes and singular modes with singular modes, we have that the image of $\Ap_0$ is the singular sector of $\ell=5$.

Since the monopole is connected with the multipoles $\ell=4k$ for $k=1,2,4,\dots$ via the big ladder we can construct the corresponding tower of conserved charges as
\be
\mQ_{4k}\Phi_{4k}\equiv \mQ_0\Am_0\cdots\Am_{4k-4}\Phi_{4k}.
\ee
However, the non-trivial kernel of $\Am_0$ has an important consequence. Since this kernel is given by the singular sector of $\ell=4$, we can still climb all the way down to the monopole from an arbitrary $\Phi_{4k}$. This is how all the physical modes are connected although the connection is not both way. Since the regular solution for the monopole has trivial charge, we conclude that all the physical solutions for the multipoles in this tower have trivial charge. The situation is similar to the multipoles reached from the dipole by the action of the big ladder except that here the connection is downwards only and there it was exclusively upwards.

Similarly, the monopole is connected with the multipoles $\ell=4k+3$ by going to $\ell=3$ with $\ap_0$ and then climbing up with the big ladder. The tower of charges for these multipoles is then constructed as
\be
\mQ_{4k-1}\Phi_{4k-1}\equiv \mQ_0\am_0\Am_3\cdots\Am_{4k-5}\Phi_{4k-1},
\ee
with the understanding that the big ladder operators only appear for $k\geq2$. For this tower of multipoles there are no obstructions from the kernels of the ladders so it is a safe two way path. We can now apply again an argument based on the conservation of these charges to relate the decaying mode with the physical solutions. We now have that, asymptotically, $\am_0\Am_3\cdots\Am_{4k-5}\Phi_{4k-1}\sim z^{-1/2}$, which is not annihilated by $\mQ_{0}$. This means that a decaying mode at infinity requires a non-trivial charge. Since, as we have seen, the regular mode requires a trivial charge, we conclude that the physical solutions for multipoles $\ell=4k-1$ do not contain decaying modes. This shows again how the physical condition selecting a trivial charge for the monopole relates to the vanishing of the polarisability for the odd multipoles $\ell=4k-1$, i.e., $\ell=3,7,11,\cdots$ as we obtained more directly in the preceding sections.

We can see the explicit construction of the above general argument for the lowest multipoles. With $\varphi_1$ we can generate the $\ell=3$ modes by acting with the small ladder
\be
\varphi_3=\ap_0\varphi_0\,.
\ee
It is straightforward to confirm that this solutions coincides with the one obtained in \eqref{eq:solPsigen}. The corresponding conserved charge is given by
\be
\mQ_{3}[\Phi_3]\equiv \mQ_{0}\am_0\Phi_3
\ee
that is on-shell conserved for the $\ell=3$ multipoles by construction.\footnote{Let us notice that the conservation equation $\mQ_{3}[\Phi_3]=Q_3$ is now higher order so we are introducing spurious solutions that must be eliminated from the physical space (by imposing the solution to belong to the kernel of $H_3$ for instance). This subtlety is not relevant for us here because we want to connect the conserved charge of $\ell=0$ with that of $\ell=3$. A similar situation occurs for the higher multipoles.} For the solution $\varphi_3=\ap_0\varphi_0$ we find
\be
Q_3\equiv \mQ_{3}[\varphi_3]=Q_{0}\am_0\ap_0\varphi_0=Q_{0}\varepsilon_0\varphi_0=\frac{5}{18}Q_0.
\ee
Examining  the explicit solution
\be
\varphi_3=\ap_0\varphi_0=\frac{c_0}{12}z^{1/4}(3z-5)+\frac{3}{8}Q_0 \;_2F_1\left(-\frac52,\frac{1}{2},\frac{3}{4}, z\right)
\ee
we conclude that the regular solution for $\ell=3$ (given by the mode $c_0$)  is connected to the trivial charge of the monopole sector. This can be justified on physical grounds because the monopole solution simply corresponds to a shift of the background charge so it can be eliminated from the perturbative sector. Furthermore, we observe that  regularity selects the  solution whose hypergeometric solution is polynomial so its polarisability vanishes. Applying the same reasoning as above, the solutions obtained by raising the regular $\varphi_3$ solution with the big ladder $\Ap_\ell$ will always generate purely growing solutions and, therefore, all multipoles with $\ell=4k-1$ for $k=1,2,3,\dots$ will have vanishing polarisability. 

In summary, we have shown how the vanishing of the polarisability for odd multipoles can be connected to the vanishing of the conserved charges of the $\ell=0$ and $\ell=1$ sectors. While the tower with $\ell=4k+1$ have vanishing charges because the image of $\Am_1$ is the singular sector with trivial charge (although the physical dipolar mode has non-trivial charge), the multipoles with $\ell=4k-1$ have vanishing charge because the physical monopolar solution has vanishing charge. In the previous section we obtained the two towers of multipoles with vanishing polarisability by simply considering the multipoles for which the regular solution at the origin are expressed in terms of hypergeometric functions that reduce to polynomials.  Now, we have seen that these two series can be associated to the vanishing of the conserved charges $Q_0$ and $Q_1$, although for different reasons.

\begin{figure}
\includegraphics[width=0.99\linewidth]{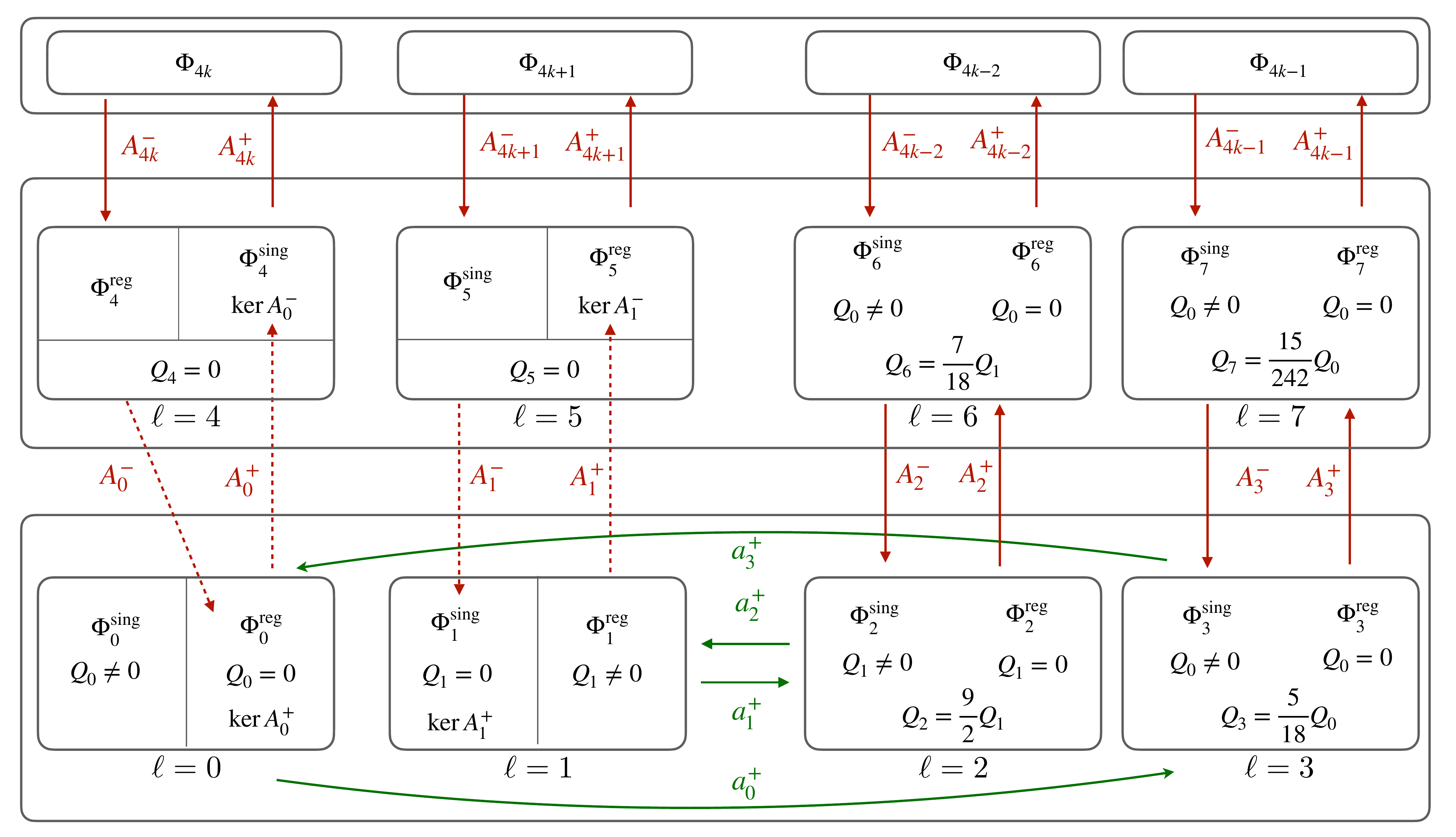}
\caption{This diagram summarises how the solutions for the different multipoles in the polar sector
are related via the ladder operators as well as their relation to the conserved charges discussed in the
main text. The dashed lines denote the relations where non-trivial kernels are present. Since the ladder operators connect regular modes with regular modes and singular modes with singular modes, the fact that a given ladder operator has one of these sectors in its kernel implies that its image must live on the kernel of its complementary ladder operator. In this diagram we observe that the physical solutions in the towers of odd multipoles (second and fourth columns) have vanishing charges. However, this happens for different reasons. In the second column, this occurs because the kernel of $\Am_1$ annihilates the regular solution so the charge is associated to the singular solution of the dipole that is trivial. On the other hand, the fourth column does not contain kernels for the ladder, but the regular solution in that tower selects a vanishing charge for the monopole. These are precisely the multipoles with vanishing polarisability.}
\label{Fig:PolarLadderQ}
\end{figure}

\begin{figure}
\includegraphics[width=0.99\linewidth]{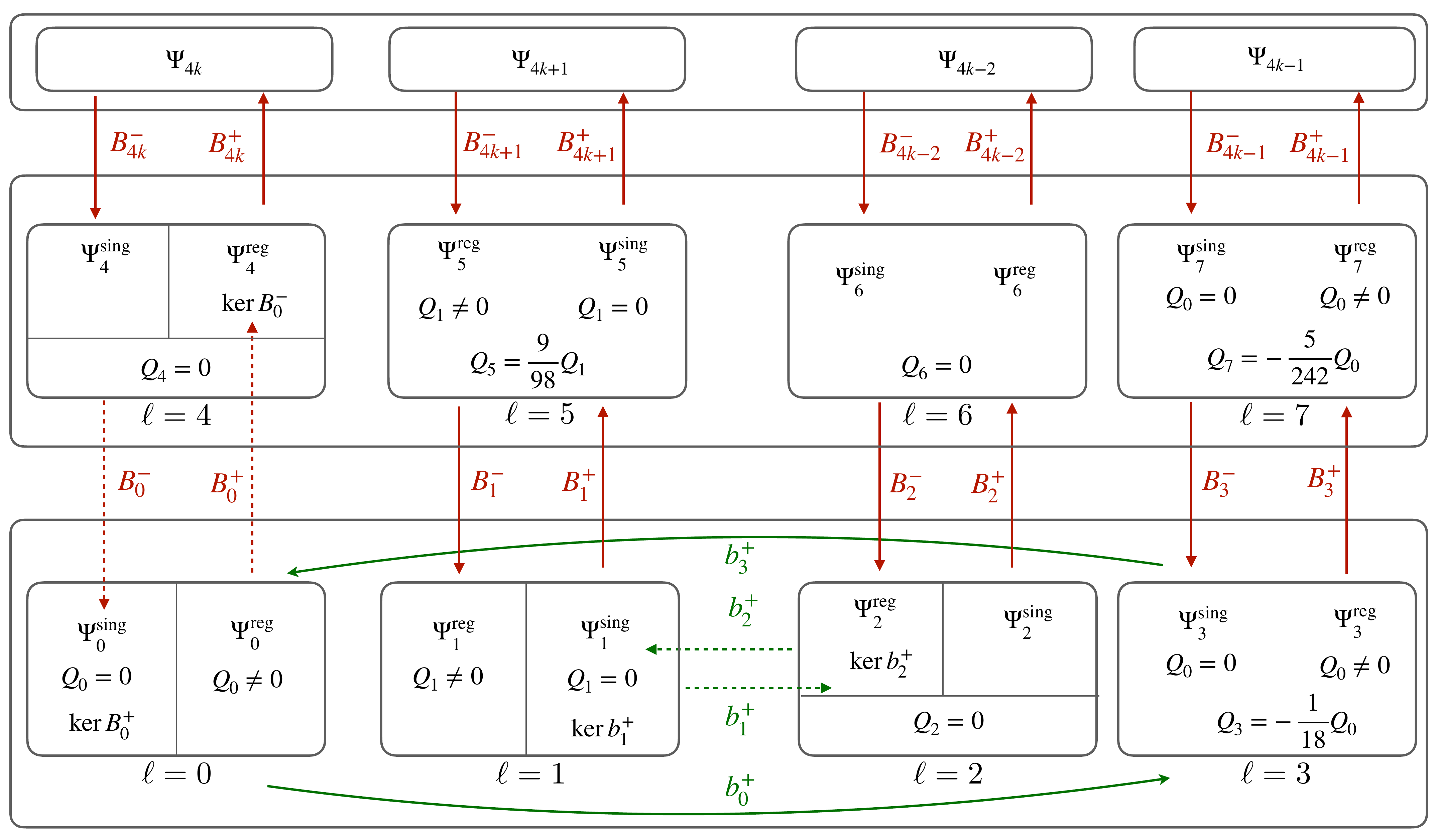}
\caption{This is the analogous diagram of \ref{Fig:PolarLadderQ} for the axial sector. The main difference with the polar sector is that the small ladder also exhibits non-trivial kernels in the axial sector. An analogous discussion applies here for the relation between the charges of the monopole and the dipole and the vanishing magnetisation, this time for the even multipoles.}
\label{Fig:AxialLadderQ}
\end{figure}

\subsubsection{Axial sector}

The story for the axial sector closely resembles the polar sector, although with some differences which, in turn, lead to the absence of asymptotically decaying modes for the physical solutions of even multipoles instead of odd multipoles as occurred in the polar sector. The axial sector also exhibits an obvious conserved charge, but this time for the monopole. The dipole also contains a conserved charge although not so evidently. Let us start with the obvious case.

\vspace{0.5cm}
\noindent\bf{Monopole}\\

The equations \eqref{eq:AxialHyper} with $\ell=0$ can be recast in the form
\be
\frac{\dd}{\dd z}\left[z^{3/4}\sqrt{1-z}\frac{\dd\Psi_{0}}{\dd z}\right]=0
\ee
so we have the conserved quantity generated by \footnote{One can check that this generator satisfies $\left[\mQ_{0},\frac{H_0}{z^{3/2}(1-z)^3}\right]=0$. We can also introduce the coordinate $\dd\rho\equiv\frac{1}{z^{3/4}\sqrt{1-z}}\dd z$ and the Hamiltonian $\tilde{H}_0\equiv z^{3/4}\sqrt{1-z}H_0$ so $[\mQ_0,\tilde{H}_0]=0$ and $\mQ_0$ generates $\rho-$translations.} 
\be
\mQ_{0}[\Psi_0]=z^{3/4}\sqrt{1-z}\Psi_{0}'(z).
\ee
Near the origin we have $\Psi_0\simeq c_0+c_1 z^{3/4}$ with $c_0$ and $c_1$ the singular and regular modes respectively. We then have that $\mQ_0$ exactly annihilates $c_0$ so the physical solution has a non-vanishing charge. We can then write the solution for the monopole in terms of  $Q_0\equiv\mQ_0[\Psi_0]$ to obtain
\be
\psi_0=c_0+\frac{4}{3}Q_0z^{3/4}\;_2F_1\left(\frac12,\frac34,\frac{7}{4}, z\right).
\ee
By acting with the small ladder we can construct the solution for $\ell=3$ as
\be
\psi_3=\bp_0\psi_0,
\ee
and from here we can climb up with the big ladder to generate all the multipoles $\Psi_{4k+3}$. All the involved operators do not have non-trivial kernels, so all these multipoles are nicely connected. On the other hand, if we raise the monopole solution using the big ladder operator directly, we obtain
\be
\psi_4=\Bp_0\psi_1=Q_0 z^{1/4}\sqrt{1-z}.
\ee
We see again here that the raising operator projects out the sector with a trivial monopolar charge (the mode $c_0$). In this case, the mode with the trivial charge corresponds to the singular mode so $\Bp_0$ is able to raise the physical solution and, furthermore, its image also belongs to  the regular sector of $\ell=4$. This situation is analogous to what occurred for the dipole of the polar sector. Once more, we have obtained that the physical mode of the monopole (with a non-trivial charge) is raised to the physical solution for $\ell=4$ that is a purely growing function so it has no decaying mode at infinity. The same reasoning applied for the polar sector then shows that, by raising to higher multipoles, we thus establish that the vanishing of the magnetic susceptibility for the multipoles $\Phi_{4k}$ is associated to a non-trivial charge for the monopole. 

The hierarchy of conserved charges is constructed as in the polar sector. For the two towers connected with the monopole we have
\bea
\mQ_{4k}\Psi_{4k}&\equiv& \mQ_0\Bm_0\cdots\Bm_{4k-4}\Psi_{4k},\\
\mQ_{4k-1}\Psi_{4k-1}&\equiv& \mQ_0\bm_0\Bm_3\cdots\Bm_{4k-5}\Psi_{4k-1}.
\eea
The first tower of charges encounters the same obstruction as in the polar sector due to the fact that the kernel of $\Bm_0$ coincides with the regular solutions of $\ell=4$. Thus, descending with the big ladder in the tower of multipoles $\Psi_{4k}$ we will hit $\ell=4$ where the regular mode cannot descend any further. In addition, the image of $\Bm_0$ is the singular sector of the monopole that has trivial charge, so all the charges $\mQ_{4k}$ will be trivial. Let us emphasise once again, that we can still connect all the physical solutions with the monopole via $\Bp_0$, although the connection is only in one direction. Since the regular mode of the monopole has non-trivial charge, we arrive at the conclusion that the physical modes of the tower $\Phi_{4k}$ originate from the non-trivial charge of the monopole. Furthermore, since $\psi_4$ does not have a decaying mode, we obtain that the physical modes of the multipoles $\Phi_{4k}$ will lack an asymptotically decaying mode and, thus, we establish a link between their trivial charge and the vanishing of the corresponding magnetisation.

Concerning the second tower of charges $\mQ_{4k-1}$, there are no obstructions from the kernels of the ladder operators so this route is two ways. In this case, we can resort to the conservation of the charges to show the presence of decaying modes for the physical solutions. The asymptotically decaying mode behaves as $z^{-\ell/4}$ so $\bm_0\Bm_3\cdots\Bm_{4k-5}\Psi_{4k-1}\sim z^{1/4}$ that is not annihilated by $\mQ_0$, thus showing how the mode with non-trivial charge can have a decaying tail. Since physical solutions have non-trivial charges, these will have decaying modes and, therefore, non-vanishing magnetisation.

\vspace{0.5cm}
\noindent\bf{Dipole}\\

Although less evident, the axial dipolar sector also contains a conserved quantity. This becomes apparent by noticing that the equation \eqref{eq:AxialHyper} (with $\gamma=0$) for $\ell=1$ can be written as
\be
\frac{\dd}{\dd z}\left[z^{1/4}(1-z)^{3/2}\frac{\dd}{\dd z}\frac{\Psi_{1}}{\sqrt{1-z}}\right]=0,
\ee
so that it is immediate identify the following conserved charge:
\be
\mQ_{1}[\Psi_1]\equiv z^{1/4}(1-z)^{3/2}\frac{\dd}{\dd z}\frac{\Psi_{1}}{\sqrt{1-z}},
\ee
Proceeding as before, we can express the dipole in terms of $Q_1\equiv \mQ_1[\Psi_1]$ as
\be
\psi_1=c_1\sqrt{1-z}+\frac{4}{3}Q_1z^{3/4}\;_2F_1\left(\frac14,1,\frac{7}{4}, z\right).
\ee
This expression shows that the non-physical mode has trivial charge, while the physical mode has a non-trivial charge.
We can then go to $\ell=2$ by using the small ladder
\be
\psi_2=\bp_1\psi_1=Q_1 z^{3/4}.
\ee
This in turn reproduces the regular solution and we then see that the non-trivial charge of the monopole generates the regular solution for the quadrupole, as expected since the ladders connect regular modes with regular modes. Then, we can use the big ladder to generate all higher multipoles with $\ell=4k+2$ that, in view of the above expression, will comprise purely growing functions and, therefore, will give rise to vanishing magnetization by virtue of the absence of decaying modes. We have thus recovered the result that all even multipoles have vanishing magnetic susceptibilities and we can ultimately link this property to the nature of the conserved charges of the monopole and the dipole.

As it happened for the polar sector, some ladder operators for the monopole and the dipole have non-trivial kernels and this obstructs the construction of a two way connection of all the higher multipoles with the ladders starting only from those two. The situation is analogous to the case of the polar sector already discussed, so we will spare the details for the axial sector to the reader. The situation is however illustrated in the diagram \ref{Fig:AxialLadderQ}. Let us simply emphasise that the vanishing of the magnetisation for the even multipoles is associated to the vanishing of the charges for the regular solutions, but this time due to the obstruction of non-trivial kernels for both towers.

\def\arraystretch{2}
\begin{table}[]
\resizebox{\textwidth}{!}{
\begin{tabular}{|l||l||l||}
\hline 
& \multicolumn{1}{c||}{Polar sector $\Phi$}                                                                           & \multicolumn{1}{c||}{Axial sector $\Psi$}                  \\\hline\hline 
 Hamiltonian & $ -z(1-z)\left[z(1-z)\partial^2_{z}+\frac{3-z}{4}\partial_z+\frac{\ell(\ell+1)-2}{16}\right]$ & 
$ -z(1-z)\left[z(1-z)\partial^2_z+\frac{1-3z}{4}\partial_z+\frac{\ell(\ell+1)}{16}\right]$

\\\hline\hline 

\multicolumn{1}{|l||}{Susceptibility} & $\alpha_\ell=\frac{\Gamma\left(-\frac{2\ell+1}{4}\right)\Gamma\left(\frac{\ell}{4}\right)\Gamma\left(\frac{\ell+6}{4}\right)}{\Gamma\left(-\frac{\ell+1}{4}\right)\Gamma\left(\frac{5-\ell}{4}\right)\Gamma\left(\frac{2\ell+1}{4}\right)}\rs^{2\ell+1}$ & $
\chi^{\text{BI}}_\ell=
\frac{\Gamma\left(-\frac{2\ell+1}{4}\right)\Gamma\left(\frac{\ell+3}{4}\right)\Gamma\left(\frac{\ell+5}{4}\right)}
{\Gamma\left(-\frac{\ell-2}{4}\right)\Gamma\left(-\frac{\ell-4}{4}\right)\Gamma\left(\frac{2\ell+1}{4}\right)}
\rs^{(2\ell+1)/4}$\\\cline{2-3}

\multicolumn{1}{|l||}{}&$\alpha_{4k+1}=\alpha_{4k-1}=0$&$\chi^{\rm BI}_{4k}=\chi^{\rm BI}_{4k-2}=0$\\

\hline\hline
\multicolumn{1}{|l||}{Charges}&
$\mQ_0=\frac{z^{5/4}}{\sqrt{1-z}}\partial_z\frac{\Phi_0}{z^{1/4}}$& $\mQ_0=z^{3/4}\sqrt{1-z}\partial_z\Psi_0$\\ \cline{2-3}

\multicolumn{1}{|l||}{}&
$\mQ_1=\frac{z^{3/4}}{\sqrt{1-z}}\partial_z\Phi_1$& $\mQ_1=z^{1/4}(1-z)^{3/2}\partial_z\frac{\Psi_1}{\sqrt{1-z}}$

\\ \hline\hline
\multicolumn{1}{|l||}{}             & $\Ap_\ell=-z(z-1)\partial_z-\frac{\ell-1}{4}\left(z-\frac{\ell+5}{2\ell+5}\right)$ & $\Bp_\ell=-z(z-1)\partial_z-\frac{\ell}{4}\left(z-\frac{\ell+2}{2\ell+5}\right)$   \\ \cline{2-3} 
\multicolumn{1}{|l||}{Big Ladder}   & $\Am_\ell= z(z-1)\partial_z-\frac{\ell+6}{4}\left(z-\frac{\ell}{2\ell+5}\right)$   & $\Bm_\ell= z(z-1)\partial_z-\frac{\ell+5}{4}\left(z-\frac{\ell+3}{2\ell+5}\right)$ \\ \cline{2-3} 
\multicolumn{1}{|l||}{}             & $\varepsilon_\ell=\frac{\ell(\ell+6)(\ell+5)(\ell-1)}{16(2\ell+5)^2}$               & $\delta_\ell=\frac{(\ell+5)(\ell+3)(\ell+2)\ell}{16(2\ell+5)^2}$                    \\ \hline\hline
\multicolumn{1}{|l||}{}             & $\ap_\ell=-z(z-1)\partial_z+\frac{\ell+2}{4}\left(z-\frac{\ell-4}{2\ell-3}\right)$    & $\bp_\ell=-z(z-1)\partial_z-\frac{\ell+1}{4}\left(z+\frac{\ell-1}{3-2\ell}\right)$ \\ \cline{2-3} 
\multicolumn{1}{|l||}{Small Ladder} & $\am_\ell=z(z-1)\partial_z+\frac{\ell-5}{4}\left(z-\frac{\ell+1}{2\ell-3}\right)$     & $\bm_\ell= z(z-1)\partial_z-\frac{\ell-4}{4}\left(z+\frac{\ell-2}{3-2\ell}\right)$ \\ \cline{2-3} 
\multicolumn{1}{|l||}{}             & $\varepsilon_\ell= \frac{(\ell-5)(\ell-4)(\ell+2)(\ell+1)}{16(2\ell-3)^2}$           & $\delta_\ell=\frac{(\ell-4)(\ell-2)(\ell-1)(\ell+1)}{16(2\ell-3)^2}$                \\ \hline\hline
\end{tabular}}
\caption{In this table we summarise the main expressions for both the polar and the axial sectors for the pure Born-Infeld theory.}
\label{tab:SummaryLadders}
\end{table}

\subsection{Ladder supersymmetric structure.}
\label{sec:superLadder}

The ladder structure discussed in the previous section can be understood in terms of supersymmetric quantum mechanics (see e.g. \cite{Cooper:1994eh,gangopadhyaya2011supersymmetric} for an introduction). To see how the supersymmetric structure arises we will focus on the polar sector, although the axial sector can be treated in an analogous manner (see Appendix \ref{app:MM}). Let us recall the ladder operators written as
\be
\Am_{\ell}=z(1-z)\partial_z+W_{1,\ell},\quad \Ap_{\ell}=-z(1-z)\partial_z+W_{2,\ell}
\ee
where $W_{1,\ell}$ and $W_{2,\ell}$ are the functions obtained in Sec.~\ref{sec:ladder}, but whose specific form is not relevant here. The important fact is that the ladder operators factorise the Hamiltonian as
\bea
\Am_\ell A^+_\ell&=&H_\ell+\varepsilon_\ell,\\
\Ap_\ell A^-_\ell&=&H_{\ell+4}+\varepsilon_\ell.
\eea
In order to unveil the supersymmetric structure of the system, we introduce a Hamiltonian defined as the direct sum ${\mathcal H}_\ell=(\Ap_\ell \Am_\ell)\oplus(\Am_\ell \Ap_\ell)$, i.e., 
\be
{\mathcal H}_\ell=\begin{pmatrix}
\Ap_\ell \Am_\ell & 0 \\
0 &  \Am_\ell \Ap_\ell
\end{pmatrix}.
\label{eq:SHam}
\ee
When this operator acts on the vector $\phi_\ell=(\varphi_{\ell+4},\varphi_\ell)$, with $\varphi_{\ell}\in\text{Ker}\; H_\ell$ and $\varphi_{\ell+4}\in\text{Ker}\;H_{\ell+4}$, it gives ${\mathcal H}_\ell\phi_\ell=\varepsilon_\ell\phi_\ell$, i.e., $\phi_\ell$ is an eigenvector of ${\mathcal H}_\ell$ with eigenvalue $\varepsilon_\ell$. This allows to introduce the super-charge operators
\begin{equation}
\mathcal{Q}^-_{\ell}=\begin{pmatrix}
0 & 0 \\
\Am_\ell & 0 
\end{pmatrix},\quad \mathcal{Q}^+_\ell=\begin{pmatrix}
0 & \Ap_\ell \\
0 & 0 
\end{pmatrix}
\end{equation} 
that generate the Hamiltonian via anticommutation $\{\mathcal{Q}^-_\ell,\mathcal{Q}_\ell^+\}={\mathcal H}_\ell$ and commute with it $[{\mathcal H}_\ell,\mathcal{Q}^-_\ell]=[{\mathcal H}_\ell,\mathcal{Q}^+_\ell]=0$. These operators also anticommute $\{\mathcal{Q}^-_\ell,\mathcal{Q}^-_\ell\}=\{\mathcal{Q}^+_\ell,\mathcal{Q}^+_\ell\}=0$. Thus, they generate a set of supersymmetric conserved charges at each level $\ell$ and, furthermore, we have that $\big({\mathcal H}_\ell,\mathcal{Q}^-_\ell,\mathcal{Q}^+_\ell\big)$ realise the closed superalgebra $\ssl(1\vert1)$ with $\ham_\ell$ the even sector and $(\mathcal{Q}^-_\ell,\mathcal{Q}^+_\ell)$ the odd sector. Moreover, given a $\mathcal{Q}_\ell$, we can construct the operators of other levels by acting with the ladder operators. The role of this underlying supersymmetry will be unravelled in further work. We will however show how the polar sector can be recast in the form of a Schr\"odinger equation with a paradigmatic supersymmetric potential.

\subsection{P\"oschl-Teller potential for the polar sector}
\label{sec:PoshlTeller}
The nature of the modes can be further studied by having a closer look at the mode equations. We focus on the polar case and the results obtained here will be rederived from another point of view in the appendix \ref{app:xlad}. 
We can rewrite the equation for the polar modes by introducing the rapidity variable
\be
x^2\equiv\sinh\theta
\label{eq:deftheta}
\ee
and redefining the modes as
\be
\Phi_\ell\rightarrow\left(\frac{\sinh(2\theta)}{\tanh^3\theta} \right)^{1/8}\Phi_\ell\,.
\label{eq:rescalingPT}
\ee
After these two transformations, the equation for the perturbations read
\be
-\frac{\dd^2 \Phi_\ell}{\dd\theta^2}-\left(\frac{3}{16\sinh^2\theta}+\frac{2}{\cosh^2\theta}\right)\Phi_\ell=-\left(\frac{2\ell+1}{4}\right)^2\Phi_\ell
\label{newpot}
\ee
where we can recognise the form of a generalised hyperbolic P\"oschl-Teller potential \cite{Cooper:1994eh} so we can in turn identify the solutions of our equations with the bound states of this potential. Let us explore this relation in more detail. The generalised hyperbolic P\"osch-Teller potential corresponds to a 2-parameter family of Hamiltonians $H_{PT}(\alpha,\beta)$ that admit the decomposition $H_{PT}(\alpha,\beta)=A_{\alpha,\beta}^\dagger A_{\alpha,\beta}$ with
\bea
A_{\alpha,\beta}&=&\frac{\dd}{\dd\theta}+W_{\alpha,\beta}(\theta),\\
A_{\alpha,\beta}^\dagger&=&-\frac{\dd}{\dd\theta}+W_{\alpha,\beta}(\theta),
\eea
and the super-potentials
\be
W_{\alpha,\beta}=\alpha\tanh\theta-\frac{\beta}{\tanh\theta},
\ee
for some constants $\alpha$ and $\beta$. The explicit form of the Hamiltonian is then
\be
H_{PT}(\alpha,\beta)=A_{\alpha,\beta}^\dagger A_{\alpha,\beta}=-\frac{\dd^2 }{\dd\theta^2}-\frac{\alpha(\alpha+1)}{\cosh^2\theta}+\frac{\beta(\beta-1)}{\sinh^2\theta}+(\alpha-\beta)^2,
\label{HamPT}
\ee
while its super-symmetric partner is
\be
H^{(s)}_{PT}(\alpha,\beta)=A_{\alpha,\beta}A_{\alpha,\beta}^\dagger=-\frac{\dd^2 }{\dd\theta^2}-\frac{\alpha(\alpha-1)}{\cosh^2\theta}+\frac{\beta(\beta+1)}{\sinh^2\theta}+(\alpha-\beta)^2,
\label{sHamPT}
\ee
which satisfies $H^{(s)}_{PT}(\alpha,\beta)=H_{PT}(-\alpha,-\beta)=H_{PT}(\alpha+1,\beta-1)+4(\alpha-\beta+1)$. If we compare \eqref{HamPT} with our equation \eqref{newpot} we find that we must have $\alpha(\alpha+1)=2$ and $\beta(\beta-1)=-3/16$ to map the equations into a P\"osch-Teller potential. These equations have the solutions $\alpha=1,-2$ and $\beta=1/4,3/4$. The constant term $(\alpha-\beta)^2$ can always be absorbed into a shift of the eigenvalues of the Hamiltonian so we can relate our equations \eqref{newpot} to a set of four different P\"osch-Teller potentials, namely: 
\be
H_{\rm I}\equiv H_{PT}(1,3/4),\quad H_{\rm II}\equiv H_{PT}(1,1/4),\quad H_{\rm III}\equiv H_{PT}(-2,1/4),\quad
H_{\rm IV}\equiv H_{PT}(-2,3/4)\,.
\ee
In terms of these Hamiltonians, we can write the perturbation equations \eqref{newpot} in the following equivalent forms:
\bea
H_{\rm I}\;\Phi_\ell&=&-\frac{\ell(\ell+1)}{4}\Phi_\ell,\\
H_{\rm II}\;\Phi_\ell&=&-\frac{(\ell-1)(\ell+2)}{4}\Phi_\ell,\\
H_{\rm III}\;\Phi_\ell&=&-\frac{(\ell-4)(\ell+5)}{4}\Phi_\ell.\\H_{\rm IV}\;\Phi_\ell&=&-\frac{(\ell-5)(\ell+6)}{4}\Phi_\ell.
\eea
Notice that the vanishing eigenvalue of the first potential corresponds to the monopole while the zeroth-energy eigenvalue of the second one corresponds to the dipole. The third and fourth potentials have vanishing eigenvalue for $\ell=4$ and $\ell=5$ respectively, which are four $\ell-$steps away from the monopole and the dipole, i.e., they are connected via the big ladder. However, let us notice that the ladder operators connecting these multipoles have non-trivial kernels on the space of solutions. We will come back to this point later. In the language of super-symmetric quantum mechanics, the vanishing of the ground state energy is usually referred to as unbroken super-symmetry. The eigenvalues of the generalised P\"osch-Teller potentials can be obtained algebraically by exploiting the super-symmetric structure and are given by\footnote{See e.g. Eq. (245) in \cite{Cooper:1994eh}. In Appendix \ref{app:xlad} we show how to obtain the eigenvalues by algebraic methods.}
\be
E_n(\alpha,\beta)=(\alpha - \beta)^2 - (\alpha - \beta - 2 n)^2=4n(\alpha-\beta-n)
\ee
for integer values of $n$. For our particular cases we then have
\bea
&&E_{{\rm I},n}\equiv E_n(1,3/4)=n(1-4n)=-\frac{\ell(\ell+1)}{4},
\label{eq;EnI}\\
&&E_{{\rm II},n}\equiv E_n(1,1/4)=n(3-4n)=-\frac{(\ell-1)(\ell+2)}{4},\label{eq;EnII}\\
&&E_{{\rm III},n}\equiv E_n(-2,1/4)=-n(9+4n)=-\frac{(\ell-4)(\ell+5)}{4},\label{eq;EnIII}\\
&&E_{{\rm IV},n}\equiv E_n(-2,3/4)=-n(11+4n)=-\frac{(\ell-5)(\ell+6)}{4}.\label{eq;EnIV}
\eea
We can then solve these equations for $\ell$ to obtain:
\bea
{\rm I}&:&\quad\quad\ell=-4n,\qquad\qquad\ell=4n-1,\label{eq:1l(n)I}\\
{\rm II}&:&\quad\quad\ell=1-4n,\qquad\quad\ell=2(2n-1),\label{eq:1l(n)II}\\
\rm III&:&\quad\quad\ell=-(5+4n),\;\;\quad\ell=4(n+1),\label{eq:1l(n)III}\\
{\rm IV}&:&\quad\quad\ell=-2(3+2n),\quad\ell=4n+5.\label{eq:1l(n)IV}
\eea
For each case, we select the physical solutions as those with positive values of $\ell$ when $n=0,1,2,\dots$. Thus, we generate the following series of multipoles:
\bea
{\rm I}&:&\quad\quad\ell=0,3,7,\dots\\
{\rm II}&:&\quad\quad\ell=1,2,6,10,\dots\\
\rm III&:&\quad\quad\ell=4,8,12,\dots\\
{\rm IV}&:&\quad\quad\ell=5,9,13,\dots
\eea
It is interesting to note that these four cases reproduce the structure discussed in the previous section and summarised in Fig. \ref{Fig:PolarLadderQ}. For each sector we retrieve the big ladder with the action of $A_{a}$ and $A_{a}^\dagger$ where $a=$I,II,III and IV. Since these operators relate eigenfunctions with adjacent values of the quantum number $n$, we conclude from the relations \eqref{eq:1l(n)I}-\eqref{eq:1l(n)IV} that they connect multipoles separated by four $\ell-$steps.  The case I starts with the monopole (that gives the ground state of the corresponding Hamiltonian) that is obtained from the first series in \eqref{eq:1l(n)I} for $n=0$. Then we need to jump to the second series for $n>0$ that reproduces $\ell=3$ and all the multiples connected to it by a shift $\Delta \ell=4$. For the Hamiltonian II, the ground state is obtained from the first series and it corresponds to the dipole, while for $n>0$ the physical solutions are obtained from the second series that gives $\ell=2$ and those shifted by $\Delta\ell=4$. For the cases III and IV, the first series are non-physical and only the second ones are admissible, which start at $\ell=4$ and $\ell=5$. We have thus obtained all the possible multipoles. We can see that the modes that are connected by the small ladder correspond in this representation to the first multipoles of the series for I and II. On the other hand, the first multipoles for III and IV correspond to the multipoles which exhibit non-trivial kernels for the big ladder as schematised in Fig. \ref{Fig:PolarLadderQ}. Thus, we have reobtained the result that one has to provide the solutions of four multipoles in order to generate the entire space of solutions for all the multipoles via the ladder operators. In the representation in terms of the P\"osch-Teller potentials, this stems from the four different potentials. We can then understand the obstruction found from the existence of non-trivial kernels for the big ladder operators for the monopole and the dipole in terms of the four different potentials that we need in order to reproduce all the multipoles. 

The existence of a super-potential allows to write the ground state as
\be
\Phi^{(0)}_{\alpha,\beta}=N \exp\left[-\int W_{\alpha,\beta}(\theta)\dd\theta\right]=N \cosh^{-\alpha}\theta \sinh^\beta\theta,
\ee
with $N$ some constant. This state can be easily shown to be annihilated by $A_{\alpha,\beta}$ and, in fact, it arises as the solution of the equation $A_{\alpha,\beta}\Phi^{(0)}_{\alpha,\beta}=0$. Near the origin we obtain $\Phi^{(0)}_{\rm I}\sim \Phi^{(0)}_{\rm IV}\sim \theta^{3/4}$ and $\Phi^{(0)}_{\rm II}\sim \Phi^{(0)}_{\rm III}\sim \theta^{1/4}$. Taking into account the re-scaling \eqref{eq:rescalingPT} and the definition \eqref{eq:deftheta} we see that the ground states of I and IV reproduce the solutions with a regular boundary condition at the origin employed in our computation of the polarisability, while III
 and IV give the singular (constant) mode. This again relates the behaviour at the origin with the vanishing polarisability, since I and IV precisely contain the multipoles with vanishing polarisability. On the other hand, the asymptotic behaviour is $\Phi^{(0)}_{\alpha,\beta}\sim e^{(\beta-\alpha)\theta}$, which means that only $\Phi^{(0)}_{\rm I}$ and $\Phi^{(0)}_{\rm II}$ are normalisable (they have $\beta-\alpha<0$), while $\Phi^{(0)}_{\rm III}$ and $\Phi^{(0)}_{\rm IV}$ are not (they have $\beta-\alpha>0$). In this case, we can relate the normalisability of the ground state with the existence of non-trivial kernels since the ladders with non-trivial kernels correspond to the cases with a normalisable ground state.

Since the supersymmetric partners of the obtained Hamiltonians relate to P\"osch-Teller potentials as 
\be
H^{(s)}_{\rm I}=H_{PT}(2,-1/4), H^{(s)}_{\rm II}=H_{PT}(2,-3/4),H^{(s)}_{\rm III}=H_{PT}(2,-1/4), H^{(s)}_{\rm IV}=H_{PT}(2,-3/4), 
\ee
our perturbation equations \eqref{newpot} can also be expressed in terms of the supersymmetric partners in the following form:
\bea
H^{(s)}_{\rm I}\;\Phi_\ell&=&-\frac{\ell(\ell+1)}{4}\Phi_\ell,\\
H^{(s)}_{\rm II}\;\Phi_\ell&=&-\frac{(\ell-1)(\ell+2)}{4}\Phi_\ell,\\
H^{(s)}_{\rm III}\;\Phi_\ell&=&-\frac{(\ell-4)(\ell+5)}{4}\Phi_\ell.\\
H^{(s)}_{\rm IV}\;\Phi_\ell&=&-\frac{(\ell-5)(\ell+6)}{4}\Phi_\ell.
\eea
so again we can associate the solutions for the multipoles $\Phi_\ell$ to eigenfunctions of these super-symmetric partners. 

We have thus obtained four supersymmetric systems associated to the perturbation equations. The super-symmetric Hamiltonians are given by
\be
{\mathcal H}_{a}=\begin{pmatrix}
H_{a} & 0 \\
0 &  H^{(s)}_{a}
\end{pmatrix}=\begin{pmatrix}
A_{a}^\dagger A_{a} & 0 \\
0 &  A_{a} A^\dagger_{a}
\end{pmatrix}\,,
\ee
and the associated super-charges are
\bea
\mathcal{Q}_{a}=\begin{pmatrix}
0 & 0 \\
A_{a} & 0 
\end{pmatrix},\quad\quad \mathcal{Q}^\dagger_{a}=\begin{pmatrix}
0 & A_{a}^\dagger \\
0 & 0 
\end{pmatrix},
\eea 
with $a=$I, II, III, IV. We then obtain as usual for super-symmetric quantum mechanical systems that $(\Phi_n,A_a\Phi_n)$ are eigenfunctions of $\mathcal{H}_{a}$ with eigenvalues $E_{n,a}$. As explained above, the operator $A_a$ connects adjacent values of $n$ which corresponds to $\Delta\ell=4$ in full analogy with the eigenfunctions $(\phi_{\ell+4},\phi_\ell)$ of \eqref{eq:SHam}, thus showing the full correspondence of both formulations. In the more detailed treatment presented in terms of the P\"osch-Teller potential we have unveiled that the system actually exhibits four supersymmetric structures that endow the multipole equations with four copies of the $\ssl(1\vert1)$ Lie super-algebra. 

The associated super-symmetric quantum system allows to interpret the regularity conditions considered for our computation of the electric polarisability and magnetisation in terms of the normalisability of the wave functions for associated Schr\"odinger equations. Let us also mention that the axial sector also admits a map to P\"osch-Teller potentials and, in that case, Born-Infeld again stands out as a singular theory. Here we  content ourselves with showing how the perturbations for the polar sector can be mapped into the paradigmatic class of super-symmetric Hamiltonians provided by the generalised hyperbolic P\"osch-Teller potentials and a more exhaustive exploitation of the associated super-symmetric quantum system will be presented elsewhere.

\section{Discussion and conclusions}
\label{sec:conc}

Theories with non-linear kinetic interactions have a number of interesting properties one of which is the presence of  screening mechanisms based on derivative self-interactions of the $K-$mouflage type. In this work we have considered the oldest example of this type of screening in the general framework of non-linear electromagnetism. We have obtained the equations governing both polar (electric) and axial (magnetic) static perturbations around spherically symmetric screened objects and shown that the effects of the non-linearities are encoded into the corresponding anomalous propagation speeds and effective masses. Although we have obtained the perturbation equations for general non-linear electromagnetism, we have focused on the class of Born-Infeldised ModMax theories that interpolate between Born-Infeld at small distances and ModMax at large distances. This theory has exact duality invariance, while conformal invariance only arises approximately in the ModMax regime. We have shown that duality invariance leads to a non-trivial relation between the propagation speed of the axial perturbation and the screening factor so that the larger the screening factor the smaller the propagation speed. This points towards a potential strong coupling problem deep inside the screened region, which is in line with the usual strong/weak coupling regimes of dual theories. 

After obtaining the equations for the perturbations we have shown how they can be recast into the form of hypergeometric equations that allow to obtain analytical solutions. The ModMax parameter $\gamma$ only appears in the axial sector, while the equations for the polar sector are oblivious to it. We impose boundary conditions so that the perturbed electric and magnetic fields remain finite at the position of the particle. This is motivated by the regularised behaviour near the particle granted by the Born-Infeld regime that operates in that zone. In the polar sector, we have computed the electric polarisability of the object and we have found that the odd modes above the dipole have vanishing polarisability. For the axial perturbations we compute the magnetic susceptibility that now depends on  $\gamma$. We have analysed the behaviour of the magnetisation and we have shown that some values of $\gamma$ lead to the vanishing of the susceptibility for some multipoles. When reducing to the pure Born-Infeld, the perturbations have a remarkably singular behaviour that leads to the vanishing of the susceptibility for all even modes. For this theory, there is a simple expression that relates the electric polarisability and the magnetic susceptibility. In view of our results, the Born-Infeld electromagnetism emerges as the theory that presents most resistance to deformation by external perturbations. These results are reminiscent of the vanishing of the Love numbers for black holes. Our case is significantly different as there is no horizon around the point charges. We also find that the vanishing polarisabilities and susceptibilities (for Born-Infeld) are not valid for all multipoles but only for odd and even $\ell$'s respectively. Nevertheless one can surmise that there should be some loose analogy as the screening sphere around the point charge could be seen as a fuzzy boundary separating an inside region where the electric field is nearly constant from an outside region where Maxwell's theory applies. In this sense, the vanishing polarisabilities and susceptibilities could be envisaged as properties of the fuzzy ``object" of size the screening radius under external perturbations. We will return to this issue in future work.

As in the  black hole case, the vanishing of the polarisability and the susceptibility can be understood in terms of ladder operators \cite{Hui:2021vcv,BenAchour:2022uqo}. We have unveiled a structure of ladder operators that split into two ladders, namely: a big ladder connecting multipoles separated by $\Delta\ell=4$ and a small ladder that acts as an automorphism between the first four multipoles connecting $\ell\to3-\ell$. This ladder structure further shows the singular nature of Born-Infeld  since it is the only theory that allows for the existence of the ladder in both sectors. Based on the unveiled ladder and the existence of conserved charges for the monopole and the dipole in both sectors, we have constructed a hierarchy of charges for all multipoles. By using these charges and the ladder we have established a relation between the regular solutions relevant for the computation of the polarisability and magnetization and the charges. We have also discussed how the presence of non-trivial kernels for some low-$\ell$ ladder operators obstructs to raise some solutions to higher moments as well as trivialising some charges of high angular momentum. Finally, we have discussed the relation of our results with known results of supersymmetric quantum mechanism. We have written the equations for the perturbations in the form of a supersymmetric system with certain super-charges that, together with the Hamiltonian, realise the $\ssl(1\vert1)$ Lie super-algebra. Furthermore, we have explicitly shown that the equations of the polar sector can be re-written in the form of a Schr\"odinger equation with four paradigmatic P\"oschl-Teller potentials, that represents a classical example of solvable potentials using super-symmetric methods. Borrowing known results on these potentials, we have been able to reproduce the big ladder.

The results obtained in this work call for further exploration to clarify some of the intriguing relations that we have obtained. A study that is worth pursuing is how the ladder operators arise within more general theories of non-linear electromagnetism. Our results suggest that the existence of these ladders is not a generic feature of non-linear electrodynamics. Already our analysis shows that the ladder in the axial sector only seems to exist for the Born-Infeld theory. Although one might be tempted to ascribe it to its duality invariance, this cannot be the answer, since the general Born-Infeldised ModMax theory treated here is also duality invariant but we have not been able to construct an analogous ladder. Rather, it seems that the existence of the ladder structure relies on the absence of birefringence, which is a distinctive feature of Born-Infeld theory and one of the properties that make it the only exceptional non-linear electromagnetism. It would be interesting to provide an alternative characterisation of Born-Infeld theories in terms of admitting a ladder structure. In relation to this, the existence of the ladder structure and the related symmetries may be understood in term of the isometries of the effective metric that governs the dynamics of the perturbations. Another intriguing question concerns the seemingly non-standard ladder that we have obtained conformed by a small ladder and a big ladder. We are not aware of any other system where a similar structure emerges. This particular ladder structure arises in Born-Infeld electromagnetism and it would be interesting to find to what extent it can be extended to arbitrary dimensions and more general set-ups. Furthermore, we have only superficially touched the connection to super-symmetric quantum mechanics, but a deeper exploration would be worthwhile that could, for instance, exploit the non-standard ladder structure from the super-symmetric quantum mechanical system to provide new classes of solvable potentials. Finally, the duality invariance of Born-Infeld points to the possibility of having a related (dual) ladder structure for magnetic backgrounds. We have already commented how this duality can be behind the remarkably simple relations that we have found for both sectors such as the vanishing of the polarisability and the magnetisation for odd and even modes respectively. In this respect, dyons represent very interesting objects in this subject and, hence, exploring their relation to the ladder structures could unveil new phenomena. For instance, selfdual objects might exhibit a stronger resilience to external stimuli with vanishing polarisability and magnetisation for both even and odd modes above the dipole. Along these lines, a more thorough analysis of the role played by duality invariance is desirable as well as an analysis of the symmetries exhibit by the system in relation to the ladders. For instance, finding out how these symmetries relate the quasi-normal modes of both sectors or to what extent the potential problem of strong coupling found here affects the viability of the EFT. We hope to return to these issues in future work.\\


\bf {Acknowledgments:} The authors acknowledge support by Institut Pascal at Université Paris-Saclay during the Paris-Saclay Astroparticle Symposium 2022, with the support of the P2IO Laboratory of Excellence (program “Investissements d’avenir” ANR-11-IDEX-0003-01 Paris-Saclay and ANR-10-LABX-0038), the P2I axis of the Graduate School of Physics of Université Paris-Saclay, as well as IJCLab, CEA, APPEC, IAS, OSUPS, and the IN2P3 master projet UCMN. DB acknowledges support from \textit{Programa II: Contratos postdoctorales} by Salamanca University. JBJ and DB ackowledge support from Project PGC2018-096038-B-I00 and PID2021-122938NB-I00 funded by the Spanish ``Ministerio de Ciencia e Innovaci\'on" and FEDER ``A way of making Europe". This article is based upon work from COST Action CA15117, supported by COST (European Cooperation in Science and Technology).

\appendix


\section{Non-static quadratic action}\label{app:polardeco}

In this appendix,  we will derive  the general quadratic action for non-static perturbations around a screened object. This complements the discussion of the corresponding action as discussed in the main text. For this object we will study the electromagnetic perturbations that we  will split as
\be
\delta A_\mu=(a_0,\vec{a}).
\ee 
The quadratic action can be written as
\be
\mS=\frac12\int\dd^4x\left[\mK_Y(\vec{e}^2-\vec{b}^2)+
2\mK_Z\vec{e}\cdot\vec{b}+\mK_{YY}(\vec{E}\cdot\vec{e})^2+\mK_{ZZ}(\vec{E}\cdot\vec{b})^2+2\mK_{YZ}(\vec{E}\cdot\vec{e})(\vec{E}\cdot\vec{b})
\right]
\ee
where $\vec{e}=\vec{\nabla}a_0-\dot{\vec{a}}$ and $\vec{b}=\vec{\nabla}\times\vec{a}$ are the perturbed electric and magnetic fields. Let us start by considering the parity-preserving case so we will have $\mK_Z=\mK_{YZ}=0$. In that case, we can express the quadratic action as
\be
\mS=\frac12\int\dd^4x\left[\mK_Y(\vec{\nabla}a_0-\dot{\vec{a}})^2-\mK_Y(\vec{\nabla}\times\vec{a})^2+\mK_{YY}(\vec{E}\cdot(\vec{\nabla}a_0-\dot{\vec{a}}))^2+\mK_{ZZ}(\vec{E}\cdot(\vec{\nabla}\times\vec{a}))^2
\right]
\ee
The spherical symmetry of the problem allows us to use  spherical harmonics that provide representations of $SO(3)$.
The temporal component will be decomposed in spherical harmonics
\be
a_0=\sum_{\ell,m}a_{\ell,m}(t,r)Y_{\ell,m}(\theta,\phi)
\ee
while the spatial perturbations will be expanded in vector spherical harmonics
\be
\vec{a}=\sum_{a,\ell,m}a^a_{\ell,m}(t,r)\vec{\mY}^a_{\ell,m}(\theta,\phi)
\ee
with
\be
\vec{\mY}^1_{\ell,m}\equiv Y_{\ell,m}\frac{\vec{r}}{r},\quad\quad
\vec{\mY}^2_{\ell,m}\equiv r\vec{\nabla}Y_{\ell,m},\quad\quad
\vec{\mY}^3_{\ell,m}\equiv\vec{r}\times\vec{\nabla}Y_{\ell,m}.
\ee
The background configuration can be expressed as $\vec{E}=\sqrt{4\pi}E(r)\vec{\mY}^1_{00}$. The gradient of $a_0$ takes the form
\be
\vec{\nabla}a_0=\sum_{\ell,m}\left(a'_{\ell,m}\vec{\mY}^1_{\ell,m}+\frac{a_{\ell,m}}{r}\vec{\mY}^2_{\ell,m}\right)
\ee
so we obtain the perturbed electric field
\bea
\vec{e}=\sum_{\ell,m}\left[
\Big(a'_{\ell,m}-\dot{a}_{\ell,m}^{(1)}\Big)\vec{\mY}^{(1)}_{\ell,m}
+\Big(\frac{a_{\ell,m}}{r}-\dot{a}_{\ell,m}^{(2)}\Big)\vec{\mY}^{(2)}_{\ell,m}
-\dot{a}^{(3)}_{\ell,m} \vec{\mY}^{(3)}_{\ell,m}\right].
\eea
The magnetic field is given by
\be
\vec{b}=\vec{\nabla}\times \vec{a}=-\sum_{\ell,m}\left[
\frac{\LL}{r}a^3_{\ell,m}\vec{\mY}^1_{\ell,m}
+\frac{1}{r}\left(ra^3_{\ell,m}\right)'\vec{\mY}^2_{\ell,m}
+\frac{1}{r}\left(a^1_{\ell,m}-(ra^2_{\ell,m})'\right)\vec{\mY}^3_{\ell,m}
\right].
\ee
We can now compute the projection of the perturbed fields along the direction of the background electric field as:
\bea
\vec{E}\cdot\vec{e}&=&\sqrt{4\pi}E(r)Y_{00}\sum_{\ell,m}\left(a'_{\ell,m}-\dot{a}^1_{\ell,m}\right)Y_{\ell,m}, \\
\vec{E}\cdot\vec{b}&=&\sqrt{4\pi}E(r)Y_{00}\sum_{\ell,m}\frac{\LL}{r}a^{(3)}_{\ell,m}Y_{\ell,m}. 
\eea
Equipped with the above expressions and using the orthogonality relations of the vector spherical harmonics, we can express the quadratic action as
\begin{eqnarray}
\mS&&=\frac12\sum_{\ell,m}\int\dd t r^2\dd r\Big[\Big(\mK_Y+2Y\mK_{YY}\Big)\vert e^{(1)}_{\ell,m}\vert^2+\ell(\ell+1)\mK_Y\vert e_\Omega\vert^2 \nonumber \\ &&-\Big(\mK_Y-2Y\mK_{ZZ}\Big)\vert b^{(1)}_{\ell,m}\vert^2-\ell(\ell+1)\mK_Y\vert b_\Omega\vert^2\Big]
\end{eqnarray}
with $\vert e_\Omega\vert^2=\vert e^{(2)}_{\ell,m}\vert^2+\vert e^{(3)}_{\ell,m}\vert^2$.
In terms of the vector potential components we find
\bea
\mS&&=\frac{2\ell+1}{2}\sum_{\ell}\int\dd t r^2\dd r\left[\Big(\mK_Y+2Y\mK_{YY}\Big)\Big(a'_{\ell}-\dot{a}_{\ell}^{(1)}\Big)^2+\ell(\ell+1)\mK_Y\left(\Big(\frac{a_{\ell}}{r}-\dot{a}_{\ell}^{(2)}\Big)^2+(\dot{a}^{(3)}_{\ell})^2\right)\right.\nonumber\\
&&\left.-\Big(\mK_Y-2Y\mK_{ZZ}\Big)\left(\frac{\LL}{r}a^3_{\ell}\right)^2-\LL \mK_Y\left(\left(\frac{1}{r}\left(ra^3_{\ell}\right)'\right)^2+\left(\frac{1}{r}\left(a^1_{\ell,m}-(ra^2_{\ell,m})'\right)\right)^2\right)\right]\nonumber \\
\eea
 We have used the rotational symmetry of the background to perform the sum over $m$ so we have evaluated at $m=0$ and we have omitted the $m$-dependence to simplify the notation. From this action, we see that, due to the transformation properties under parity, the perturbation $a_\ell^3$ decouples from the rest. We will commence our analysis for this simpler sector.
 
 \subsection*{Axial sector}
 Let us then write down the action for the axial sector
\bea
\mS&&=\frac{2\ell+1}{2}\sum_{\ell}\int\dd t r^2\dd r\left[\ell(\ell+1)\mK_Y\left(\dot{a}^{(3)}_{\ell}\right)^2-\Big(\mK_Y-2Y\mK_{ZZ}\Big)\left(\frac{\LL}{r}a^3_{\ell}\right)^2\right.\nonumber\\
&&\left.
-\LL \mK_Y\left(\frac{1}{r}\left(ra^3_{\ell}\right)'\right)^2\right]\nonumber\\
&&=\frac{1}{2}\sum_{\ell}\int\dd t \dd r\left[\mK_Y\left(\dot{a}_{T}^2-a_T'^2\right)-\Big(\mK_Y-2Y\mK_{ZZ}\Big) \frac{\LL}{r^2}a_{T}^2\right]
\eea
 where we have defined $a_T\equiv\frac{r}{\sqrt{(2\ell+1)\LL}}a^3_{\ell}$. We can now introduce the tortoise coordinate $\dd r_\star\equiv \mK_Y\dd r$ so the action can finally be expressed as
\bea
\mS=\frac{1}{2}\sum_{\ell}\int\dd t \dd r_\star\left[\left(\dot{a}_{T}^2-\frac{1}{\mK_Y}a_T'^2\right)-\Big(\mK_Y-2Y\mK_{ZZ}\Big)\frac{\LL}{r^2(r_\star)}a_{T}^2\right]
\eea
where now the prime stands for $\partial_{r_\star}$. Alternatively, we can canonically normalise $a_T\rightarrow a_T/\sqrt{\mK_Y}$ so the action reads
\bea\label{eq:action2aT}
\mS=\frac{1}{2}\sum_{\ell}\int\dd t \dd r\left[\left(\dot{a}_{T}^2-a_T'^2\right)-m_T^2a_{T}^2\right]
\eea
 with
 \be
 m_A^2\equiv\frac{\LL}{r^2}c_A
^2+\frac14(\partial_r\log \mK_Y)^2+\frac12\partial^2_r \log \mK_Y\,,\qquad c_A^2 = 1-\frac{2Y\mK_{ZZ}}{\mK_Y}
 \ee
the effective mass and sound speed for the perturbation. In the static limit, this action reproduces the equation for $\psi_\ell$, which coincides with $a_T$.
 
 \subsection*{Polar sector}
 Let us now turn to the polar sector. We will fix a gauge with $a^{(2)}_{\ell}=0$. Furthermore, we will introduce an auxiliary field $\phi$ to linearise the action in the non-dynamical field $a_\ell$ so we have
\be
\mS=\frac{2\ell+1}{2}\sum_{\ell}\int\dd t r^2\dd r\left[2\Big(\mK_Y+2Y\mK_{YY}\Big)\Big(a'_{\ell}-\dot{a}_{\ell}^{(1)}-\frac12\phi\Big)\phi+\frac{\LL}{r^2}\mK_Y\left(a_{\ell}^2-(a_\ell^{(1)})^2\right)\right].
\ee
We can obtain the equations for $a_\ell$ and $a_\ell^{(1)}$:
\bea
\LL \mK_Ya_\ell&=&\partial_r\left[r^2\big(\mK_Y+2Y\mK_{YY}\big)\phi\right],\\
\LL \mK_Ya_\ell^{(1)}&=&r^2\big(\mK_Y+2Y\mK_{YY}\big)\dot{\phi}.
\eea
When replacing these solutions into the action, we find
\be
\mS=\frac{2\ell+1}{2}\sum_{\ell}\int\dd t \dd r\left[\left(\frac{r^2\sqrt{\mK_Y} c_P^{-2}}{\sqrt{\LL}}\dot{\phi}\right)^2-\left(\frac{1}{\sqrt{\mK_Y\LL}}\partial_r\Big(r^2\mK_Yc_P^{-2}\phi\Big)\right)^2-r^2\mK_Yc_P^{-2}\phi^2\right],
\ee
with
\be\label{eq:c_polar}
c_P^{-2}\equiv1+\frac{2Y\mK_{YY}}{\mK_Y}.
\ee
We can now introduce the field 
\be
\Phi\equiv\frac{r^2\mK_Yc^2}{\sqrt{\LL}}\phi
\ee
to express the action as
\be
\mS=\frac{2\ell+1}{2}\sum_{\ell}\int\dd t \dd r\left[\frac{1}{\mK_Y}\left(\dot{\Phi}^2-(\partial_r\Phi)^2\right)-\frac{\LL}{r^2\mK_Y}c_P^2
\Phi^2\right].
\ee
Upon canonical normalisation $\Phi\to \mK_Y^{1/2}\Phi$ we finally find
\be\label{eq:action2fi}
\mS=\frac{2\ell+1}{2}\sum_{\ell}\int\dd t \dd r\left[\dot{\Phi}^2-(\partial_r\Phi)^2-m_P^2
\Phi^2\right],
\ee
with
\be
m_P^2\equiv \frac{c_P^2\LL}{ r^2}+\frac{1}{4}(\partial_r\log \mK_Y)^2-\frac12\partial^2_r \log \mK_Y\,,
\ee
and with the speed of sound for the polar perturbation given in \eqref{eq:c_polar}. Again, this equation reproduces the equations for $\Phi_\ell$ obtained in the main text in the static limit.

\section{An alternative approach to the ladder operators}
\label{app:xlad}


In the main text we have uncovered a supersymmetric structure for the perturbation equations. In this appendix, we will give an alternative approach to the same problem. 
In fact static solutions in the axial and polar cases can be obtained as a zero-eigenvalue problem for two Hamiltonian operators.
In the axial case (+) and the polar case (-) we have
$
H_{\pm}\psi_\pm =0 \,,
$
i.e., we are looking for zero modes of 
\be 
H_\pm= -\frac{d^2}{dr^2}+ m_\pm^2 
\ee
with
\be 
m^2_\pm= \frac{\ell (\ell+1) c_\pm^2}{r^2}+ \frac{1}{4}(\partial_r \ln K_Y)^2 \pm \frac{1}{2}\partial^2_r \ln K_Y\,.
\ee
where $c_+=c_A$ and $c_-=c_P$ (see Eq. \eqref{eq:speeds_general}.
In ModMax theories, the two velocities are constant and in the Born-Infeldised ModMax they are simply proportional as can be seen form equations \eqref{eq:ModMaxSpeeds} and \eqref{eq:BIModMaxSpeeds}.

Let us introduce the two supersymmetric operators $A_\pm$ and the associated superpotential $W$
\be
A_\pm= -\frac{d}{dr} \pm W\,,\qquad W=\frac{1}{2}\partial_r \ln K_Y\,,
\ee
Then the two Hamiltonian can be written as
\begin{eqnarray}
&&H_+ = -A_-A_+ + \frac{\ell(\ell+1) c_+^2}{r^2}\nonumber \,,\\
&&H_- = -A_+A_- +\frac{\ell(\ell+1) c_-^2}{r^2}\nonumber \,.
\end{eqnarray}
When $\ell=0$, we see that the two Hamiltonians are supersymmetric conjugates. When $\ell\ne 0$, the property is lost. 

\subsection*{Eigenvalue problem}
It is useful to change coordinates and define the mapping
$
\frac{d r}{dz_\pm}= \frac{r}{c_\pm}
$
This is a differential equation and we focus on the case where the map $r(z_\pm)$ is one-to-one. The coordinate $z$ is different for the two cases as $c_\pm$ are not equal in general. 
We find that
\bea\nonumber
H_+ &=&\frac{c^2_+}{r^2}\left[-\left(B_-- \frac{d\ln \frac{dz_+}{dr}}{dz_+}\right) B_+ +  \ell(\ell+1)\right]\,,\\
H_-&=&\frac{c^2_-}{r^2}\left[-\left(C_--\frac{d\ln \frac{dz_-}{dr}}{dz_-}\right)C_+ + \ell(\ell+1)\right]\,,
\eea
where we have introduced two pairs of new operators
\be
B_\pm = -\frac{d}{dz_+}\pm U\,,\qquad
C_\pm = -\frac{d}{dz_-}\mp V\,,
\ee
as a function of $z_\pm$ respectively where
\be\label{eq:UandV}
U=\frac{1}{2}\frac{d \ln K_Y}{dz_+}\,, \qquad\ V= \frac{1}{2}\frac{d\ln K_Y}{dz_-}\,.
\ee
Notice the change of $\pm$ to $\mp$ between $B_\pm$ and $C_\pm$.
The zero modes are now solutions to
\bea\nonumber\label{eq:eigenvalueequations}
\left(B_--\frac{d\ln \frac{dz_+}{dr}}{dz_+}\right) B_+\psi_+ &=& \ell(\ell+1)\psi_+\,,\\
\left(C_--\frac{d\ln \frac{dz_-}{dr}}{dz_-}\right)C_+\psi_- &=& \ell(\ell+1)\psi_-\,.
\eea
This is simply an eigenvalue problem for two factorised operators. 

A first and natural approach to  the eigenvalue problem is in two steps, i.e., we decompose the eigenvalue problem for the factorised operators in \ref{eq:eigenvalueequations} into two eigenvalue problems. 
So we define the eigenstates and eigenvalues
\begin{align}
&B_+ \phi_+ = \lambda_+ \phi_+\nonumber\,,  && C_+ \varphi_+ = \mu_+ \varphi_+\nonumber \,,\\
&\left(B_-  -\frac{d\ln \frac{dz_+}{dr}}{dz_+}\right)\phi_- = \lambda_-\phi_- \nonumber\,,  && \left(C_-  -\frac{d\ln \frac{dz_-}{dr}}{dz_-}\right)\varphi_- = \mu_-\varphi_- \nonumber\,, \\
\end{align}
where the eigenvalues are not determined and will be specified by imposing that the wave function vanishes at the origin. 
Let us now assume that the pairs of operators are diagonalisable in the same basis of eigenfunctions $\phi_\lambda$ and $\varphi_\mu$ respectively. This implies that in such a basis
\begin{align}
&B_+ \phi_\lambda = \lambda_+ \phi_\lambda\nonumber\,, && C_+ \varphi_\mu = \mu_+ \varphi_\mu\nonumber\,,\\
&\left(B_- -\frac{d\ln \frac{dz_+}{dr}}{dz_+}\right)\phi_\lambda = \lambda_-\phi_\lambda \nonumber\,, &&\left(C_-  -\frac{d\ln \frac{dz_-}{dr}}{dz_-}\right)\varphi_\mu = \mu_-\varphi_\mu \,. \\
\label{eq:eigenvaluediagonal}
\end{align}
In both cases, the pairs of operators in the equations above are diagonalisable in the same basis if they commute. 
This happens when
\be 
U+\frac{1}{2} \frac{d\ln \frac{dz_+}{dr}}{dz_+} = -c_+^{-1}+\frac{d\ln c_+}{dz_+}= u\,,\quad V-\frac{1}{2} \frac{d\ln \frac{dz_-}{dr}}{dz_-}=-c_-^{-1}+ \frac{d\ln c_-}{dz_-}=v \,,
\ee
 with $u$ and $v$ two constants and where we have used that 
$
\frac{d\ln \frac{dz_\pm}{dr}}{dz_\pm}=-c_\pm^{-1}+\frac{d\ln c_\pm}{dz_\pm}.
$
When these two conditions are satisfied, the spectral problem can be easily analysed.

\subsection*{The spectrum of duality invariant theory}

In duality invariant theories (see Sec. \ref{sec:duality}) we have the condition
\be \label{eq:duality_eigenmodes}
c_+ K_Y=1
\ee
which allow to rewrite the operators $U$ and $V$ in Eq. \eqref{eq:UandV} as a function of $c_+$. Hence, the pair of operators $B_+$ and $ B_-  -\frac{d\ln \frac{dz_+}{dr}}{dz_+}$ commute when $c_+$ is, or can be treated as a, constant. For the ModMax theories, the two speeds $c_\pm$ are constant and therefore the two pairs of operators can be diagonalised in the same basis.
For the  Born-Infeldised ModMax theories, the two speeds are proportional and therefore the two pairs of operators commute when $c_\pm$ is nearly constant, i.e., around the origin and at infinity. In this case, we will generalise the setting and allow for space-dependent eigenvalues. This will allow us to analyse the spectrum in terms of new supersymmetric operators.

\subsubsection*{ModMax}
\label{app:MM}
The ModMax models are   duality invariant and such that  $c_\pm$ are constant. Hence, we can diagonalise the pairs of operators simultaneously, i.e.,
\begin{align}
& -\frac{d \phi_\lambda}{dz_+} = \lambda_+ \phi_\lambda\nonumber\,,  && -\frac{d\varphi_\mu}{dz_-} = \mu_+ \varphi_\mu\nonumber \,,\\
& \left(-\frac{d}{dz_+}+c_+^{-1}\right)\phi_\lambda = \lambda_-\phi_\lambda \nonumber \,, &&\left(-\frac{d}{dz_-}  +c_-^{-1}\right)\varphi_\mu = \mu_-\varphi_\mu \,,\nonumber\\
\end{align}
from which we deduce that the eigenvalues are such that
\be 
\lambda_-= \lambda_+ +c_+^{-1}\,,\qquad \mu_-= \mu_+ + c^{-1}\,.
\ee
Then, from \eqref{eq:eigenvalueequations} we need to solve the pair of quadratic equations
\be 
\lambda_+(\lambda_+ +c_+^{-1})= \ell(\ell+1)\,,\qquad \mu_+( \mu_+ + c_-^{-1})= \ell(\ell+1)
\ee
corresponding to the eigenmodes with $z_\pm = c_\pm \ln r$
\be 
\phi_+= \alpha_+ e^{-\lambda_+^{+}z_+} + \alpha_- e^{-\lambda_+^{-}z_+}\,,\quad \varphi_+= \beta_+ e^{-\mu_+^{+}z_-} + \alpha_- e^{-\mu_+^{-}z_-}\,,
\ee
where we have
\be 
\lambda_+^\pm= \frac{-1 \pm \sqrt{ 1+4 c_+^2 l(l+1)}}{2c_+}\,,\quad \mu_+^\pm= \frac{-1\pm  \sqrt{1+4 c_-^2 l(l+1)}}{2c_-}\,,
\ee
which coincides with Eq. \eqref{eq:ModMax_solutions} taking into account the relation between $r$ and $z_\pm$ variables. 

\subsubsection*{Generalised eigenvalue problem}
In this more complex family of models,  the two eigensystems \eqref{eq:eigenvaluediagonal} are now
\begin{align}
&\left(-\frac{d }{dz_+}+U\right)\phi_\lambda = \lambda_+ \phi_\lambda\nonumber \,, && \left(-\frac{d}{dz_-}-V\right) \varphi_\mu= \mu_+ \varphi_\mu\nonumber\,,\\
&\left(-\frac{d}{dz_+}+U +c_+^{-1}\right)\phi_\lambda = \lambda_-\phi_\lambda. \,, &&\left(-\frac{d}{dz_-}+3V +c_-^{-1}\right)\varphi_\mu = \mu_-\varphi_\mu \,. \nonumber \\
\label{eq:tric1}
\end{align}
where we have used the duality invariance of the theory Eq. \eqref{eq:duality_eigenmodes} and the fact that
for theories like the Born-Infeldised ModMax one the ratio $c_+/c_-$ is constant, see Eq. \eqref{eq:BIModMaxSpeeds}.
We can immediately see that when $c_\pm $ are constant, the spectrum can be obtained in the same way as already explained  for the ModMax model.

When the velocities $c_\pm$ are not constant anymore, we can in fact adapt the method to find exact solutions by simple integration. This is achieved by 
requesting that the eigenvalues become radius dependent instead of constant. In a sense this method ressembles the variation of the constant way of solving first order differential equations applied to second order differential equations with factorised operators. In the following we will obtain new differential equations for the eigenvalues which are exact and valid for any duality invariant theories for which $c_+/c_-$ is constant. Eventually these equations will be equivalent to Schr\"odinger equations for supersymmetric operators which will be directly related to the P\"oschl-Teller potentials obtained in the main text.

So we impose that $\lambda_\pm$ and $\mu_\pm$ become functions of space. 
First of all we have the identities
\be 
\lambda_-= \lambda_+ +c_+^{-1}\,,\quad  \mu_-= \mu_+ +4V + c_-^{-1}\,,
\ee
which are still valid even when the eigenvalues are space-dependent.  The eigenmode equations become now  a pair of differential equations for the eigenvalues 
\bea\nonumber
&&-\frac{d\lambda_+}{dz_+} + \left(\lambda_+ +c_+^{-1}\right)\lambda_+ = \ell(\ell+1)\,,\\
&&-\frac{d\mu_+}{dz_-} + \left(\mu_+ +4V + c_-^{-1}\right)\mu_+ = \ell(\ell+1)\,,
\label{eq:eigenvaluesdiffeq}
\eea
where we have used Eq.~\eqref{eq:tric1}.
Once these equations have been solved, the modes themselves are simply obtained by integration
$
\phi_\lambda \propto e^{\int dz_+ (U-\lambda_+)}
$
and 
$
\varphi_\mu \propto e^{\int dz_- -(V+\mu_+)}.
$
In general there are two solutions for $\lambda_+$ and $\mu_+$ implying two solutions for the modes. As the space of solutions is a vector space of dimension two, this is enough to obtain the complete solutions. So we can write the modes as
\be 
\psi_+= a_+  e^{\int dz_+ (U-\lambda_+^+)}+ a_- e^{\int dz_+ (U-\lambda_+^-)}\,,
\label{mood1}
\ee
where $\lambda_+^\pm$ are the two solutions to the eigenvalue problem and $a_\pm$ are constant coefficients. Similarly we have
\be 
\psi_-= b_+  e^{\int dz_- -(V+\mu_+^+)}+ b_- e^{\int dz_- -(V+\lambda_-^-)}\,,
\label{mood2}
\ee
where $\mu^\pm_+$ are also the two eigenvalues.

\subsubsection*{Supersymmetric eigenvalue problem}

The two differential equations for the eigenvalues~\eqref{eq:eigenvaluesdiffeq} satisfy a Riccati equation which can be linearised by defining
$ 
\lambda_+=- {\frac{d \ln l_+}{dz_+}},\  \mu_+= - {\frac{d \ln m_+}{dz_-}}
$
which gives two second order and linear differential equations
\bea
&& \frac{d^2 l_+}{d^2z_+}- c_+^{-1} \frac{d l_+}{dz_+}- \ell(\ell+1) l_+=0 \,,\\
&&\frac{d^2 m_+}{d^2z_-}- (4V +c_-^{-1}) \frac{d m_+}{dz_-}- \ell(\ell+1) m_+=0\,.
\eea
By further redefining the functions as
$ 
l_+= u_+ f_+,\ \ m_+= u_-f_-,
$
where
$ 
u_+=e^{\frac{1}{2}\int dz_+ c_+^{-1}}, u_-=e^{\frac{1}{2}\int dz_- (c_-^{-1}+4V})
$
and introducing
\be
W_+=\frac{1}{2c_+}\,,\qquad  W_-= \frac{1}{2c_-}+ 2V\,,
\ee
it is possible to write the eigenvalue equations as Schr\"odinger equations
\be 
-\frac{d^2 f_\pm }{dz_\pm^2} + \left( \ell(\ell+1) + W_\pm^2 - \frac{dW_\pm}{dz_\pm}\right)f_\pm=0\,.
\ee
We recognise two pairs of supersymmetric quantum mechanics problems 
and the associated supersymmetric operators 
\be 
Q_\pm= -\frac{d}{dz_\pm} - W_\pm \,, \qquad 
Q^\dagger_\pm= \frac{d}{dz_\pm} - W_\pm \,.
\ee
The Hamiltonian is given in terms of the supersymmetric operators
\be 
{\cal H}_\pm= Q^\dagger_\pm Q_\pm\,,
\ee
such that 
\be 
{\cal H}_\pm f_\pm = -\ell(\ell+1) f_\pm\,,
\ee
i.e., we are looking for bound states of supersymmetric quantum mechanics. We will see explicitly below how this is linked to the P\"oschl-Teller potentials in the Born-Infeld case. 

\subsubsection*{Born-Infeldised ModMax}

Let us now focus on models like the Born-Infeldised ModMax. 
In this case we have
\be 
c_+= e^{-\gamma}\frac{x^2}{\sqrt{1+x^4}}\,,\qquad c_-=\frac{x^2}{\sqrt{1+x^4}}\,,
\ee
so the relation between the $z$ and radial variable $r$ is
\be 
\frac{dz_+}{dx}= e^{-\gamma}\frac{x}{\sqrt{1+x^4}}\,,
\ee
where $x = r/r_{\rm s}$,  $r_{\rm s}$ being the screening radius.
The last equation can be integrated 
 which allows to obtain the velocities as
\be 
c_+= e^{-\gamma}\tanh (2 e^{\gamma} z_+)\,, \quad c_-= \tanh (2 z_-)\,,
\ee
where the variables $z_\pm$ play an analogous role to rapidities 
and the superpotentials $W_\pm$ 
as
\be
W_+ = \frac{e^\gamma}{2} \frac{1}{\tanh(2e^\gamma z_+)}\,,\qquad W_- = -\frac32 \frac{1}{\tanh(2z_-)} + 2 \tanh(2z_-)\,.
\ee
We can now study the spectrum of the Born-Infeldised ModMax theories. To do so, 
we shall introduce a  family of supersymmetric quantum mechanics which generalises the usual reflectionless models.

\subsubsection*{Natanzon potentials}

As a mathematical aside, let us notice that the two potentials $W_\pm$ belong to the general family of superpotentials
\be 
{\cal W}_{a,b}= \frac{a}{t} + b t\,,\qquad t=\tanh \kappa z\,.
\ee
The Born-Infeldised ModMax superpotentials can be obtained by setting
$
a_+=\frac{e^\gamma}{2},\ b_+=0, \ a_-=\frac{1}{2}-\kappa_-, \ b_-=\kappa_-
$
and 
$
\kappa_+=2e^\gamma, \ \kappa_-=2
$. 
It is then possible to define the  ladder operators 
\be 
Q_{a,b} = -\frac{d}{dz}- {\cal W}_{a,b}\,,\quad  Q^\dagger_{a,b}= \frac{d}{dz}- {\cal W}_{a,b}\,,
\ee 
such that
\be \label{eq:HamNat}
{\cal H}_{a,b}\equiv Q^\dagger_{a,b}Q_{a,b}=-\frac{d^2}{dz^2}+ {\cal W}_{a,b}^2 -\frac{d{\cal W}_{ab}}{dz}=-\frac{d^2}{dz^2}+ V_{a,b}\,,
\ee
where we have introduced the potential $V_{a,b}$ that has the explicit form
\be 
V_{a,b}=  (a+b)^2 + \frac{a(a+\kappa)}{s^2}-\frac{b(b+\kappa)}{c^2}\,,
\ee
where $c=\cosh \kappa z$ and $s=\sinh \kappa z$. 

There is an interesting set of symmetries enjoyed by this potential. These symmetries correspond to changing $a\to -a-\kappa$ and $b\to -b-\kappa$. These transformations only change the potential by a constant and we can obtain the following families of related potentials:
\bea
V_{a,b}&=&V_{-a-\kappa,b}+(2b-\kappa)(2a+\kappa),\\
V_{a,b}&=&V_{a,-b-\kappa}+(2a-\kappa)(2b+\kappa),\\
V_{a,b}&=&V_{-a-\kappa,-b-\kappa}-4\kappa(a+b+\kappa).
\eea
which allows to obtain eigenvectors of the Hamiltonian defined by $V_{a,b}$ from eigenvectors of the related potentials with the corresponding substitutions. In the following we denote by
$
\tilde a= -\kappa -a, \ \tilde b= -\kappa -b
$
and we obtain the new eigenvalues $\tilde c(a,b)$ from the eigenvalues $c(a,b)$ of ${\cal H}_{a,b}$. We then find the following families of eigenvalues:
\bea
c_{\rm I}(a,b)&=&c(a,b),\\
c_{\rm II}(a,b)&=& c(\tilde a, b) + (2b-\kappa)(2a+\kappa)\,,\\
 c_{\rm III}(a,b)&=& c(\tilde a,\tilde b) - 4\kappa(\kappa+a+b)\,,\\
c_{\rm IV}(a,b)&=& c(a,\tilde b)+ (2a-\kappa)(2b+\kappa)\,.
\eea
This constructs four sets of eigenvalues and eigenvectors for ${\cal H}_{a,b}$.

For  arbitrary values of $a$ and $b$, we will find  eigenvalues and eigenvectors for bound states of the Hamiltonian ${\cal H}_{a,b}$.
From Eq. \eqref{eq:HamNat}
we obtain  the explicit ladder identity
\be 
Q_{a,b}Q_{a,b}^\dagger= 4\kappa(a+b-\kappa) + {\cal H}_{a-\kappa,b-\kappa}\,.
\label{recur1}
\ee
Notice that the action of $Q_{a,b}Q^\dagger_{a,b}$ lowers the parameters of the Hamiltonian by $\kappa$. 

As usual in supersymmetric systems, we introduce the vacuum state as being in the kernel of the supersymmetric operator $Q_{\alpha,\beta}$ for a given choice of the  indices $(\alpha,\beta)$. Here we introduce the vacuum state by the property
\be 
Q_{a-n\kappa,b-n\kappa}\vert f_0^{\rm I}\rangle =0\,,
\ee
where $n$ is an  integer which is not specified yet. 
Explicitly the  wave function reads
$ 
f_0^{\rm I}(x)= e^{-\int dx {\cal W}_{a-n\kappa,b-n\kappa}(x)}\,.
$
We can also introduce the 
excited states using the ladder operators
\be 
\vert f_n^{\rm I}\rangle= Q^\dagger_{a,b} Q^\dagger_{a-\kappa,b-\kappa}\dots Q^\dagger_{a -(n-1)\kappa,b-(n-1)\kappa}\vert f_0^{\rm I}\rangle\,.
\ee
Using the recursion relation (\ref{recur1}) we find that this excited state is an eigenstate of ${\cal H}_{a,b}$, i.e.  we have
\be 
{\cal H}_{a,b}\vert f_n^{\rm I}\rangle  = c_n(a,b) \vert f_n^{\rm I}\rangle\,,
\ee
with 
\be 
c_n(a,b)= 4\kappa\sum_{j=0}^{n-1}( (a-j\kappa) + (b-j\kappa) -\kappa)=4n\kappa (a+b-\kappa n)\,.
\ee
Finally notice that  the wave function  is given explicitly by
\be 
f_n^{\rm I}(z)= (\vert \sinh \kappa z\vert )^{-(a-n\kappa)/\kappa}
( \cosh \kappa z)^{-(b-n\kappa)/\kappa}\,,
\ee
which is an even function. As the ladder operators are odd, the excited states are either odd or even depending on the parity of $n$.
Using the constancy of the Wronskian of the mode equation, we can always construct a  second independent solution as
\be 
\tilde f_n^{\rm I}(z) = D f_n^{\rm I} (z)  \int dz_\pm' (f_n^{\rm I}(z'))^{-2} dz'
\ee
where $D$ is the Wronskian. 
The two set of functions $f_n^{\rm I}=f_n$ and $\tilde f_n^{\rm I}$ form a basis for the space of solutions.

We can now then construct three more series of eigenvectors and eigenvalues defined by
\bea 
Q_{\tilde a-n\kappa, b-n\kappa}\vert  f_0^{\rm II}\rangle &=&0\,,\\
Q_{\tilde a-n\kappa,\tilde b-n\kappa}\vert f_0^{\rm III}\rangle &=&0\,,\\
Q_{ a-n\kappa,\tilde b-n\kappa}\vert  f_0^{\rm IV}\rangle &=&0\,,
\eea
where $n$ is an  integer which is not specified yet. Analogously, we can also introduce the
excited states using the ladder operators constructed out of the tilded quantities
\bea
\vert  f_n^{\rm II} \rangle&=& Q^\dagger_{\tilde a, b} Q^\dagger_{\tilde a-\kappa, b-\kappa}\dots Q^\dagger_{\tilde a -(n-1)\kappa, b-(n-1)\kappa}\vert  f_0^{\rm II}\rangle\,,\\
\vert  f_n^{\rm III}\rangle &=& Q^\dagger_{\tilde a,\tilde b} Q^\dagger_{\tilde a-\kappa,\tilde b-\kappa}\dots Q^\dagger_{\tilde a -(n-1)\kappa,\tilde b-(n-1)\kappa}\vert \tilde f_0^{\rm III}\rangle\,,\\
\vert  f_n^{\rm IV}\rangle &=& Q^\dagger_{\tilde a,\tilde b} Q^\dagger_{\tilde a-\kappa,\tilde b-\kappa}\dots Q^\dagger_{\tilde a -(n-1)\kappa,\tilde b-(n-1)\kappa}\vert  f_0^{\rm IV}\rangle\,,
\eea
whose eigenvalues are simply
\bea
c_{\rm II}(a,b)&=& -(2a+\kappa(2n+1))(-2b+\kappa(2n+1))\,,\\
 c_{\rm III}(a,b)&=&-4\kappa (n+1) (a+b +(n+1)\kappa)\,,\\
c_{\rm IV}(a,b)&=&(2a-\kappa(2n+1))(2b+\kappa(2n+1))\,.
\eea
As a result we have four ladders of eigenstates for the Natanzon potentials.

\subsubsection*{The eigenvalues of the Born-Infeldised ModMax theories}
We can now apply this formalism to the Born-Infeldised Mod-Max theories. In this case we have
$
a_++b_+=\frac{e^\gamma}{2}, \  a_-+ b_-= \frac{1}{2}
$
and we get the relation between the eigenvalues
\be 
c_{I,II,III,IV}(a_+,b_+)= e^{2\gamma}c_{I,II,II,IV}(a_-,b_-)
\ee
where the ones for the polar case are simply
\bea
c_{\rm I}(a_-,b_-)&=&4n(1-4n)\,,\\
c_{\rm II}(a_-,b_-) & =& -2(2n-1)(4n-1)\,,\\
c_{\rm III}(a_-,b_-)&=& -4(n+1)(4n+5)\,,\\
c_{\rm IV}(a_-,b_-) &=& -2(2n+3)(4n+5)\,,
\eea
which solves the eigenvalue problem with
\bea
{\rm I}&:&\quad\quad\ell=-4n,\qquad\qquad\ell=4n-1,\label{eq:l(n)Ib}\\
{\rm II}&:&\quad\quad\ell=1-4n,\qquad\quad\ell=2(2n-1),\label{eq:l(n)IIb}\\
\rm III&:&\quad\quad\ell=-(5+4n),\;\;\quad\ell=4(n+1),\label{eq:l(n)IIIb}\\
{\rm IV}&:&\quad\quad\ell=-2(3+2n),\quad\ell=4n+5.\label{eq:l(n)IVb}
\eea
For each case, we select the physical solutions as those with positive values of $\ell$ when $n=0,1,2,\dots$. Thus, we generate the following series of multipoles:
\bea
{\rm I}&:&\quad\quad\ell=0,3,7,\dots\\
{\rm II}&:&\quad\quad\ell=1,2,6,10,\dots\\
\rm III&:&\quad\quad\ell=4,8,12,\dots\\
{\rm IV}&:&\quad\quad\ell=5,9,13,\dots
\eea
We have therefore obtained the spectrum of the Born-Infeldised Mod-Max theories in agreement with the results presented in section \ref{sec:PoshlTeller}.

Let us comment briefly on the behaviour of the solutions close to the origin. We have explicitly 
\be 
f_n^{I,IV}(z_-;a_-,b_-) \sim z_-^{-a_-/\kappa_-}
\ee
and 
\be 
f^{II,III}_n(z_-;a_-,b_-)\sim z^{-\tilde a_-/\kappa_-}.
\ee
where
$
\frac{a_-}{\kappa_-}= -\frac{3}{4},\ \frac{\tilde a_-}{\kappa_-}=-\frac{1}{4} .
$
The associated  solutions  (\ref{mood2}) scale like
\be 
\psi_{-n}^{I,II,III,IV}(z_-)\sim {c_-^{1/2}}{u_- f_{n}^{I,II,III,IV}}
\ee
where $c_-\simeq z_-$ and $u_-\sim z_-^{-3/4}$ implying
\be 
\psi_{-n}^{I,IV}(z_-)\sim z^{1/2}, \ \psi_{-n}^{II,III}(z_-)\sim {\rm constant}.
\ee
As $z_-\sim x_-^2$, we retrieve that the modes vanish linearly at the origin or are constant. This selects the spectrum $I$ and $IV$ as the physical ones. This corresponds to the odd values of $\ell$ and corresponds to the vanishing polarisabilities.


Let us now turn to the axial case. In this case the eigenvalue problem can be rewritten as
\be 
c_{I,II,III,IV}(a_-,b_-)=-\ell(\ell+1)
\ee
or equivalently
\be 
\ell_{\rm eff}= c_{I,II,III,IV}(a_-,b_-).
\ee
We only consider the cases I and IV as they lead to regular solutions. In the case I
by putting $n=m-1$ we retrieve $ \ell_{\rm eff}= 4(4m^2-9m+5)$ which is one of the series of values where the susceptibility vanishes. This can be achieved only when
\be 
\gamma_{m,\ell}=\frac{1}{2}\ln \frac{\ell(\ell+1)}{4(m-1)(4m-5)}
\ee
 The second series of vanishing susceptibilities are obtained by solving in the case IV
 for $n=m-2$ giving
 \be 
 \gamma_{m,\ell}=\frac{1}{2}\ln \frac{\ell(\ell+1)}{2(2m-1)(4m-3)}
 \ee
 as found in the main text.

\bibliography{BibKmouflage}

\end{document}